\documentclass[useAMS,usenatbib]{mnras}

\usepackage{mathptmx}
\usepackage{mathtools}
\usepackage[T1]{fontenc}
\usepackage{txfonts}
\usepackage{ae,aecompl}
\usepackage{soul}
\usepackage{array}
\usepackage{amssymb}
\usepackage{pdflscape}
\usepackage{CJKutf8}

\usepackage[T1]{fontenc}
\usepackage{ae,aecompl}
\usepackage{soul}
\usepackage{indentfirst}
\usepackage{array}
\usepackage{times}

\usepackage{float}
\usepackage{graphicx}	

\usepackage{amsmath}	
\usepackage{amssymb}	
\usepackage{threeparttable}
\usepackage{subfigure}
\usepackage[all]{hypcap}

\usepackage[usenames]{color}
\usepackage{pdflscape}
\def\mnras{MNRAS}
\def\apjl{ApJL }
\def\aj{AJ }
\def\apj{ApJ }
\def\pasj{PASJ}
\def\pasp{PASP }

\def\apjs{ApJS }
\def\araa{ARAA }
\def\aap{A\&A }

\hypersetup{colorlinks=true,citecolor=blue,linkcolor=blue,filecolor=black,runcolor=black,breaklinks=true}
\definecolor{purple}{RGB}{160,32,240}
\definecolor{Cerulean}{RGB}{0,123,167}
\usepackage{etoolbox}

\newcommand{\mbh}{$M_{\bullet}$}

\newcommand{\fdime}{$f(\eta_\mathrm{rad} > 0.1)$}
\newcommand{\fcent}{$f(\eta_\mathrm{rad} > 0.01)$}
\newcommand{\Msun}{M_{\odot}}

\newcommand{\mpeak}{$M_\mathrm{peak}$}

\newcommand{\shadedregions}{The shaded regions show the 68\% confidence intervals inferred from the model posterior distribution.}
\newcommand{\halocurves}{The white solid lines are the average mass growth curves of haloes with $M_{\rm peak}=10^{12},10^{13},10^{14}$, and $10^{15} M_{\odot}$ at $z=0$.}
\newcommand{\smhm}{$M_*$--$M_\mathrm{peak}$}
\newcommand{\bhbm}{$M_\bullet$--$M_\mathrm{bulge}$}
\newcommand{\bmsm}{$M_\mathrm{bulge}$--$M_*$}
\newcommand{\bhsm}{$M_\bullet$--$M_*$}

\newcommand{\eff}{$\epsilon_\mathrm{tot}$}

\newcommand{\sfcurve}{The yellow dashed line shows the halo mass at which the galaxy star-forming fraction $f_\mathrm{SF}$ is $0.5$ as a function of $z$.}

\title[\textsc{Trinity}: QLF decomposition from $z=0-10$]{\textsc{Trinity} III: Quasar Luminosity Functions Decomposed by Halo, Galaxy, and Black Hole Masses as well as Eddington ratios from $z=0-10$}

\author[H. Zhang et al.]{
Haowen Zhang (\begin{CJK*}{UTF8}{gbsn}
张昊文
\end{CJK*}),$^{1}$\thanks{E-mail: hwzhang0595@email.arizona.edu}
Peter Behroozi,$^{1,2}$
Marta Volonteri,$^{3}$
Joseph Silk,$^{3,4,5}$
\newauthor{
Xiaohui Fan,$^{1}$
James Aird$^{6,7}$,
Jinyi Yang (\begin{CJK*}{UTF8}{gbsn}
杨锦怡
\end{CJK*}),$^{1,\ast}$
and Philip F. Hopkins$^{8}$
}
\\
$^{1}$University of Arizona, 933 N Cherry Ave., Tucson, AZ 85721, USA, \\
$^{2}$Division of Science, National Astronomical Observatory of Japan, 2-21-1 Osawa, Mitaka, Tokyo 181-8588, Japan\\
$^{3}$Institut d'Astrophysique de Paris (UMR 7095: CNRS \& Sorbonne Universite), 98 bis Bd. Arago, F-75014, Paris, France\\
$^{4}$Department of Physics and Astronomy, Johns Hopkins University, Baltimore, MD 21218, USA\\
$^{5}$BIPAC, Department of Physics, University of Oxford, Keble Road, Oxford OX1 3RH, UK\\
$^{6}$Institute for Astronomy, University of Edinburgh, Royal Observatory, Edinburgh EH9 3HJ, UK\\
$^{7}$Department of Physics and Astronomy, University of Leicester, University Road, Leicester LE1 7RH, UK\\
$^{8}$Theoretical Astrophysics, California Institute of Technology, Pasadena, CA 91125, USA\\
$^{\ast}$Strittmatter Fellow
}

\date{Accepted XXX. Received YYY; in original form ZZZ}

\pubyear{2021}

\begin{document}
\label{firstpage}
\pagerange{\pageref{firstpage}--\pageref{lastpage}}
\maketitle

\begin{abstract}
We present the redshift evolution of quasar luminosity functions decomposed by halo mass, galaxy mass, supermassive black hole (SMBH) mass, and Eddington ratio, as well as SMBH kinetic/radiative energy output ratios from \textsc{Trinity}, a flexible empirical model that self-consistently infers the halo--galaxy--SMBH connection that match observational data. Key findings include: 1) The normalization of QLF increases by $\sim 3-4$ dex from $z\sim10$ to $z\sim4$, due to the fast mass build-up of different SMBH populations; 2) From $z\sim4$ to $z\sim 1$, less massive galaxies and SMBHs make up bigger and bigger fractions of QLFs, due to the AGN downsizing effect; 3) At $z\sim 0$, massive haloes/galaxies/SMBHs are responsible for most bright quasars due to low Eddington ratios among all SMBHs; 4) The bright ends of quasar luminosity functions (QLFs) are dominated by SMBHs that are at least 0.3 dex over-massive relative to the median SMBH mass--galaxy mass relation;  5) QLFs at $z\sim 6-7$ are dominated by SMBHs accreting at Eddington ratios $0.1 < \eta_\mathrm{rad} < 1$, but super-Eddington AGNs contribute more significantly to QLFs towards $z\sim 9-10$.
\end{abstract}




\section{Introduction}
\label{s:introduction}

Supermassive black holes (SMBHs) are found in the centres of many galaxies \citep{Kormendy1995,Magorrian1998,Ferrarese2000,Gebhardt2000,Tremaine2002,Ho2008,Gultekin2009,Kormendy2013,Heckman2014}. When actively accreting, SMBHs release huge amounts of energy in radiative and/or kinetic forms. This energy can regulate SMBHs' own growth and the star formation of their host galaxies \citep[e.g.,][]{Silk1998,Bower2006,Somerville2008,Sijacki2015}. On the other hand, galactic physical processes such as supernova feedback and galaxy mergers can also regulate SMBH growth by, e.g., modulating gas fueling and/or leading to SMBH mergers. Hence,  galaxy--SMBH interactions (and potential coevolution) are a critical component of galaxy formation and evolution theories \citep[see, e.g.,][]{Hopkins2006,Ho2008,Alexander2012,Kormendy2013,Heckman2014,Brandt2015}. 

Different lines of evidence support a coevolution scenario. First, there are tight galaxy--SMBH scaling relations (e.g., the black hole mass--velocity dispersion relation or black hole mass--bulge mass relation) in the local Universe. \citep[see][]{Haring2004,Gultekin2009,Kormendy2013,McConnell2013,Savorgnan2016}. Second, studies have found a roughly constant ratio of $10^{-4}-10^{-3}$ between the cosmic SMBH accretion rate (CBHAR) density and the cosmic star formation rate (CSFR) density \citep{Merloni2004, Silverman2008, Shankar2009, Aird2010, Delvecchio2014, Yang2018,Zhang2023a,Peca2023} over $0<z<4$. However, it has been traditionally hard to examine other predictions of coevolution, like a tight SMBH--galaxy mass connection at higher redshifts. This is due to: 1) systematic uncertainties in SMBH and galaxy mass measurements at high redshifts (see \citealt{Shen2013} and references therein, also \citealt{Izumi2019}); 2) difficulty in measuring masses for non-active SMBHs, galaxy mass, and velocity dispersion at $z>0$, due to the limits in sensitivity and resolution (see the summary by, e.g., \citealt{Reines2015}); and 3) the bias that overmassive SMBHs are overrepresented in luminosity-selected samples \citep[][]{Lauer2007}.

Several theoretical models have invoked kinetic energy as the primary driver of SMBH--galaxy coevolution \citep[e.g.,][]{Croton2006,Somerville2008,Fanidakis2012,Hirschmann2012,Costa2014,Volonteri2016,Weinberger2017,Choi2017}.
When SMBHs are actively accreting, gravitational energy from accreted matter can be converted into both radiation and kinetic energy. At high Eddington ratios, geometrically thin but optically thick accretion disks form around AGNs, which converts accreted matter into radiation effectively (see, e.g., \citealt{Shakura1973}). At low Eddington ratios, accretion flows becomes radiatively inefficient, and converts more accreted matter into kinetic energy (see theoretical studies like \citealt{Narayan1994} and observational studies like \citealt{Nagar2005}, \citealt{Ho2002} and references therein). Hence, theoretical models often assumed that the fraction of kinetic energy output dominates in the low Eddington ratio regime \citep{Merloni2008}. However, the measurement of AGN kinetic energy is often indirect, relying on observations of X-ray cavities in gaseous haloes \citep{Hlavacek-Larrondo2015,McDonald2021} or loose scaling relations between kinetic energy and radio luminosities \citep{Merloni2007,Cavagnolo2010,Smolcic2017}. As a result, the kinetic energy fraction is less well quantified when SMBHs are accreting at higher Eddington ratios. \citet[][]{Shankar2008} assumed a simple linear relation between the AGN kinetic and radiative luminosities, with scatter. By fitting to the AGN optical and radio luminosity functions from \citet{CMC05}, they obtained a median kinetic-to-radiative luminosity ratio of $0.1^{+0.05}_{-0.01}$, and a log-normal scatter of $0.38^{+0.04}_{-0.09}$. Based on the scaling relations between radio luminosities (the proxy for AGN kinetic energy output), X-ray luminosities (the proxy for radiative energy output), and black hole masses, \citet[][]{Merloni2008} estimated AGN kinetic vs. radiative energy outputs as functions of redshift. \citet{LaFranca2010} took this a step further by modeling the \emph{full} probability distribution of AGN radio luminosity as a function of X-ray luminosity and redshift, and obtained the cosmic AGN kinetic power density as a function of redshift.

This paper builds on many past studies that have estimated the buildup of the SMBH population and quasar luminosity functions. Beyond the local Universe, the buildup of the SMBH population is studied mostly by observing actively accreting SMBHs, i.e., active galactic nuclei (AGNs). The total cosmic mass in SMBHs can grow only via accretion, even though individual SMBHs can grow via mergers. Based on the So\l{}tan argument \citep{Soltan1982}, the cosmic SMBH mass density ($\rho_\mathrm{BH}(z)$) is the integral of the quasar luminosity function ($\Phi(L,z)$) along the dimensions of luminosity and cosmic time, modulated by the 
radiative efficiency, $\epsilon_\mathrm{rad}$ of AGNs:

\begin{equation}
    \rho_\mathrm{BH}(z) = \int_{0}^{t(z)}\int_{0}^{\infty}\frac{1 - \epsilon_\mathrm{rad}}{\epsilon_\mathrm{rad}c^2} L\Phi(L,z)dLdt\ .
\label{e:soltan}
\end{equation}

To calculate the temporal and luminosity-weighted average $\epsilon_\mathrm{rad}$, one compares the integral with the local SMBH black hole mass density $\rho_\mathrm{BH}(z=0)$. This is traditionally found by convolving the distribution function of a certain galaxy property $X_\mathrm{gal}$ and the $X_\mathrm{gal}$--SMBH mass scaling relation, with the implicit assumption that SMBHs exist in the centres of all galaxies. Based on different AGN luminosity functions, different studies generally find average $\epsilon_\mathrm{rad}\sim 0.1$ \citep{Yu2002,Marconi2004,Merloni2004,Shankar2009,Shankar2013}. 

In \textsc{Trinity}, we apply the So\l{}tan argument to different halo/galaxy populations, rather than treating \emph{all the galaxies} in the Universe as a single population. This enables us to decompose QLFs by many halo/galaxy/SMBH properties, and helps us better understand: 1) The contributions from different halo/galaxy/SMBH populations to total QLFs; 2) Typical AGN activity levels throughout the host galaxy growth histories; 3) How selection effects impact observed galaxy--SMBH mass connections at $z>0$; and 4) The amplitude of AGN clustering.  We also predict QLFs beyond $z\sim 6$, and investigate the dominant specific SMBH accretion rates and plausible formation mechanisms of SMBHs with $M_\bullet\gtrsim 10^{9} M_\odot$ at high redshifts ($z\gtrsim 6$).

In this paper, we continue to present findings from \textsc{Trinity}, an empirical model that infers self-consistent halo--galaxy--SMBH connections from $z=0-10$.  Compared to previous empirical models, \textsc{Trinity} is equipped with a larger compilation of galaxy and AGN data. To constrain the halo--galaxy connection, we included galaxy stellar mass functions (SMFs), galaxy UV luminosity functions (UVLFs), galaxy quenched fractions (QFs), galaxy specific star formation rates (SSFRs), and cosmic star formation rates (CSFRs). The information on the galaxy--SMBH connection comes from quasar luminosity functions (QLFs), quasar probability distribution functions (QPDFs), active black hole mass functions (ABHMFs), and the local bulge mass--black hole mass relations. The inclusion of QPDFs is crucial, because they constrain AGN duty cycles and SMBH growth histories of different halo/galaxy/SMBH populations. Several previous studies made assumptions about Eddington ratio distributions, such as a mass-independent Eddington ratio that may change with redshift \citep[][]{Marconi2004,Shankar2009,Caplar2015,Caplar2018}, or a fixed Eddington ratio distribution \citep[][]{Shankar2013}. With these assumptions, AGN duty cycles and SMBH growth trajectories can be obtained. Nonetheless, these models represent a simplified picture of the Universe, as Eddington ratios are observed to vary with both redshift and host stellar mass \citep[e.g.,][]{Aird2018}.

The joint constraining power of this unique dataset allows \textsc{Trinity} to have a more flexible parametrization than past studies, which includes, e.g., the redshift evolution and host mass dependence of AGN Eddington distribution shapes, and duty cycles.

\textsc{Trinity} inherits the empirical modeling framework of \cite{Behroozi2013}. \textsc{Trinity} makes a guess for the halo--galaxy--SMBH connection over time and constructs a mock universe by applying this guess to  simulated halo populations.  This mock universe is compared with the real Universe in terms of the observables above, and a Markov Chain Monte Carlo (MCMC) algorithm is used to obtain the posterior distribution of allowed halo--galaxy--SMBH connections. Based on this redshift-dependent halo--galaxy--SMBH connection, \citet{Zhang2023a} (Paper I hereafter) reconstructed average SMBH growth histories as a function of host halo/galaxy mass, and found that: 1) the median \bhsm{} relation experiences minimal evolution for galaxies above $5\times 10^{9} M_\odot$, and 2) SMBH activity experiences downsizing, in the sense that more massive SMBHs experience decline in Eddington ratios earlier than less massive SMBHs. \citet{Zhang2023b} (Paper II hereafter) utilized the modeled random scatter around the \bhsm{} relation and SMBH Eddington ratio distributions to quantify the luminosity-dependent bias in the observed \bhsm{} relation for $z\sim 6$ AGNs (i.e., Lauer bias, \citealt{Lauer2007}).

As the third in the \textsc{Trinity} paper series, this work discusses quasar luminosity functions and the buildup of supermassive black holes across cosmic time.  The fourth paper (Paper IV) makes predictions for SMBHs at $z>6$.  The fifth paper (Paper V) investigates the variation of the SFR--BHAR correlation with halo/galaxy/SMBH mass and redshift. In the sixth paper (Paper VI) we calculate black hole merger statistics and make predictions for gravitational wave instruments including NANOGrav and LISA. The seventh (Paper VII) and eighth (Paper VIII) papers present the AGN autocorrelation functions and AGN--galaxy cross-correlation functions from \textsc{Trinity}, respectively. They also discuss whether/how models like \textsc{Trinity} can be constrained by AGN clustering signals.

The paper is organized as follows. \S \ref{s:method} covers methodology. In \S \ref{s:sims_and_data}, we describe the dark matter simulations and galaxy/SMBH observations used to constrain \textsc{Trinity}.  \S \ref{s:results} presents the results of our model, followed by the discussion and the comparison with other models in \S \ref{s:discussion}. Finally, we present the conclusions in \S \ref{s:conclusions}.  In this work, we adopt a flat $\Lambda$CDM cosmology with parameters ($\Omega_m=0.307$, $\Omega_{\mathrm{\Lambda}}=0.693$, $h=0.678$, $\sigma_8=0.823$, $n_s=0.96$) consistent with \textit{Planck} results \citep{Planck2016}. We use datasets that adopt the Chabrier stellar initial mass function \citep[IMF, ][]{Chabrier2003}, the \citet{Bruzual2003} stellar population synthesis model, and the Calzetti dust attenuation law \citep{Calzetti2000}. Halo masses are calculated following the virial overdensity definition from \citet{Bryan1998}.

\section{Methodology}
\label{s:method}

In \S\ref{ss:why_trust_trinity} we explain: 1) why \textsc{Trinity} is able to infer the halo--galaxy--SMBH connection from $z=0-10$ uniquely, robustly, and self-consistently; and 2) why such inference can be done by fitting observations alone, without assuming any particular physcical mechanisms for, e.g., halo--galaxy--SMBH interactions. \S\ref{ss:overview} gives an overview of the \textsc{Trinity} implementation, and the calculation of AGN observable properties are laid out in \S\ref{ss:agn_observables}. 

\subsection{Why observations alone can constrain the halo--galaxy--SMBH connection}
\label{ss:why_trust_trinity}

The So\l{}tan argument \citep[][]{Soltan1982} has traditionally been applied as a cosmic average: the ratio of the total luminosity output of SMBHs to their $z=0$ mass density gives the cosmic average radiative efficiency, and this in turn allows inferring the cosmic average growth history of SMBHs from the redshift evolution of the total luminosity in QLFs.

More recently, studies including \citet{Shankar2020} have recognized that the So\l{}tan argument applies for any sub-population of SMBHs. That is, the average growth history of SMBHs in a chosen set of galaxies at $z=0$ is proportional to the net AGN luminosity of those galaxies' progenitors at earlier times.  \citet{Shankar2020} chose to estimate SMBH histories in bins of host galaxy stellar mass by combining AGN luminosity measurements as a function of galaxy mass and redshift with average galaxy growth histories (i.e., approximating each galaxy as having only a single progenitor at earlier times).  In \textsc{Trinity}, we forward model AGN luminosities across the distribution of galaxy growth histories to: 1) improve the treatment of scatter in galaxy accretion histories and mergers with other galaxies; 2) self-consistently include selection effects; and 3) self-consistently incorporate additional galaxy and SMBH data sources.

In both \citet{Shankar2020} and this paper, a well-constrained halo--galaxy relation is essential to inferring the halo--galaxy--SMBH connection, as well as galaxy assembly histories. Fortunately, such tight constraints are provided in recent empirical models of the halo--galaxy connection, e.g., \citet{Moster2013,Moster2018} and \citet{Behroozi2013,Behroozi2019}. In these models, the galaxy--halo relation is constrained by the number density and clustering of galaxies as a function of redshift.  Since halo formation histories are well-quantified in dark matter simulations, a constraint on the halo--galaxy connection automatically gives a constraint on galaxy star formation histories.  Incorrect galaxy formation histories (e.g., formation times for Milky Way-mass galaxies that are too short) lead to tension with observations (e.g., number densities for Milky Way progenitor galaxies that are too low compared to observations, as well as specific star formation rates that are too high compared with observations)--meaning that constraints on galaxy formation histories are strong and data-driven.

Last but not least, the robustness of \textsc{Trinity} predictions is well tested. We experimented with an extensive series of model variants by changing: 1) the parametrization of the galaxy--SMBH mass connection; 2) accretion rate distribution shapes; 3) SMBH merger prescriptions; 4) AGN obscuration corrections; 5) AGN bolometric corrections, and found that our predictions do not change qualitatively when the input observations are self-consistent. For full details, we refer readers to the Appendices of \citet[][]{Zhang2023a}.

\subsection{Implementation overview}
\label{ss:overview}

\textsc{Trinity} is an empirical model that infers the statistical connection between dark matter haloes, galaxies, and supermassive black holes from $z=0-10$ \citep[][]{Zhang2023a}. Unlike the \textsc{UniverseMachine}, which follows individual haloes across cosmic time, \textsc{Trinity} keeps track of typical galaxy and SMBH masses of different halo populations as functions of redshift. In this section, we summarize the main aspects of the model. For full details about the \textsc{Trinity} parametrization, we refer readers to \S 2 of \citet[][]{Zhang2023a}. For clarity, ``host haloes'' of galaxies and SMBHs in this work (and in all \textsc{Trinity} papers) refers to individual subhaloes that host single galaxies at their centers, rather than the parent haloes that host multiple satellite haloes/galaxies.

\textsc{Trinity} connects dark matter haloes and galaxies by parameterizing the average galaxy SFR as a function of peak historical host halo mass ($M_\mathrm{peak}$)\footnote{In detail, \textsc{Trinity} parameterizes $\mathrm{SFR}( v_\mathrm{Mpeak}(M_\mathrm{peak},z), z)$, where $v_\mathrm{Mpeak}$ is the maximum circular velocity when the halo reaches its peak mass, \mpeak{}.} and redshift. Any guess for the SFR$(M_\mathrm{peak}, z)$ function allows populating a simulated halo population with galaxy SFRs across cosmic time. By integrating these SFRs over halo assembly histories, we establish the stellar mass--halo mass (\smhm{}) relation as a function of redshift. At a given halo mass and redshift, galaxy stellar mass is converted into bulge mass based on a redshift-dependent bulge mass--total stellar mass (\bmsm{}) relation from SDSS \citep[][]{Mendel2014} and CANDELS \citep[][]{Lang2014} observations. With galaxy bulge masses in place, median and average black hole masses are assigned to haloes with a parametrized black hole mass--bulge mass (\bhbm{}) relation that is allowed to evolve with time. Every combination of the \smhm{} and \bhbm{} relations thus fully specifies the average galaxy stellar and SMBH masses given the host halo mass and redshift. Although SMBH occupation fraction is formally a free parameter in \textsc{Trinity}, we find in practice that all haloes with $M_\mathrm{peak} > 10^{11} M_\odot$ are constrained to host SMBHs to match $z=0$ constraints (see also \citealt{Tremmel2017PhDT,Habouzit2017,Zhang2023a}). Specifically, quasar luminosity functions (QLFs) and quasar probability distribution functions (QPDFs) constrain the total amount of SMBH mass growth for different galaxy mass bins. A small occupation fraction will lead to too few SMBHs to account for this mass growth, resulting in a small number of very massive BHs rather than many less massive BHs. As such, a small occupation fraction will overproduce low-redshift active black hole mass functions (ABHMFs) at the massive end.

Average \emph{total} SMBH growth rates at fixed halo mass are calculated by taking the difference of SMBH masses assigned to halo populations between two consecutive snapshots. Given continuing uncertainties in SMBH merger rates, we assume that the fractional merger contribution to total SMBH growth is proportional to the fractional merger contribution to total \emph{galaxy} growth, with a redshift-dependent proportionality factor.

To convert average SMBH accretion rates to observable AGN luminosity distributions, we parametrize the AGN total energy efficiency ($\epsilon_\mathrm{tot}$, including radiative$+$kinetic energies, redshift-independent), duty cycles, and Eddington ratio distribution shapes. Many of our observational constraints are on AGN \emph{radiative} power, so we adopt a scaling relation between the radiative ($\epsilon_\mathrm{rad}$), kinetic ($\epsilon_\mathrm{kin}$), and total ($\epsilon_\mathrm{tot}\equiv \epsilon_\mathrm{rad} + \epsilon_\mathrm{kin}$) energy efficiencies based on previous studies \citep[][]{Mineshige2000,Merloni2008}. The scaling between $\epsilon_\mathrm{rad}$ and $\epsilon_\mathrm{kin}$ is a function of SMBH Eddington ratio:
\begin{equation}
    \eta = \frac{\epsilon_\mathrm{tot} \mathrm{BHAR}/(M_\odot/\mathrm{yr}) \times 4.5 \times 10^8}{(1 - \epsilon_\mathrm{tot}) M_\bullet/M_\odot}\ ,
\label{e:eta}
\end{equation}
where BHAR is SMBH accretion rate. This scaling relation is are inspired by the observational study by \citet{Merloni2007}. We refer the readers to Appendix \ref{a:rad_kin_edd} for the full details of the $\epsilon_\mathrm{rad}-\epsilon_\mathrm{kin}$ scaling relation.

To account for the systematic uncertainties induced by our assumed $\epsilon_\mathrm{rad}$--$\epsilon_\mathrm{tot}$ relation, we also reran \textsc{Trinity} with two different alternative models: 1) the one where no kinetic energy is produced from SMBH accretion (the ``no kinetic'' model hereafter), and 2) the one where more kinetic energy is generated in SMBH accretion (the ``more kinetic'' model hereafter). These two variants represent the two extremes of kinetic energy output in SMBH accretion, but our main conclusions remain invariant when these alternative models are adopted. Full details are introduced in Appendix \ref{a:rad_kin_edd}. We assume that SMBHs in the same halo mass bin share the same Eddington ratio distribution. 

Given that observational studies usually calculate Eddington ratios based on radiative luminosities rather than total energy output, we define ``kinetic'' and ``radiative'' Eddington ratios:

\begin{equation}
\label{e:eta_kin_rad}
    \eta_\mathrm{kin/rad} = \frac{\epsilon_\mathrm{kin/rad}(\eta)}{\epsilon_\mathrm{tot}} \eta\ .
\end{equation}

Systematic effects are also included as nuisance parameters in this work, including systematic offsets in stellar mass and SFR measurements with SED fitting, and the systematic offsets in Eddington ratios when predicting different AGN observables. We also adopt empirical corrections from observational studies for other systematic effects such as Compton-thick AGN obscuration fractions. After correcting for systematic effects, we forward model the joint distribution of galaxy and AGN observables and compare it with real observations (summarized in \S 3 of \citealt{Zhang2023a}). Such a comparison yields the likelihood of the model parameters given the real data constraints. To calculate the posterior distribution of the model parameters, we adopt a custom Metropolis Markov Chain Monte Carlo (MCMC) algorithm \citep[based on][]{Haario2001}. The Markov chain length is set to 2 million steps to ensure convergence, which is approximately 50 auto-correlation lengths. From this posterior distribution, we obtain both: 1) the best-fitting halo--galaxy--SMBH connection; and 2) the uncertainties in this connection allowed by current observations. The best-fitting model parameters, down-sampled MCMC chain, and correlations between model parameters can be found at \href{https://github.com/HaowenZhang/TRINITY/tree/main/stats}{https://github.com/HaowenZhang/TRINITY/tree/main/stats}.

\subsection{Calculating black hole statistics}
\label{ss:agn_observables}

In this work, we discuss the build-up of SMBHs. This build-up process is reflected in different statistics, whose calculations are covered here.

\textsc{Trinity} effectively parameterizes AGN bolometric luminosity ($L_\mathrm{bol}$), and black hole mass ($M_\bullet$) distributions as functions of halo mass and redshift (i.e., $P(L_\mathrm{bol}|M_\bullet,M_\mathrm{peak},z)$ and $P(M_\bullet|M_\mathrm{peak},z)$),  from which all other observables are computed.  The cosmic density of a certain AGN quantity, X, at each redshift is calculated by adding up the contributions from all the haloes in the Universe:
\begin{equation}
    \rho_\mathrm{X}(z) = \int_{0}^{\infty}\overline{\mathrm{X}}(M_\mathrm{peak},z)\phi_\mathrm{h}(M_\mathrm{peak},z)dM_\mathrm{peak}\ ,
\end{equation}
where ``X'' can be either black hole accretion rate (BHAR) or bolometric luminosity ($L_\mathrm{bol}$, \S\ref{ss:agn_rho_rad}), etc., and $\phi_\mathrm{h}(M_\mathrm{peak}, z)$ is the dark matter halo mass function at redshift $z$.

QLFs (\S\ref{ss:results_qlf}) are given by the number density of haloes hosting SMBHs with a given luminosity:
\begin{equation}
\label{e:qlf_1}
\begin{aligned}
\phi_{\mathrm{L}}\left(L_{\mathrm{bol}}, z\right) &= \int_0^\infty \phi_{\mathrm{h}}\left(M_{\mathrm{peak}},z\right)P\left(L_{\rm bol}|M_{\rm peak}, z\right)d M_{\mathrm{peak}}\ ,
\end{aligned}
\end{equation}
where $P\left(L_{\rm bol}|M_{\rm h}, z\right)$ is calculated by counting the number density of black holes with the corresponding Eddington ratio:

\begin{equation}
\begin{aligned}
P\left(L_{\rm bol}|M_{\rm peak}, z\right) = \int_0^\infty &P\left(L_{\rm bol},|M_{\bullet},M_{\rm peak},z\right)\times\\
&P\left(M_{\bullet}|M_{\rm peak},z\right)d M_{\bullet}\label{e:qlf_2}\ ,
\end{aligned}
\end{equation}

The AGN duty cycle (\S\ref{ss:results_duty_cycle_eff}) can be defined in several different ways. If defined as the fraction of black holes above a certain luminosity threshold, $L_\mathrm{min}$, then the duty cycle $f(>L_\mathrm{min})$ is:
\begin{equation}
    f(>L_\mathrm{min} | M_\mathrm{peak}, z) = \int_{L_\mathrm{min}}^{\infty} P(L_\mathrm{bol} | M_\mathrm{peak}, z) dL_\mathrm{bol}\ .
\end{equation}
Similarly, if defined as the fraction of black holes above a certain Eddington ratio threshold, $\eta_\mathrm{min}$, then the duty cycle $f(>\eta_\mathrm{min})$ is:
\begin{equation}
\begin{aligned}
    f(>\eta_\mathrm{min} | M_\mathrm{peak}, z) &=& \int_{\eta_\mathrm{min}}^{\infty} P(\eta_\mathrm{rad} | M_\mathrm{peak}, z) d \eta_\mathrm{rad}\\
    &=& \int_{\eta_\mathrm{min}}^{\infty} P(\eta | M_\mathrm{peak}, z) \frac{d \eta}{d \eta_\mathrm{rad}} \eta_\mathrm{rad}\\ 
\label{e:duty_cycle_eta}
\end{aligned}
\end{equation}
where $P(\eta | M_\mathrm{peak}, z)$ is the (conditional) Eddington ratio distribution as a function of host halo mass and redshift.

\section{Simulations and Data Constraints}
\label{s:sims_and_data}

\subsection{Dark Matter Halo Statistics}
\label{ss:dm_sims}

Unlike \citet{Behroozi2019}, \textsc{Trinity} does not follow individual haloes over cosmic time, but only uses halo statistics from N-body simulations. Specifically, we use halo mass functions from \citet{Behroozi2013} for the cosmology specified in the introduction. These mass functions are similar to the central halo mass functions from \citet{Tinker2008}, but small adjustments were made to account for satellite haloes as well as to use \mpeak{} (the historical peak halo mass) rather than the current mass. These adjustments were based on the Bolshoi \& Consuelo simulations \citep{Klypin2011}. The full details of these halo mass functions can be found in Appendix G of \citet{Behroozi2013}. Note that the \textit{Bolshoi} simulation was based on the \textit{WMAP7} cosmology \citep{Jarosik2011} rather than the \textit{Planck} cosmology, which is used everywhere else in \textsc{Trinity}. As such, we made further corrections to these halo mass functions to account for the difference between the cosmologies. This correction factor is calculated by comparing the total (i.e., central+satellite) halo mass functions from \citet{Behroozi2013} to those from the \textit{Bolshoi--Planck} simulation \citep{RodriguezPuebla2016}, as a function of halo mass and redshift. With this correction, these calibrated halo mass functions are suitable for studying the evolution of halos from $10^{10} M_\odot$ to $10^{15} M_\odot$. Of note, the halo mass functions used here are for individual subhaloes that host single galaxies in their centers, rather than the parent haloes that host multiple galaxies and/or SMBHs.

Both accretion and mergers contribute to halo growth. We use the fitting formulae in Appendix H of \citet{Behroozi2013} to describe the average halo accretion histories. As for halo mergers, the merger rates are fitted from the \textsc{UniverseMachine} \citep{Behroozi2019}. The fitting formulae for halo mergers are available in Appendix B of \citet{Zhang2023a}.

\subsection{Observational Data Constraints}
\label{ss:obs_data}

To constrain the halo--galaxy connection, we have compiled the following galaxy observables: stellar mass functions (SMFs), galaxy quenched fractions (QFs), average specific star formation rates (SSFRs), cosmic star formation rates (CSFRs), and galaxy UV luminosity functions (UVLFs). Collectively, these data cover a redshift range of $z=0-10$. The galaxy--SMBH connection is constrained by the following AGN observables: quasar luminosity functions (QLFs, from \citealt{Ueda2014}), quasar probability distribution functions (QPDFs, from \citealt{Aird2018}), active black hole mass functions (ABHMFs, from \citealt{Schulze2010,Schulze2015,Kelly2013}), the local \bhbm{} relation, and the observed \mbh{} distribution of high redshift bright quasars. The redshift coverage of these AGN data is from $z=0-6.5$. All the datasets are summarized in Table \ref{t:summary_of_obs}, and we refer readers to \citet{Behroozi2019} and \citet{Zhang2023a} for the full details about galaxy and AGN observables, respectively.

\begin{figure}
\includegraphics[width=0.48\textwidth]{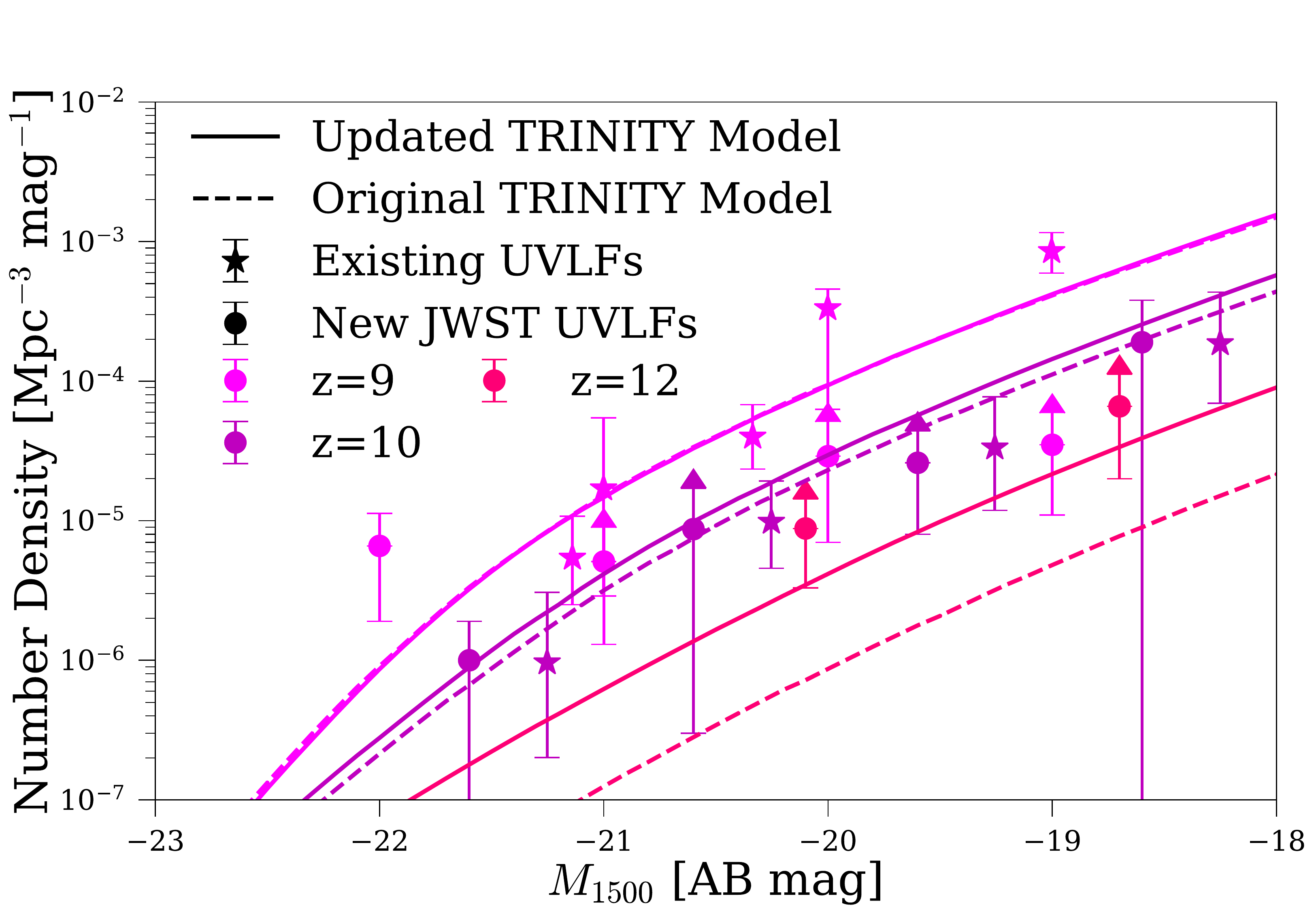}
\caption{The comparison of luminosity functions between the original \textsc{Trinity} model (dashed lines, \citealt{Zhang2023a}), the updated \textsc{Trinity} model (solid lines), the existing galaxy UVLF data (stars, \citealt{Ishigaki2018,Oesch2018,Bouwens2019}, also see Table 8 of \citealt{Zhang2023a}), and the new galaxy UVLFs from JWST \citep{Harikane2023b}. The difference between the original and updated \textsc{Trinity} models increases with redshift. See \S\ref{ss:obs_data}.}
\label{f:jwst_uvlf}
\end{figure}

JWST has detected more high-redshift UV-bright galaxies compared to the predictions of many theoretical models, e.g., the \textsc{UniverseMachine}. This discrepancy between new observations and existing model predictions means that these latest data provide new information to constrain the high-redshift behaviors of our theoretical models. In light of this, we added the the $9 \lesssim z \lesssim 13$ sepctroscopically confirmed galaxy UVLFs from \citet{Harikane2023b} to our data compilation. In \textsc{Trinity}, we calculate galaxy UV magnitudes with a scaling relation between UV magnitudes and SFR, to get statistically equivalent results to the Flexible Stellar Population Synthesis code (FSPS, \citealt{Conroy2013}). The full details of calculating galaxy UV magnitudes can be found in Appendix B of \citet{Zhang2023a}.

Adding $9 \lesssim z \lesssim 13$ galaxy UVLFs does not induce any qualitative changes in the results presented by \citet{Zhang2023a} or \citet{Zhang2023b}. Quantitatively, this addition leads to more UV-bright galaxies at $z\gtrsim 9$ and thus increased galaxy mass at a given halo mass. The redshift-dependent \bhsm{} relation is unchanged with newly added galaxy UVLFs, and thus there will be more SMBH mass at fixed halo mass as well. Fig.\ \ref{f:jwst_uvlf} shows the change in galaxy number density, which increases towards higher redshifts. This is because the new JWST UVLFs (solid circles) are largely consistent with existing $z\lesssim 10$ UVLFs in our data compilation (solid stars) But the original \textsc{Trinity} underproduced UV-bright galaxies without the higher-redshift UVLFs, leading to the larger difference at these redshifts. The results presented in this work are based on the \emph{updated} \textsc{Trinity} model after adding these $9 \lesssim z \lesssim 13$ galaxy UVLFs to our data constraints.

We note that since the SMBH data only cover a redshift range of $z=0-6.5$, all the predictions for $z\gtrsim 6$ in this work are effectively extrapolations based on lower redshift data constraints and our parameterization. Future observations by, e.g., JWST, Roman, or Euclid will provide critical tests of our $z\gtrsim 6$ predictions. Recently, JWST has detected many new SMBHs at $z\gtrsim 6$. We have compared some of the JWST detections in another paper focused on $z\gtrsim6$ predictions (Paper IV), and will defer more thorough comparisons to future studies when the systematic and selection effects in observations are better quantified.

\begin{table*}
\caption{Summary of Observational Constraints}
\begin{center}
\begin{tabular}{lcc}
\hline
Type & Redshifts & Primarily Constrains\\
\hline
Stellar mass functions & 0-8 & SFR--$M_\mathrm{peak}$ relation\\
Galaxy quenched fractions & 0-4 & Quenching--$M_\mathrm{peak}$ relation\\
Cosmic star formation rates & 0-10 & SFR--$M_\mathrm{peak}$ relation\\
Specific star formation rates & 0-9 & SFR--$M_\mathrm{peak}$ relation\\
Galaxy UV luminosity functions & 9-10 & SFR--$M_\mathrm{peak}$ relation\\
Quasar luminosity functions & 0-5 & Total SMBH accretion\\
Quasar probability distribution functions & 0-2.5 & AGN duty cycle, BHAR distributions\\
Active black hole mass functions & 0-5 & Scatter around the galaxy--SMBH mass connection\\
Black hole mass -- bulge mass relation & 0 & Galaxy--SMBH connection, AGN energy efficiency\\
Observed black hole mass distribution of bright quasars & 5.8-6.5 & Galaxy--SMBH connection, Eddington ratios at high redshifts\\
\hline
\end{tabular}
\end{center}
\label{t:summary_of_obs}

\parbox{17cm}{\textbf{Notes.} BHAR is the black hole accretion rate.}

\end{table*}

\begin{figure}
\includegraphics[width=0.48\textwidth]{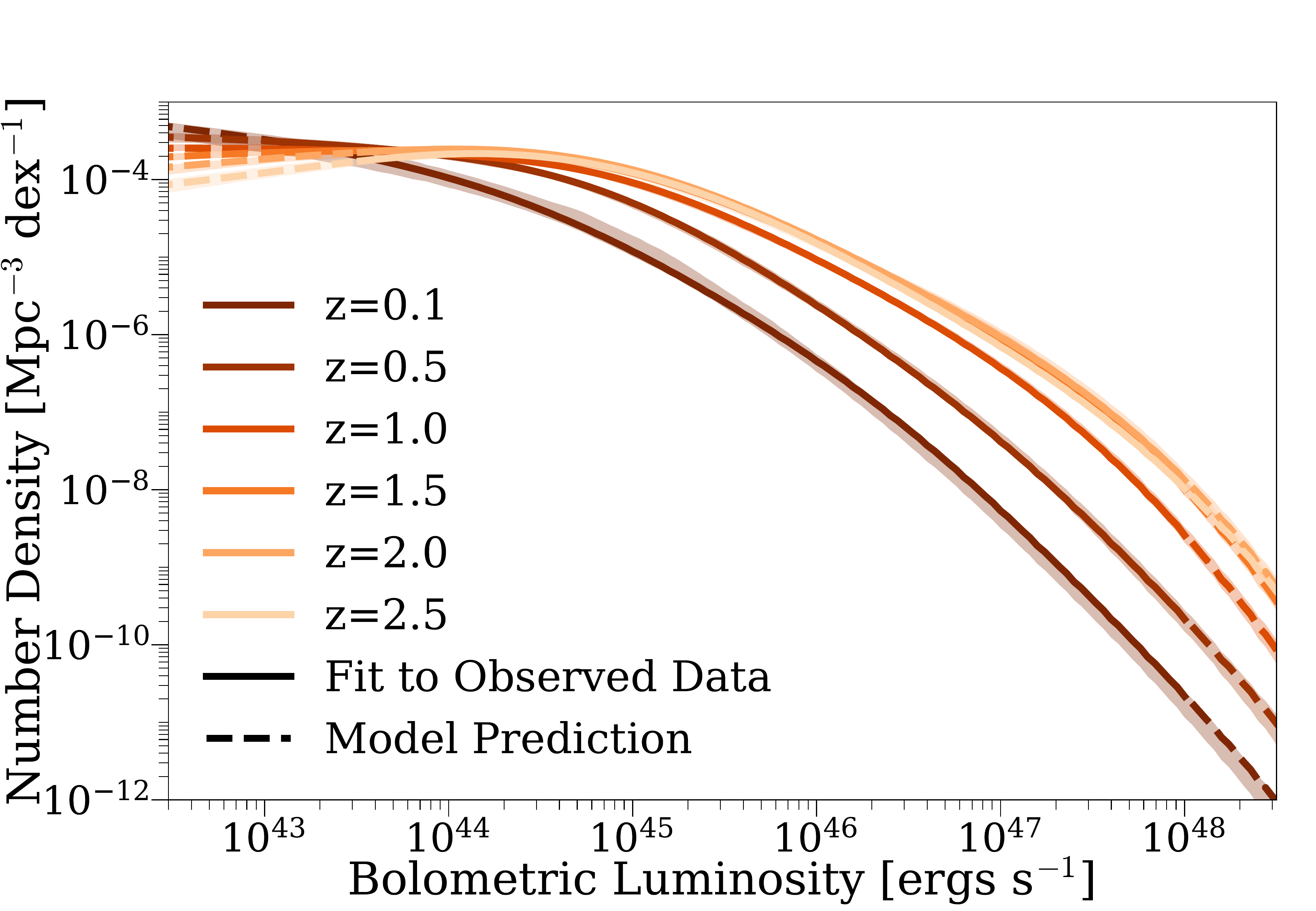}
\includegraphics[width=0.48\textwidth]{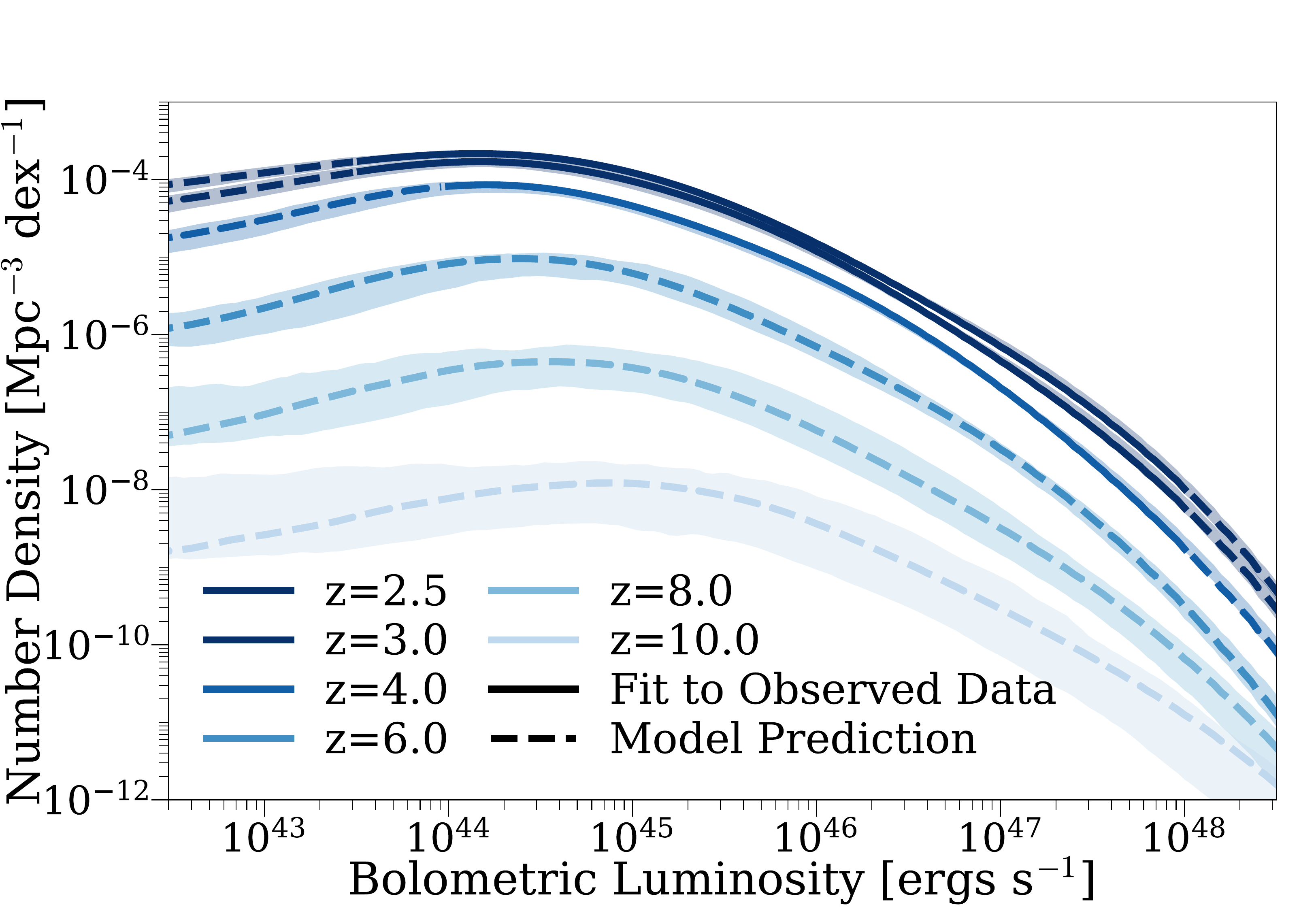}
\caption{The evolution of the bolometric quasar luminosity function from $0.1\leq z \leq 2.5$ (top panel) and from $2.5\leq z \leq 10$ (bottom panel). The solid parts of the curves represent regions where QLF data is available, and the dashed parts denote the predictions beyond the redshift or luminosity ranges covered by observed QLFs. \shadedregions{} See \S\ref{ss:qlf_evol}.}
\label{f:high_z_qlf}
\end{figure}

\section{Results}
\label{s:results}

\S\ref{ss:qlf_evol} shows the redshift evolution of quasar luminosity functions (QLFs) from $z=0-10$. Unless otherwise noted, all the QLFs shown in this work include both non-obscured and obscured (including Compton-thick obscured) quasars. In \S\ref{ss:results_qlf}, we present the QLFs split into different bins of SMBH mass, galaxy mass, halo mass, deviation from the intrinsice \bhsm{} relation, and Eddington ratio. AGN radiative power densities across cosmic time are shown in \S\ref{ss:agn_rho_rad}. AGN duty cycles from \textsc{Trinity} are presented in \S\ref{ss:results_duty_cycle_eff}. We then show the fraction of super-Eddington AGNs in \S\ref{ss:results_super_eddington}. For the comparisons of the best-fitting model to observations, we refer readers to \citet{Zhang2023a}.

\subsection{Redshift evolution of quasar luminosity functions from $z=0-10$}
\label{ss:qlf_evol}

The top panel of Fig.\ \ref{f:high_z_qlf} shows the evolution of quasar bolometric luminosity function (QLF) from $0.1\leq z\leq 2.5$. Towards $z\sim 2.5$, there is increasing AGN activity in the universe. We thus see the decrease(increase) in the faint(bright) end of QLF, and an increase in the typical quasar luminosity from $z=0.1$ to $z=2.5$. These trends are qualitatively consistent with \citet{Shen2020}. The bottom panel shows bolometric QLF at $2.5\leq z\leq 10$. In Fig.\ \ref{f:high_z_qlf}, the solid parts of the curves represent regions where QLF data is available, whereas the dashed parts denote the predictions beyond the redshift or luminosity ranges covered by observed QLFs. As was found by \citet{Shen2020},  the normalization of QLF drops quickly towards higher redshifts at $z>4$, by $\sim 3-4$ dex from $z\sim 4$ to $z\sim 10$. This is simply due to the lack of massive SMBHs to drive bright AGNs in the early universe. We also see increasing uncertainties around the QLF at higher redshifts, because black hole accretion rates at higher redshifts are smaller, have less impact on the model predictions at lower redshifts, and thus are not as well constrained by existing observations.

\subsection{Quasar luminosity function decompositions}
\label{ss:results_qlf}

\subsubsection{Quasars binned by $M_\bullet$}
\label{ss:qlf_mbh}

\begin{figure*}
\subfigure{
\includegraphics[width=0.48\textwidth]{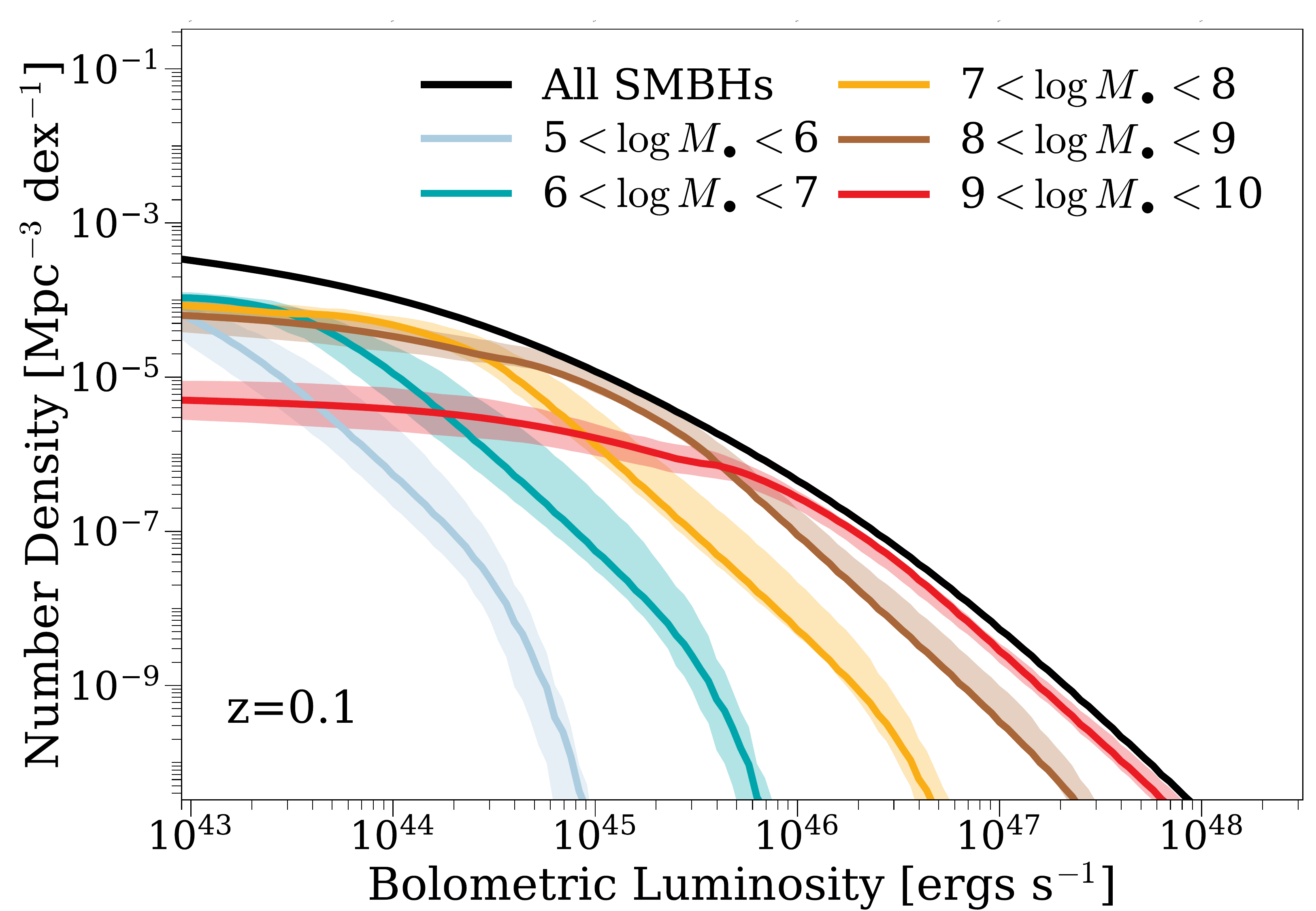}
}
\subfigure{
\includegraphics[width=0.48\textwidth]{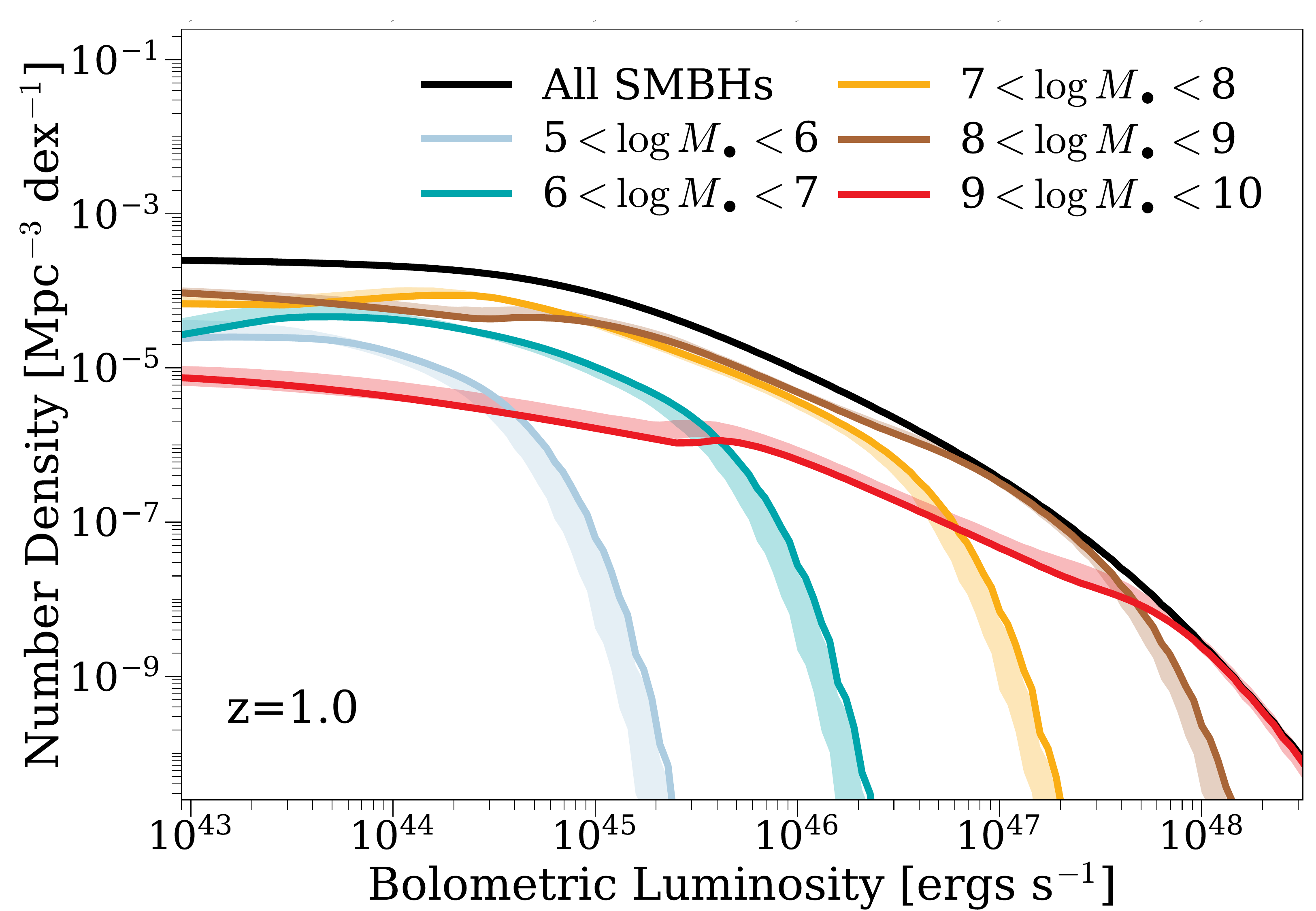}
}
\subfigure{
\includegraphics[width=0.48\textwidth]{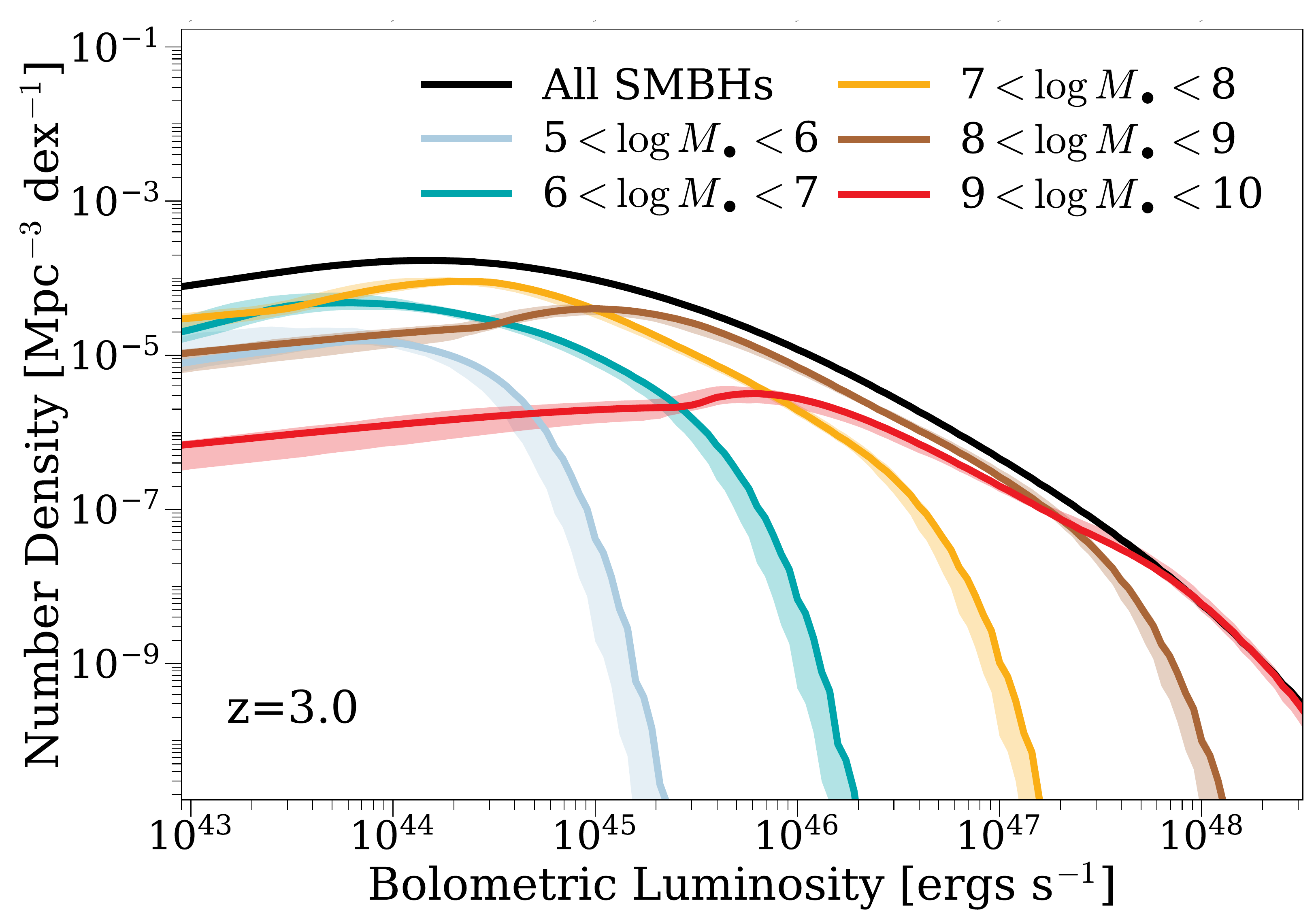}
}
\subfigure{
\includegraphics[width=0.48\textwidth]{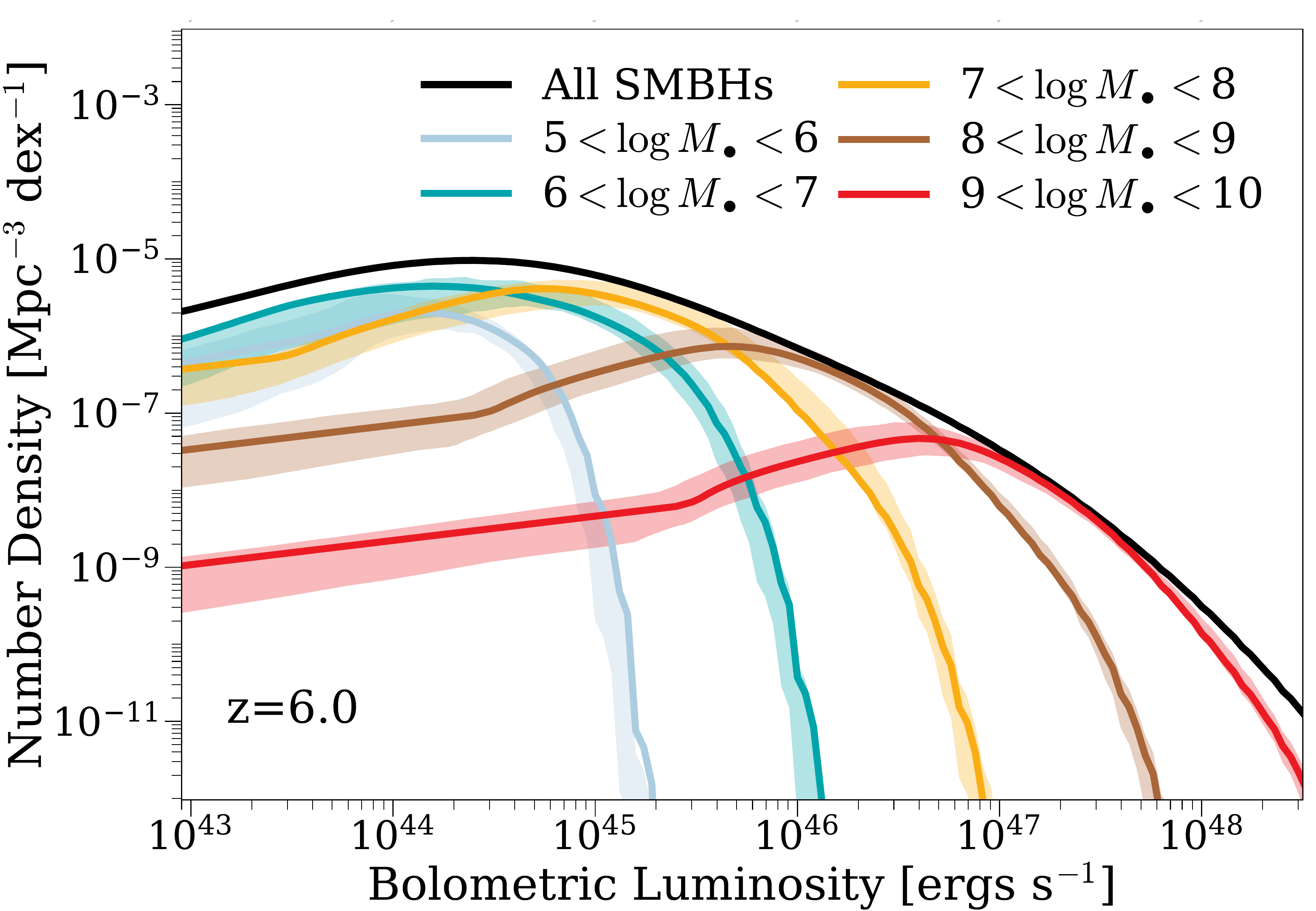}
}
\subfigure{
\includegraphics[width=0.48\textwidth]{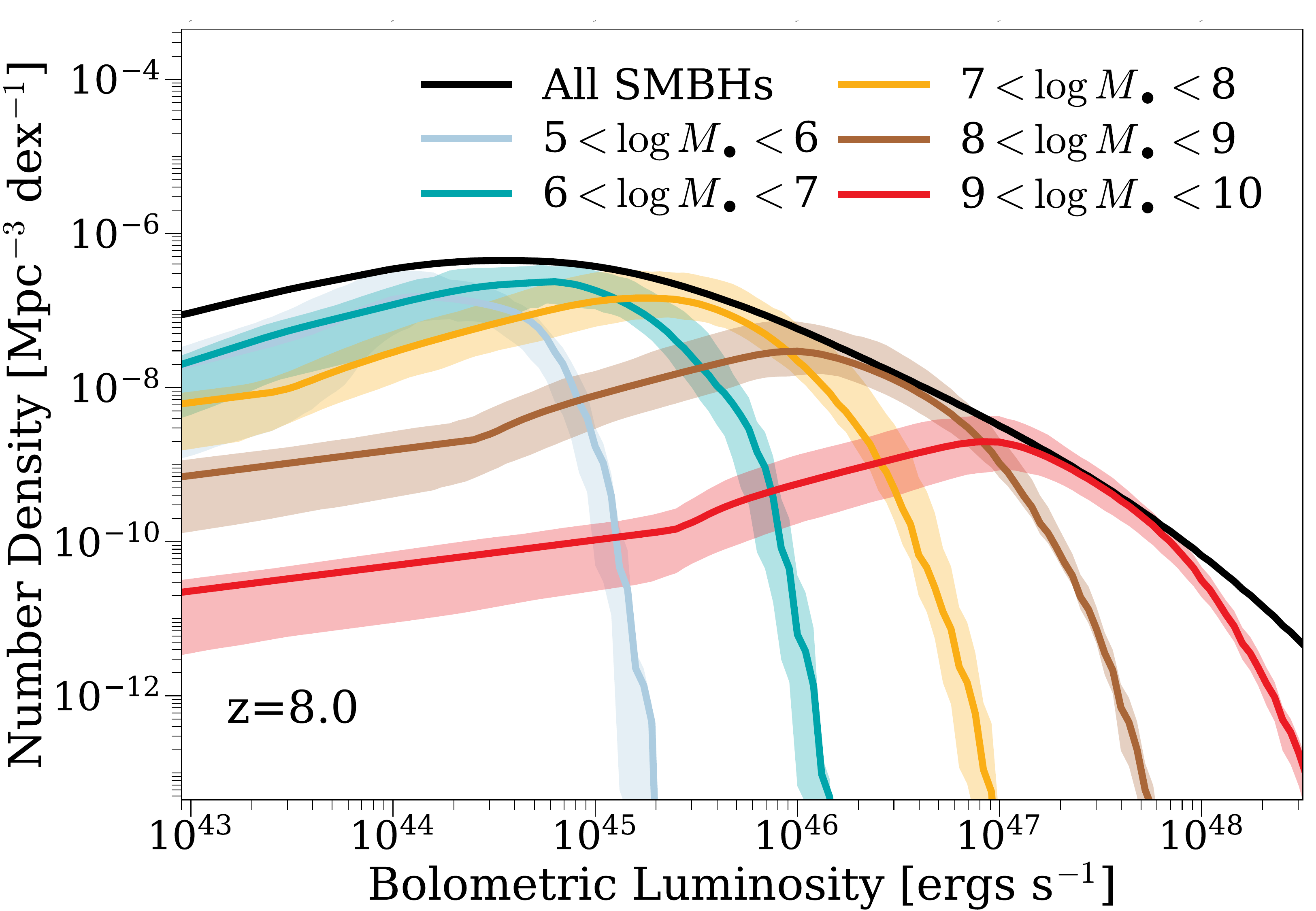}
}
\subfigure{
\includegraphics[width=0.48\textwidth]{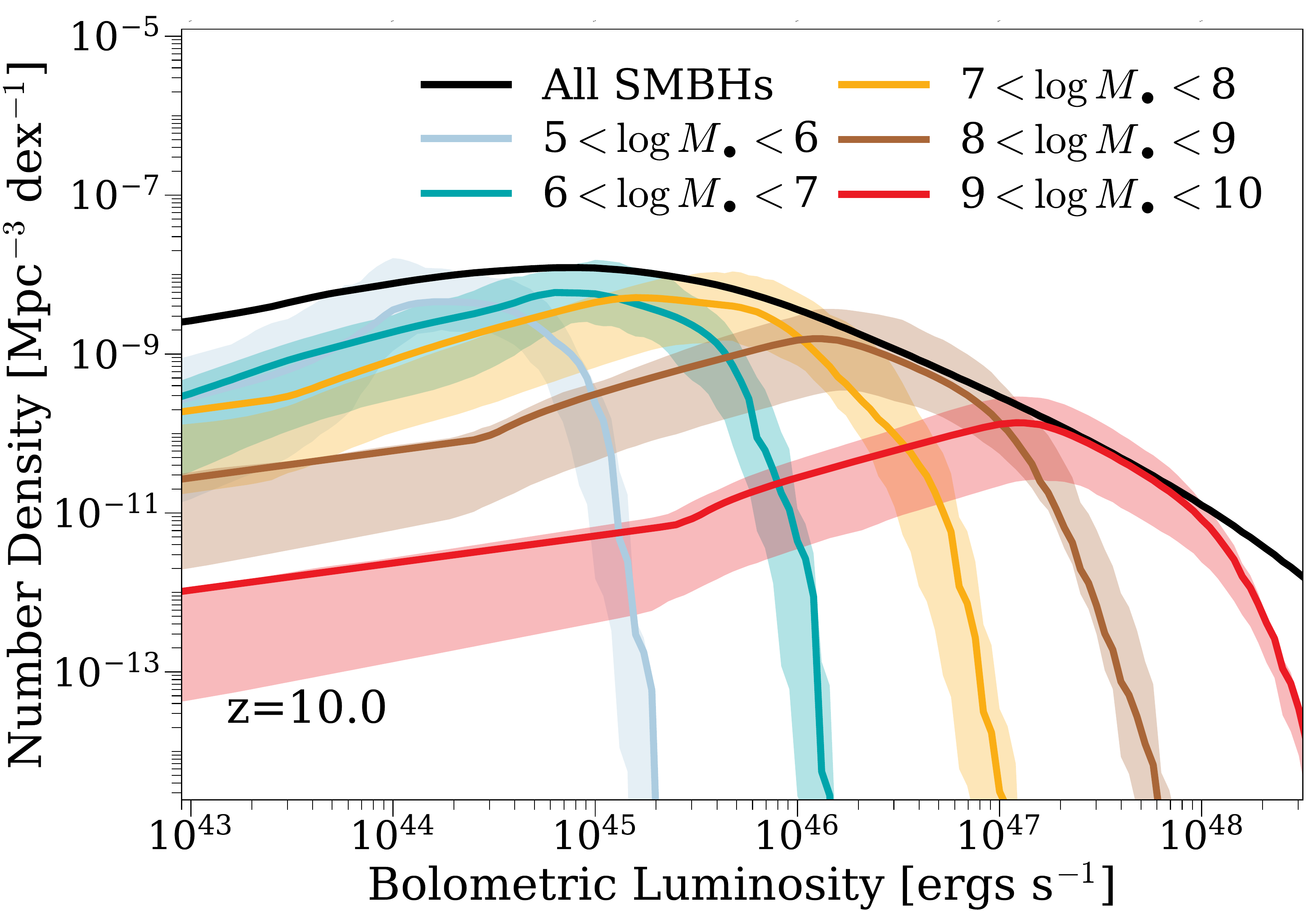}
}
\caption{Quasar luminosity functions in bins of black hole mass at $z=0.1$ (upper left), $z=1.0$ (upper right), $z=3.0$ (middle left), $z=6.0$ (middle right), $z=8.0$ (lower left), and $z=10.0$ (lower right). \shadedregions{} See \S\ref{ss:qlf_mbh}.}
\label{f:qlf_mbh}
\end{figure*}

Fig.\ \ref{f:qlf_mbh} shows quasar bolometric luminosity functions (QLFs) in bins of black hole mass at different redshifts. There are kinks at the faint end of QLFs for different mass bins. They are due to the change in the scaling of $\eta_\mathrm{rad}$ with $\eta_\mathrm{tot}$ at $\eta_\mathrm{tot}=0.03$ (see Appendix \ref{a:rad_kin_edd}). 

At lower redshifts, e.g., $z\lesssim 1$, the QLFs of different black hole populations decrease monotonically with luminosity; at higher redshifts, however, the QLFs start to show a peak. This results from the narrowing of Eddington ratio distributions towards higher redshifts. These higher and narrower Eddington ratio distributions are required by the combination of QPDFs from \citet{Aird2018} and the ABHMFs from \citet{Kelly2013}. Specifically, the QPDFs specifies the average SMBH accretion histories in different galaxies (by specifying the average SMBH accretion rates via numerical integrals) as well as AGN duty cycles (i.e., fraction of SMBHs accreting at certain luminosities). With such constraints, lower Eddington ratios and/or broader Eddington ratio distributions cannot reproduce the ABHMFs at high-z. At $6\lesssim z\lesssim 8$ ($9\lesssim z \lesssim 10$), all black holes in haloes above $\sim10^{11.5} M_\odot$ are accreting at around (above) the Eddington rate. As a result, different black hole mass populations dominate the QLF at different luminosities. 

\subsubsection{Quasars binned by host galaxy and host halo mass}
\label{ss:qlf_mstar_mh}

\begin{figure*}
\vspace{-2cm}
\subfigure{
\includegraphics[width=0.48\textwidth]{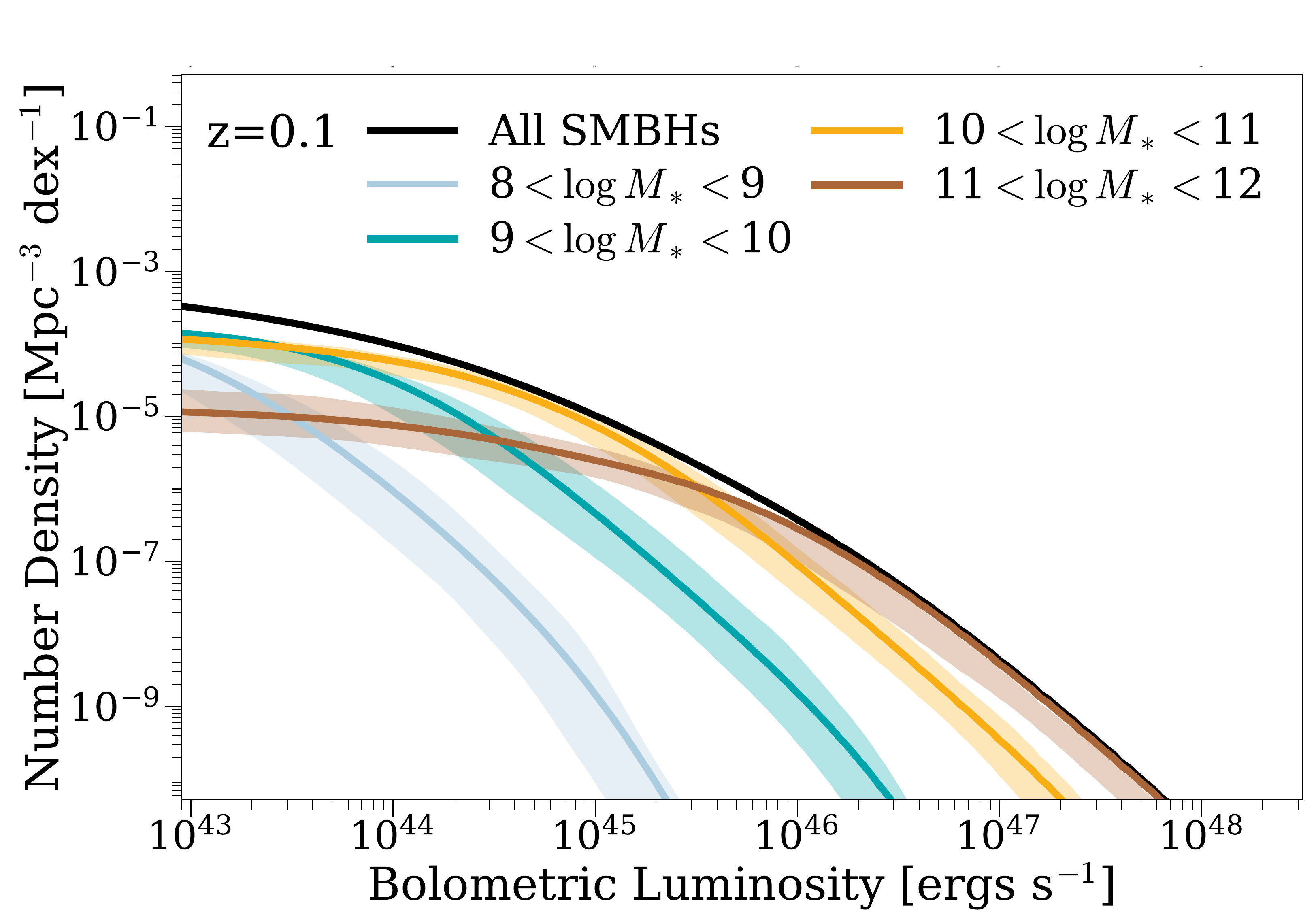}
}
\subfigure{
\includegraphics[width=0.48\textwidth]{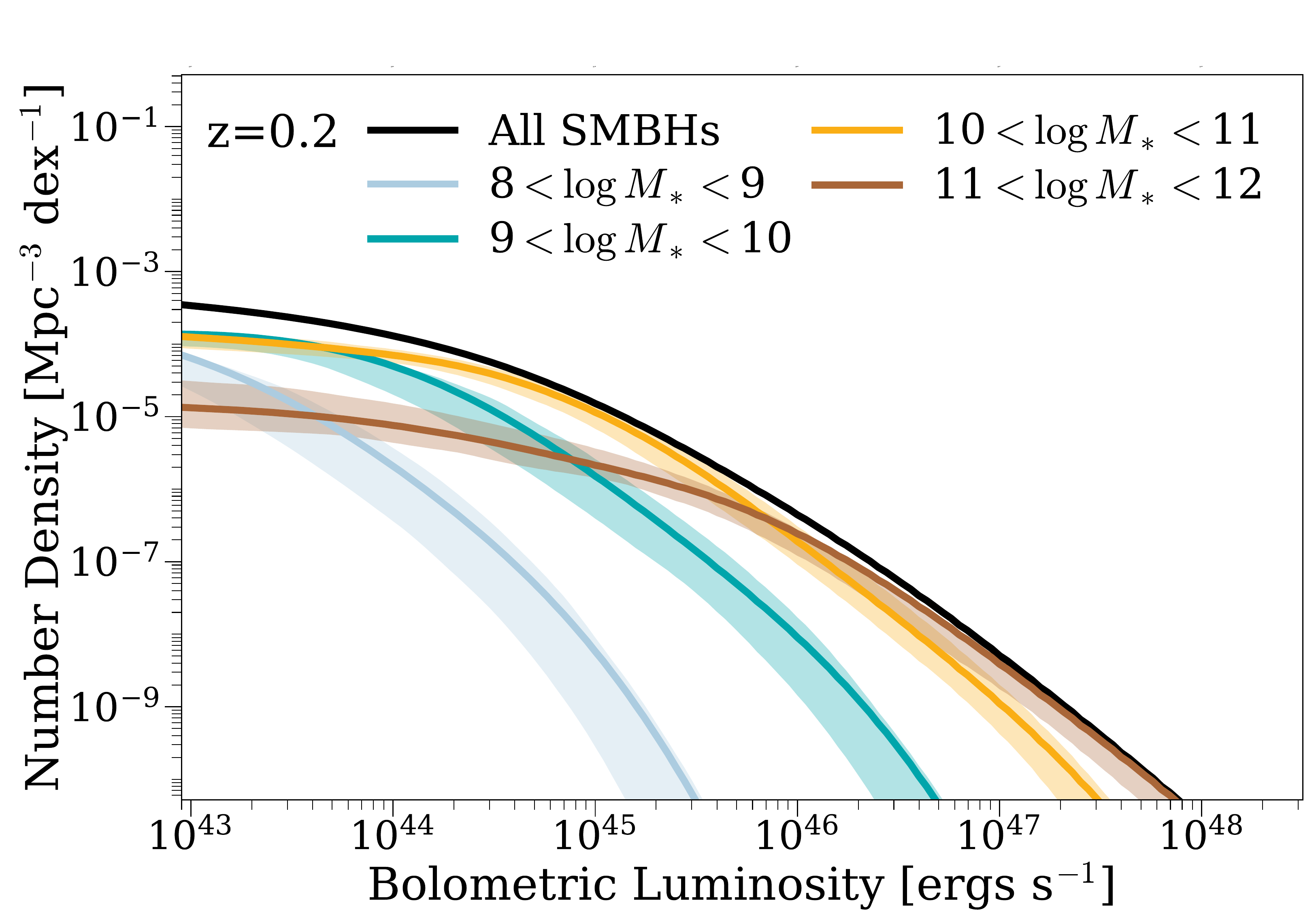}
}
\subfigure{
\includegraphics[width=0.48\textwidth]{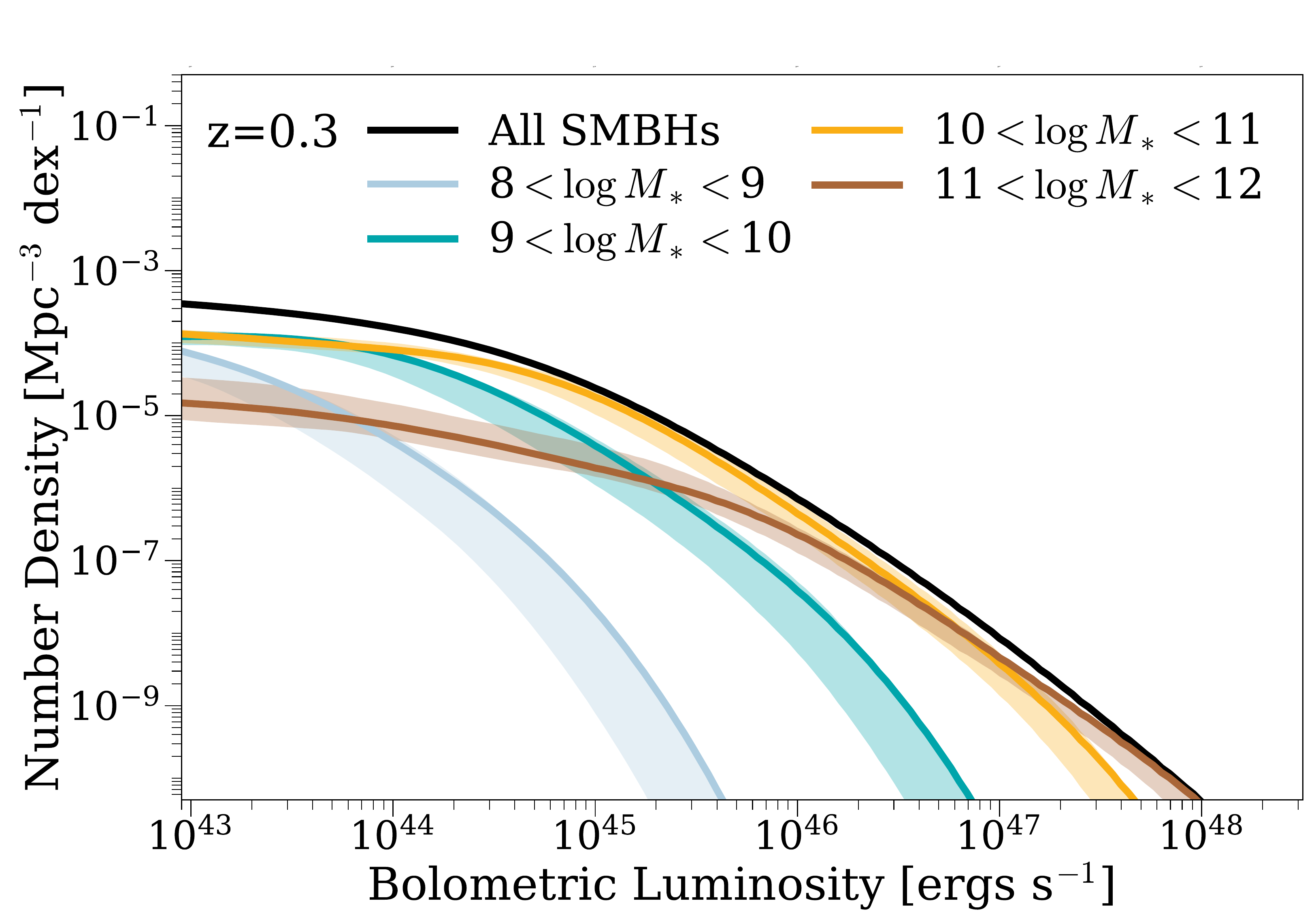}
}
\subfigure{
\includegraphics[width=0.48\textwidth]{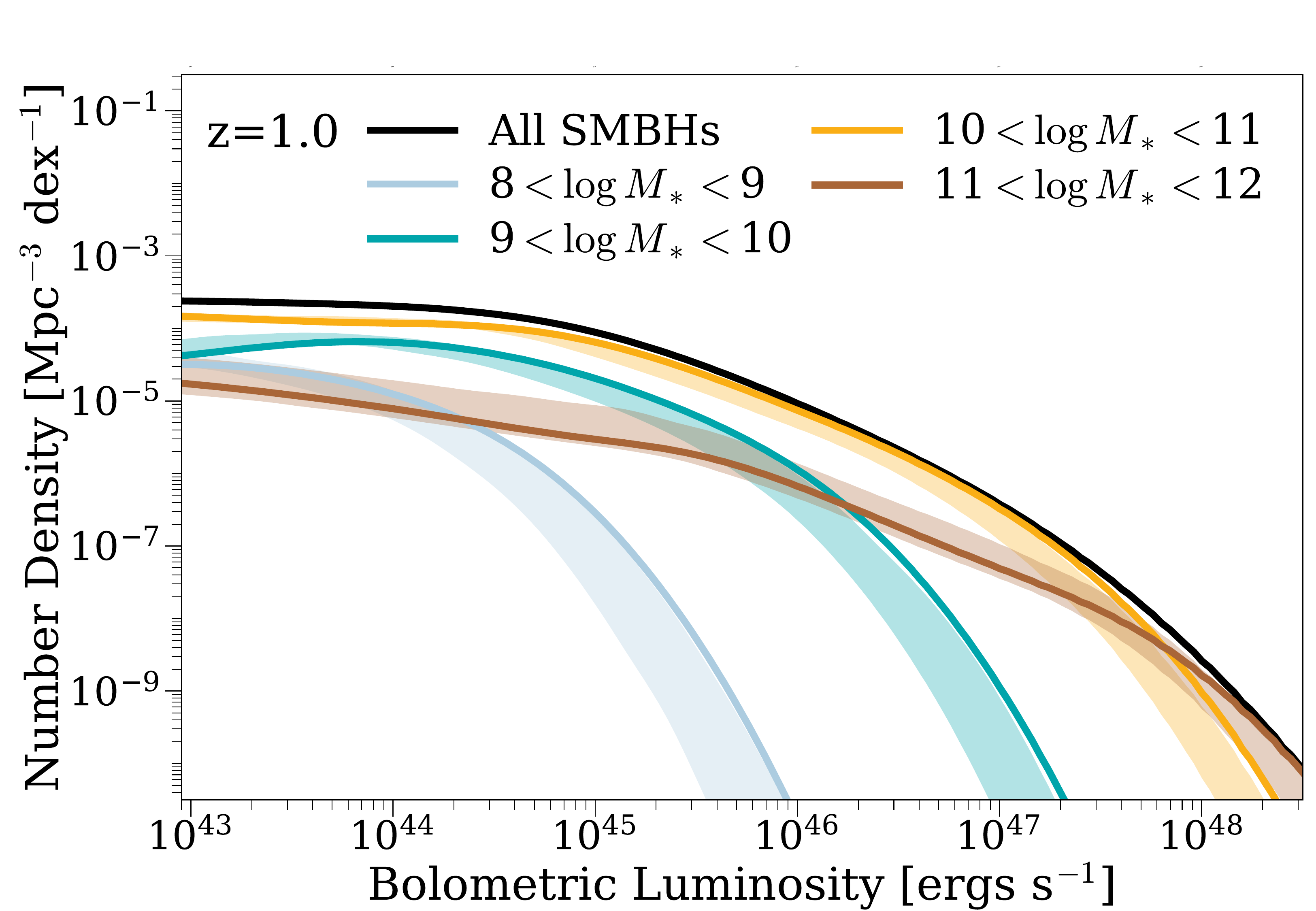}
}
\subfigure{
\includegraphics[width=0.48\textwidth]{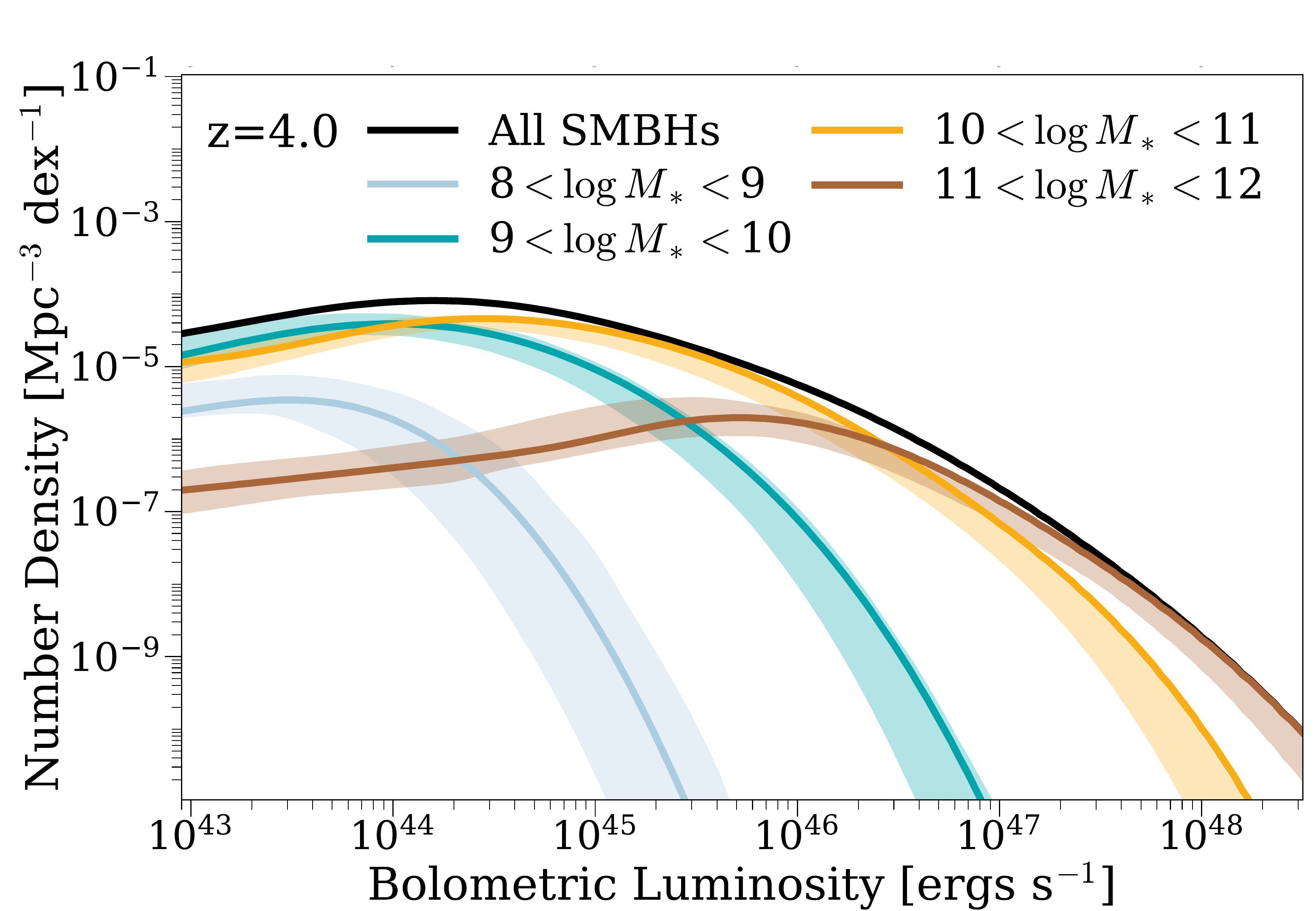}
}
\subfigure{
\includegraphics[width=0.48\textwidth]{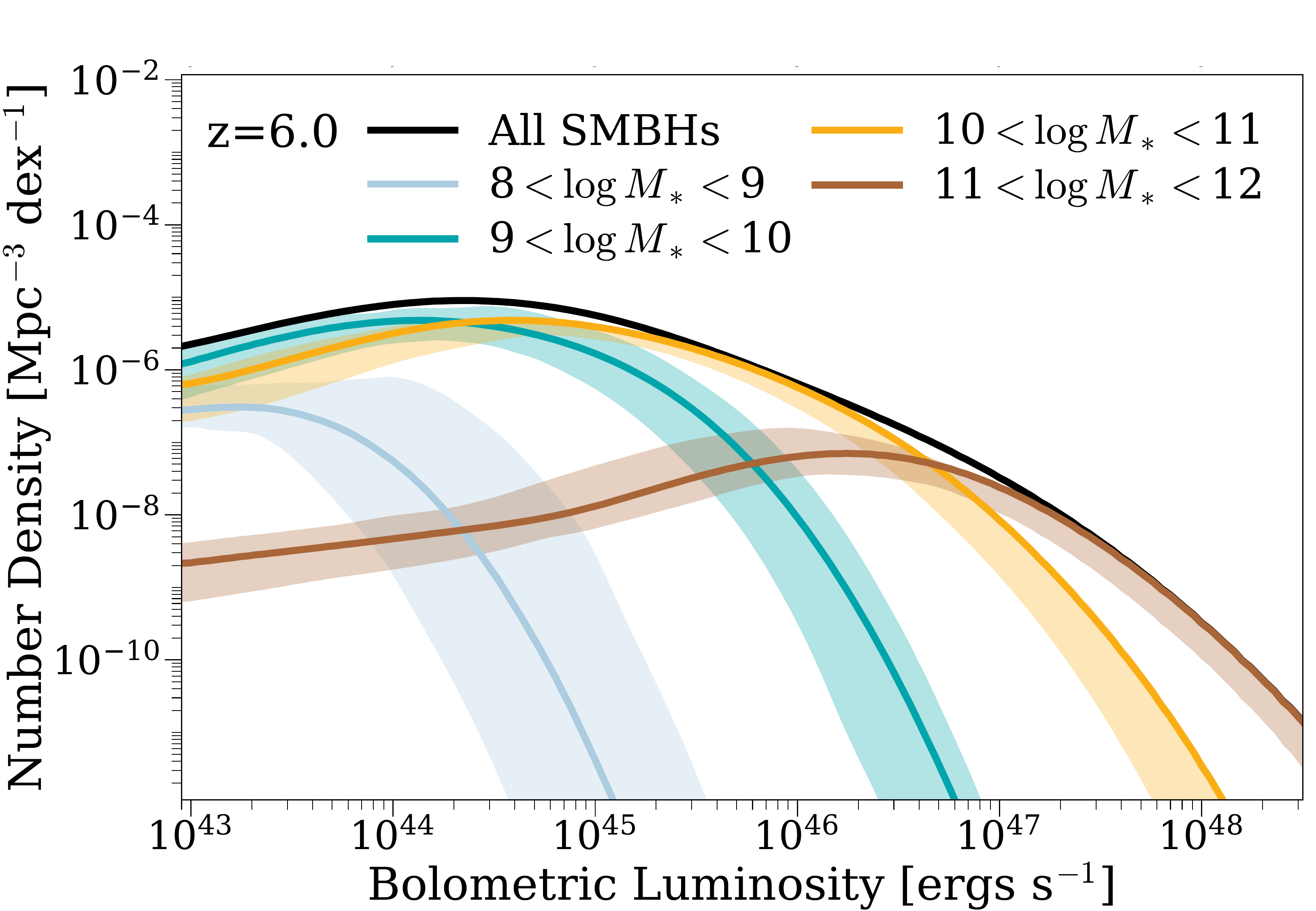}
}
\subfigure{
\includegraphics[width=0.48\textwidth]{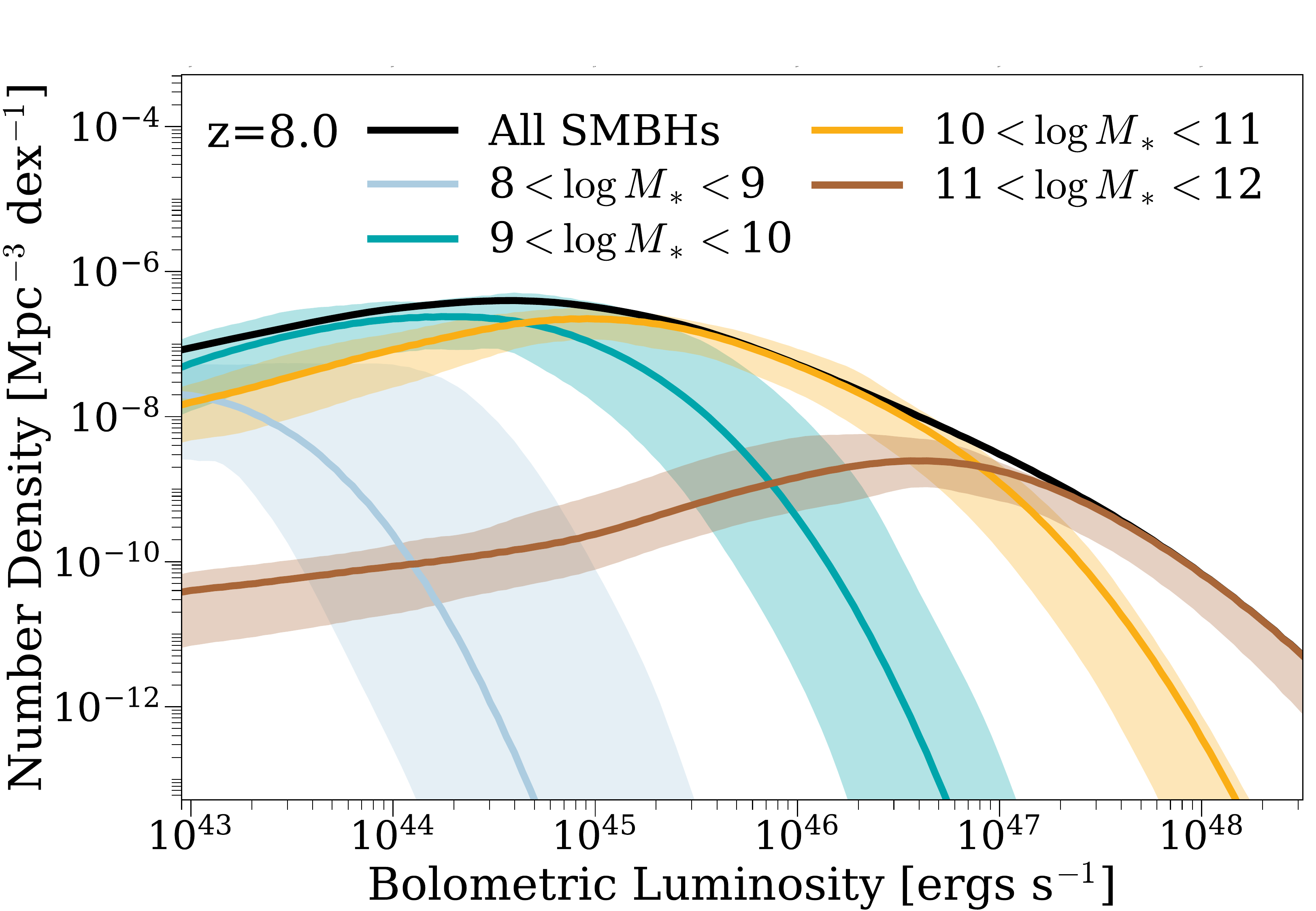}
}
\subfigure{
\includegraphics[width=0.48\textwidth]{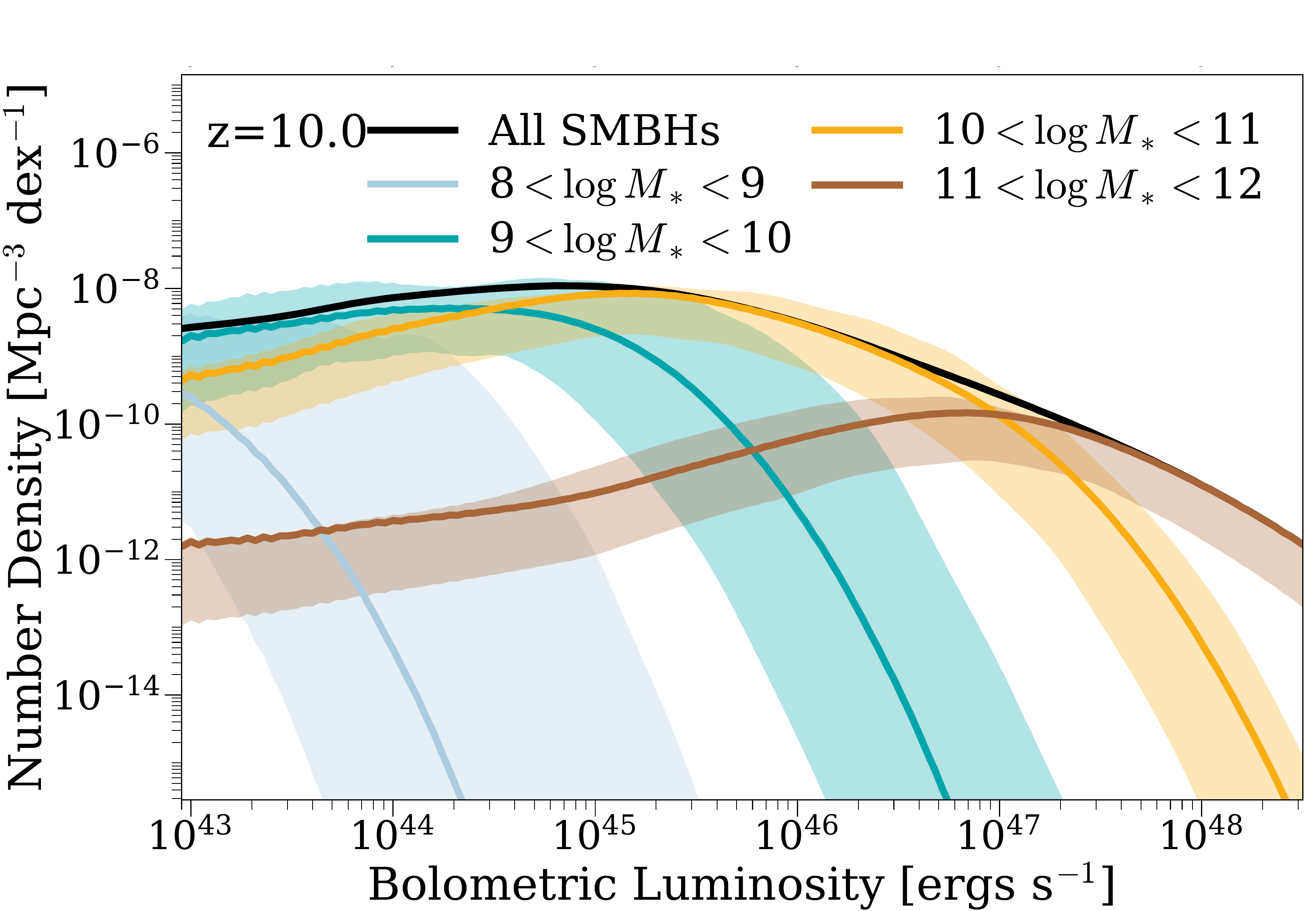}
}
\caption{Quasar luminosity functions in bins of galaxy stellar mass at $z=0.1$ (first row, left), $z=0.2$ (first row, right),
$z=0.3$ (second row, left), $z=1.0$ (second row, right), $z=4.0$ (third row, left), $z=6.0$ (third row, right), $z=8.0$ (bottom row, left), and $z=10.0$ (bottom row, right). \shadedregions{} See \S\ref{ss:qlf_mstar_mh}.}
\label{f:qlf_mstar}
\end{figure*}

\begin{figure}
\includegraphics[width=0.48\textwidth]{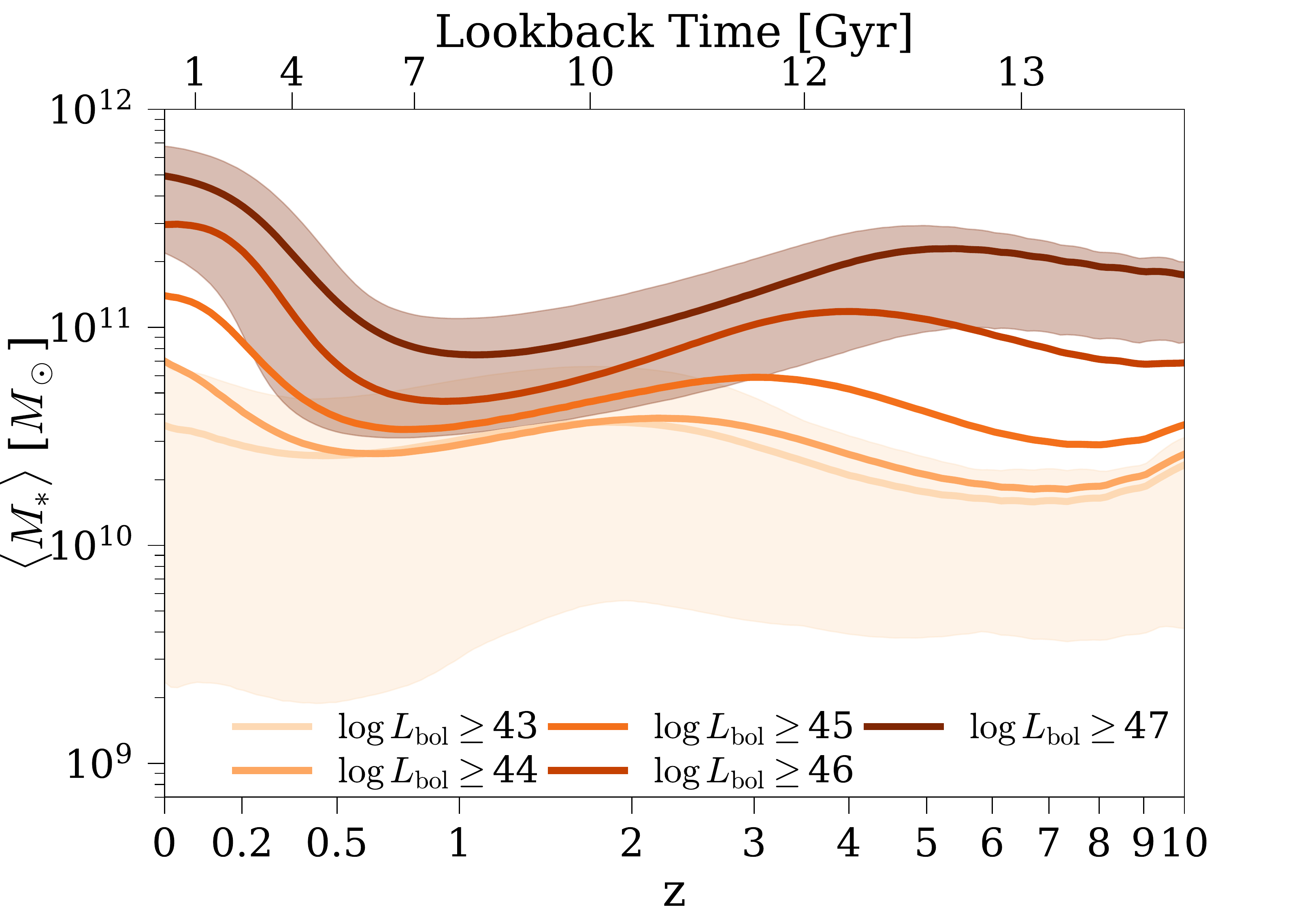}
\caption{The average host galaxy masses ($\langle M_*\rangle$) for AGNs above different bolometric luminosity limits, as functions of redshift. The lighter(darker) shaded region is the 68$^{\mathrm{th}}$ percentile range of host $M_*$ for AGNs with $\log L_\mathrm{bol} \geq 43$(47). See \S\ref{ss:qlf_mstar_mh}.}
\label{f:mstar_avg_quasars}
\end{figure}

Fig.\ \ref{f:qlf_mstar} shows QLFs in bins of galaxy stellar mass at different redshifts. At $z\gtrsim 6$, galaxies of different masses dominate different parts of the QLFs. This is again due to the fact that all active SMBHs in haloes above $\sim10^{11.5} M_\odot$ are accreting at around the Eddington rate (see \S\ref{ss:qlf_mbh}). The spreads in luminosity for each stellar mass bin are broader than those for each bin in SMBH mass. This is due to: 1) the higher-than-unity \bhsm{} slope at high redshifts; and 2) scatter in the \bhsm{} relation. From $z\sim 4$ to $z\sim 1$, galaxies with $10 < \log_{10} M_* < 11$ become more and more prevalent compared to those with $11 < \log_{10} M_* < 12$. This is mainly due to the stronger decline in the activity level of SMBHs in more massive galaxies with time, i.e., the AGN downsizing effect. From $z\sim 1$ to $z\sim 0.1$, galaxies with $10 < \log_{10} M_* < 11$ lose dominance in the bright end of QLFs. At $z=0.1$, galaxies with $9 < \log_{10} M_* < 11$ dominate QLFs below $L_\mathrm{bol} \sim 3\times 10^{45}$ erg/s; above this luminosity threshold, more massive objects with $11 < \log_{10} M_* < 12$ are dominant. This over-representation of massive galaxies at the bright end of the QLF is primarily due to the similar or higher Eddington ratios for high-mass SMBHs/haloes than their lower-mass counterparts, as required by the low-redshift QPDFs from \citet{Aird2018}. These $M_*$ predictions for different quasar populations will help us better understand the expected host galaxy properties of different quasars in current and future observations, e.g., with \textit{JWST} and \textit{ALMA}.

Fig.\ \ref{f:mstar_avg_quasars} shows the average host galaxy masses ($\langle M_*\rangle$) for AGNs above different bolometric luminosity limits as functions of redshift. In general, $\langle M_*\rangle$ increases with cosmic time from $z=10$ to $z\sim 5$, where active SMBHs are growing at high Eddington ratios with their host galaxies. $\langle M_*\rangle$ then peak and decrease as the ``AGN downsizing'' effect becomes significant: SMBHs in more massive haloes/galaxies become less active, and those in smaller galaxies take over to drive AGNs. Below $z\lesssim 0.5$, the AGN activity level further decreases, such that bright AGNs must be driven by massive enough SMBHs. As a result, we see the increase in $\langle M_*\rangle$ towards $z=0$, which is the more significant for brighter AGN populations. We also show the 68$^\mathrm{th}$ percentile range of host $M_*$ for AGNs with $\log L_\mathrm{bol} \geq 43$ (lighter shaded region) and $\log L_\mathrm{bol} \geq 47$ (darker shaded region). Since only massive galaxies can host bright AGNs, the $M_*$ range for bright AGNs is in general narrower than that for fainter AGNs.

\begin{figure*}
\subfigure{
\includegraphics[width=0.48\textwidth]{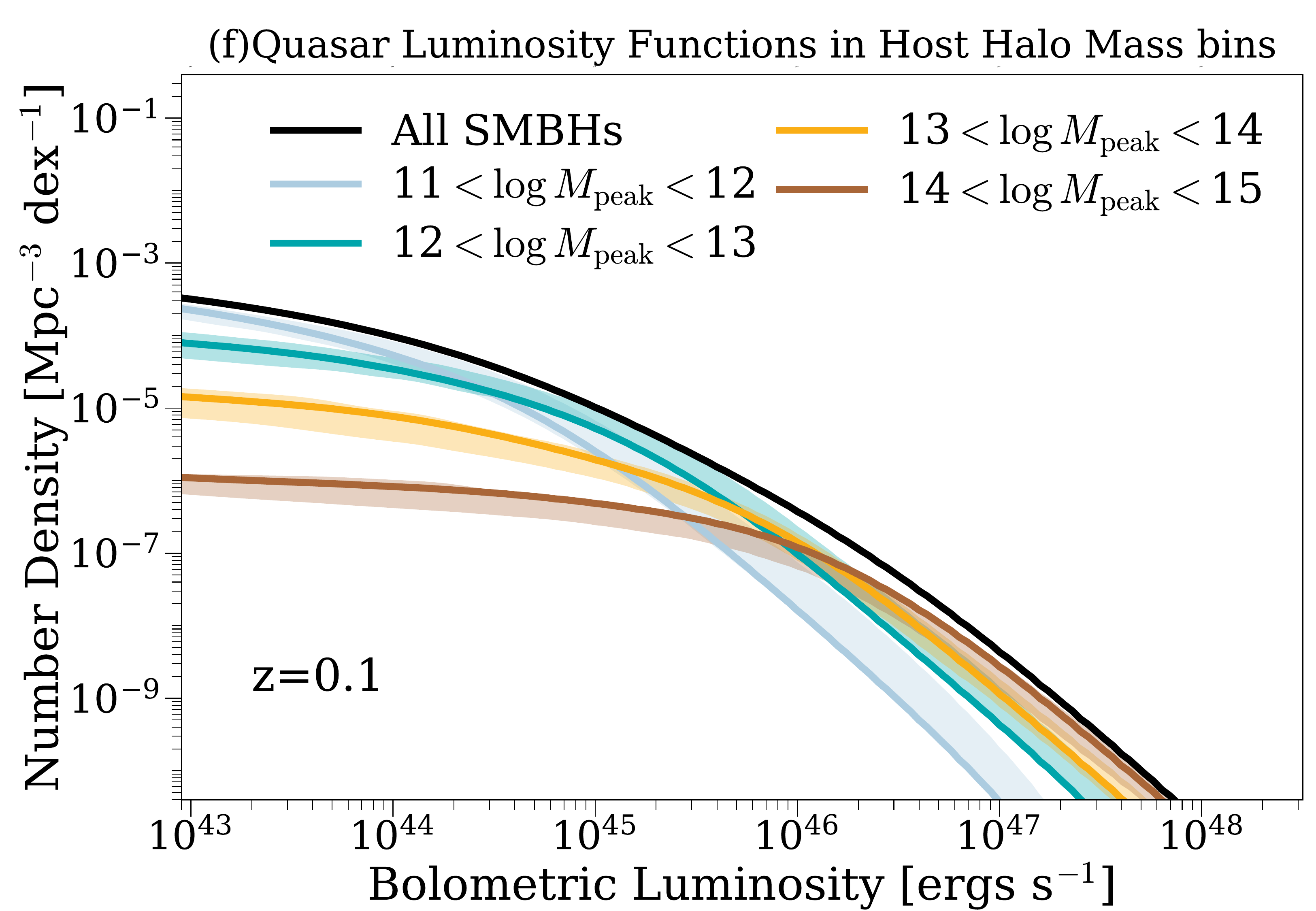}
}
\subfigure{
\includegraphics[width=0.48\textwidth]{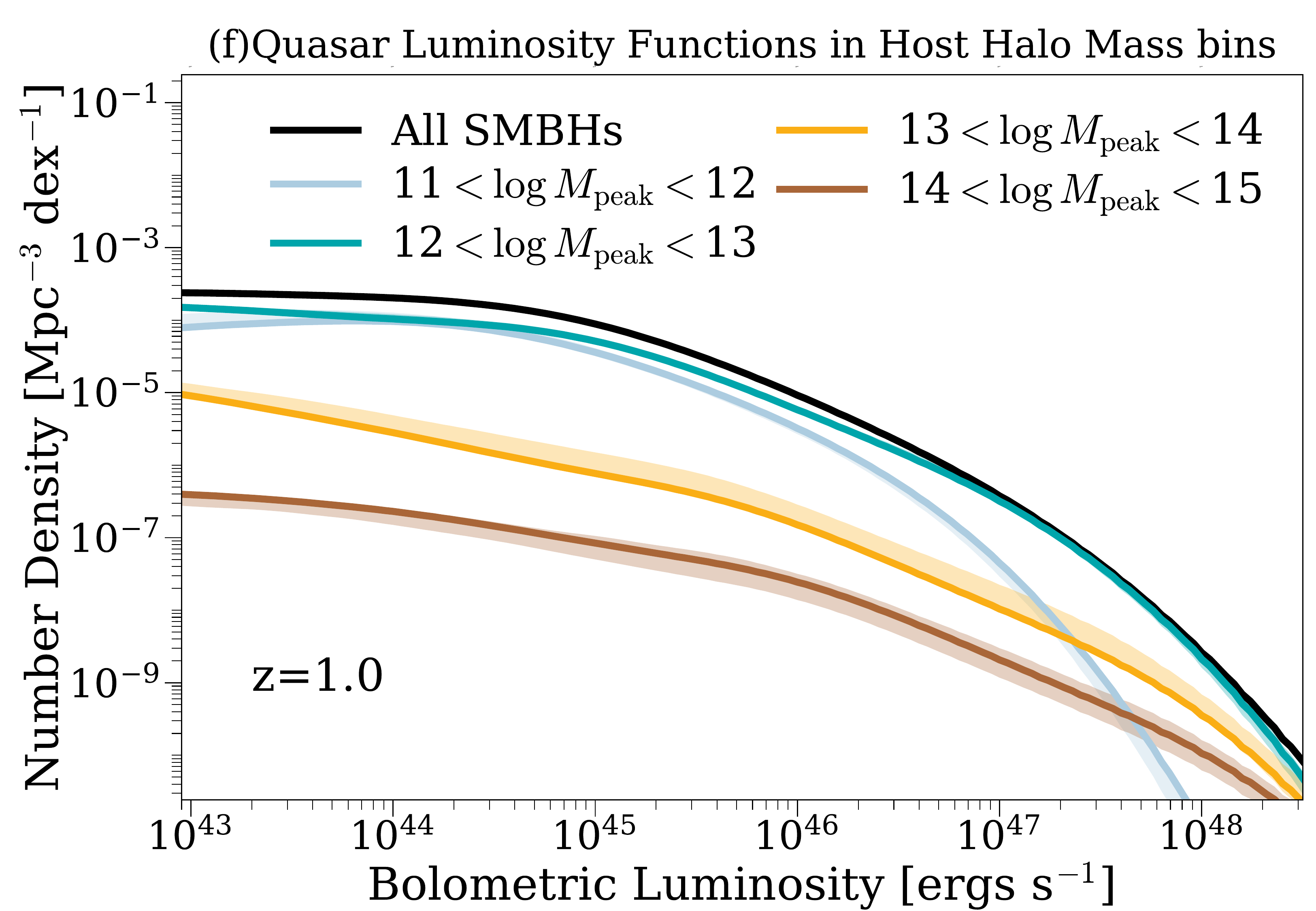}
}
\subfigure{
\includegraphics[width=0.48\textwidth]{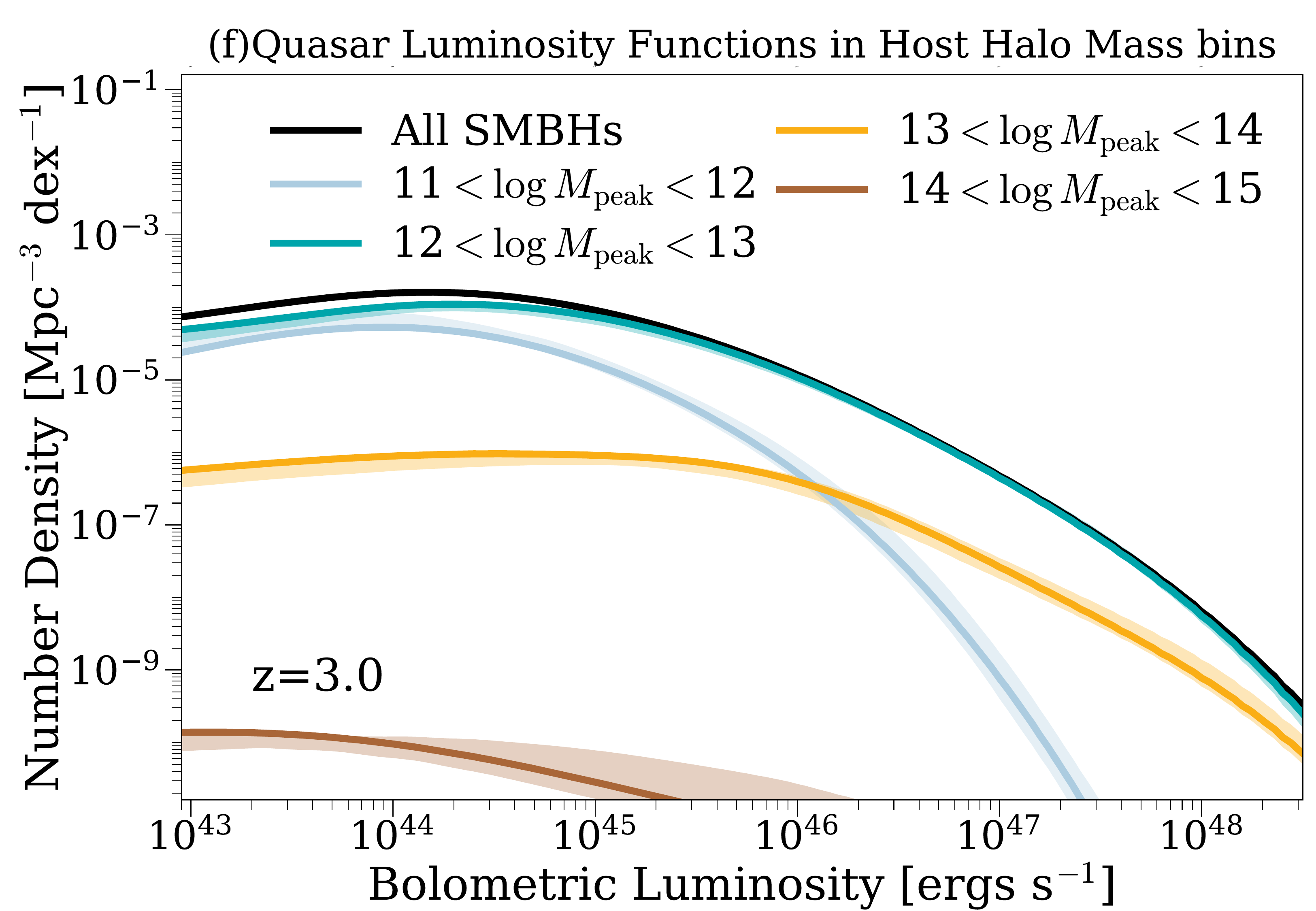}
}
\subfigure{
\includegraphics[width=0.48\textwidth]{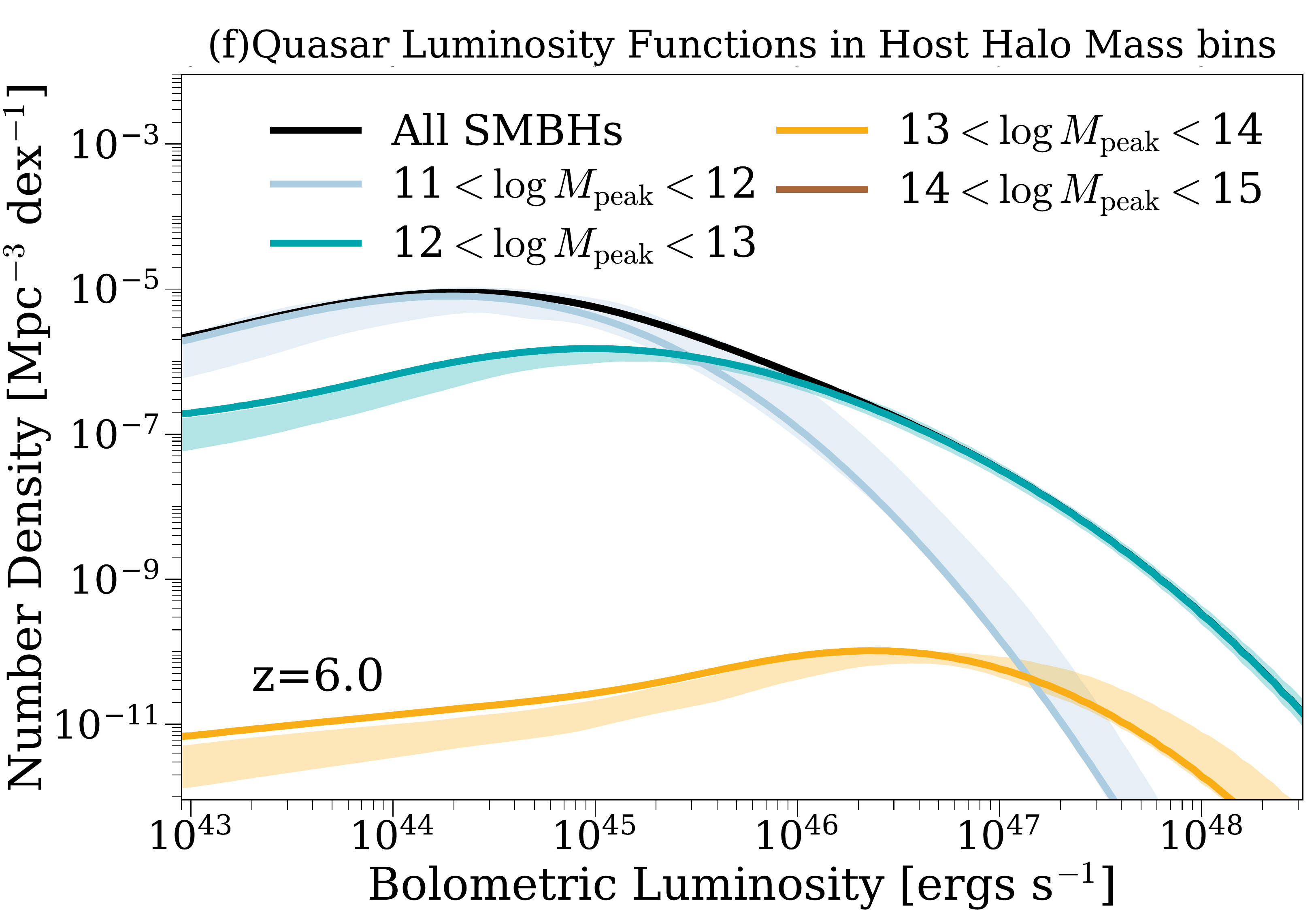}
}
\subfigure{
\includegraphics[width=0.48\textwidth]{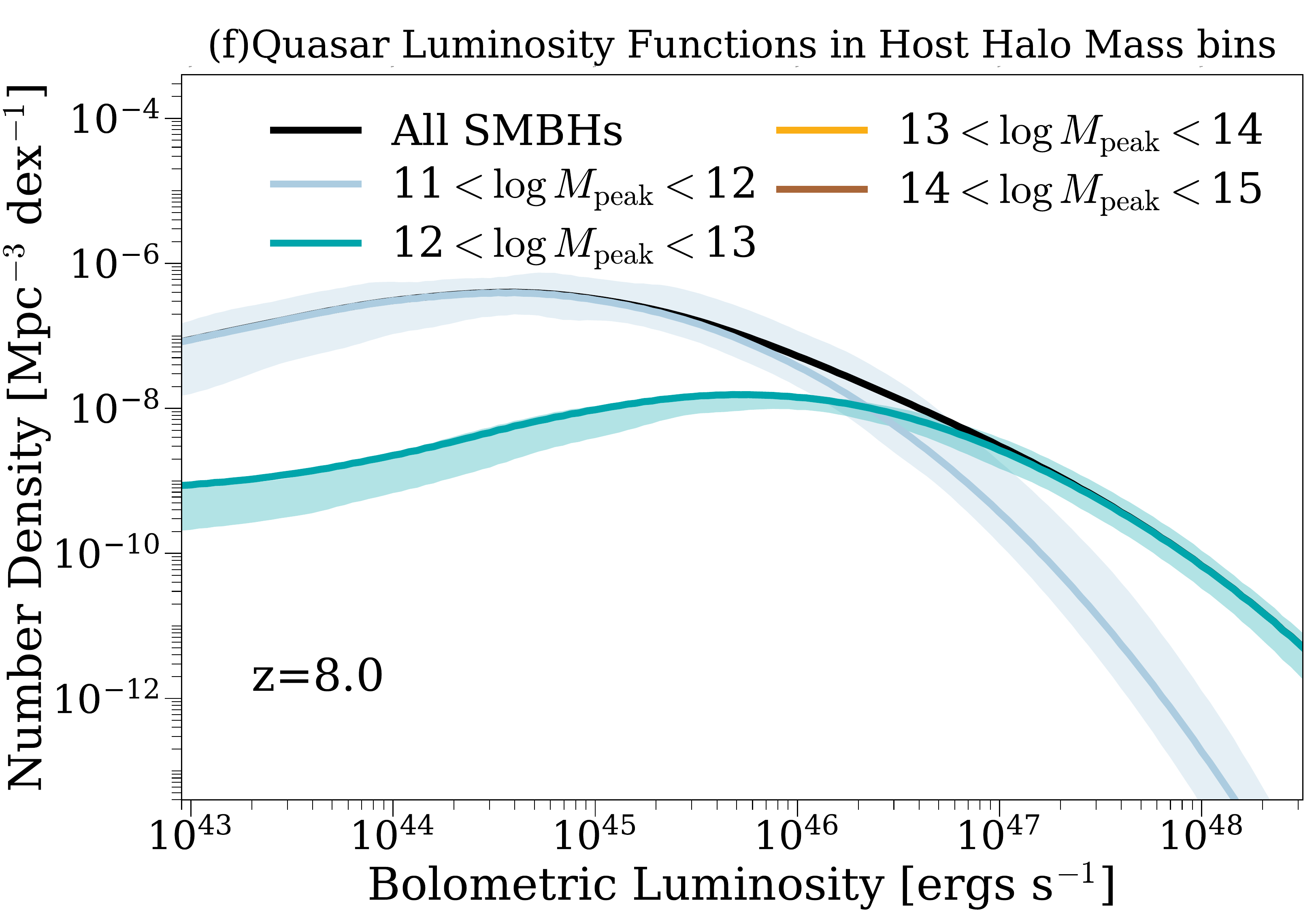}
}
\subfigure{
\includegraphics[width=0.48\textwidth]{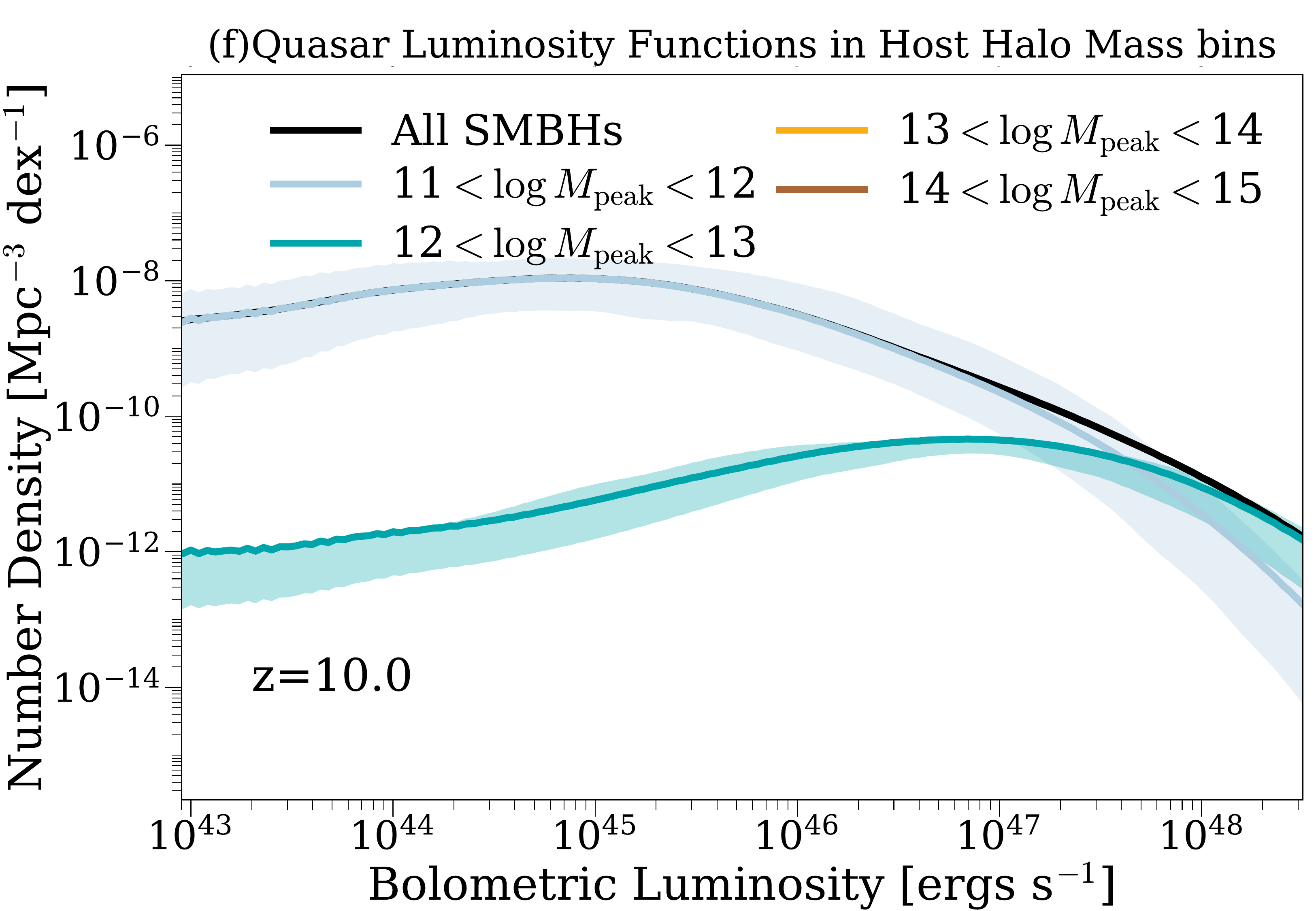}
}
\caption{Quasar luminosity functions in bins of host halo mass at $z=0.1$ (upper left), $z=1.0$ (upper right), $z=3.0$ (middle left), $z=6.0$ (middle right), $z=8.0$ (lower left), $z=10.0$ (lower right). \shadedregions{} See \S\ref{ss:qlf_mstar_mh}.}
\label{f:qlf_mh}
\end{figure*}

Fig.\ \ref{f:qlf_mh} shows QLFs in bins of host halo\footnote{the ``host halo'' here refers to individual subhaloes that host single galaxies in their centers, rather than the parent haloes that host multiple galaxies and/or SMBHs.} mass at different redshifts. At $z\gtrsim 6$, haloes with $10^{11} < M_\mathrm{peak} < 10^{12} M_\odot$ ($10^{12} < M_\mathrm{peak} < 10^{13} M_\odot$) dominate the faint (bright) end of QLFs. The critical luminosity where the contributions from these two halo mass bins are comparable decreases towards lower redshifts. This is because at higher redshifts, there are fewer massive haloes to dominate QLFs, except for the brightest end. From $z=3$ to $z=1$, haloes with $10^{12} < M_\mathrm{peak} < 10^{13} M_\odot$ dominate the whole QLF. Towards lower redshifts, they are gradually taken over by lower mass haloes at the faint end, largely due to the AGN downsizing effect. From $z=1$ to $z=0.1$, $10^{12} < M_\mathrm{peak} < 10^{13} M_\odot$ haloes are joined by $M_\mathrm{peak} > 10^{13} M_\odot$ haloes at the bright end. This is constrained by the local ABHMF from \citet{Schulze2010} and the local \bhbm{} relation, which require that massive AGNs live in $M_\mathrm{peak} > 10^{13} M_\odot$ haloes.

The relative importance of \mpeak$ \sim 10^{12}\Msun$ haloes in the QLF arises because these haloes are most efficient at forming galaxies, which in turn results in the most black hole growth compared to haloes in other mass ranges.  As a result, studies that probe quasar clustering usually find similar inferred host halo masses, with relatively weak redshift and luminosity trends.  This conclusion is further discussed in Papers VII and VIII, which focus on AGN clustering.

\subsubsection{Quasars binned by offset from the \bhsm{} relation}
\label{ss:qlf_mstar_bias}

\begin{figure*}
\subfigure{
\includegraphics[width=0.48\textwidth]{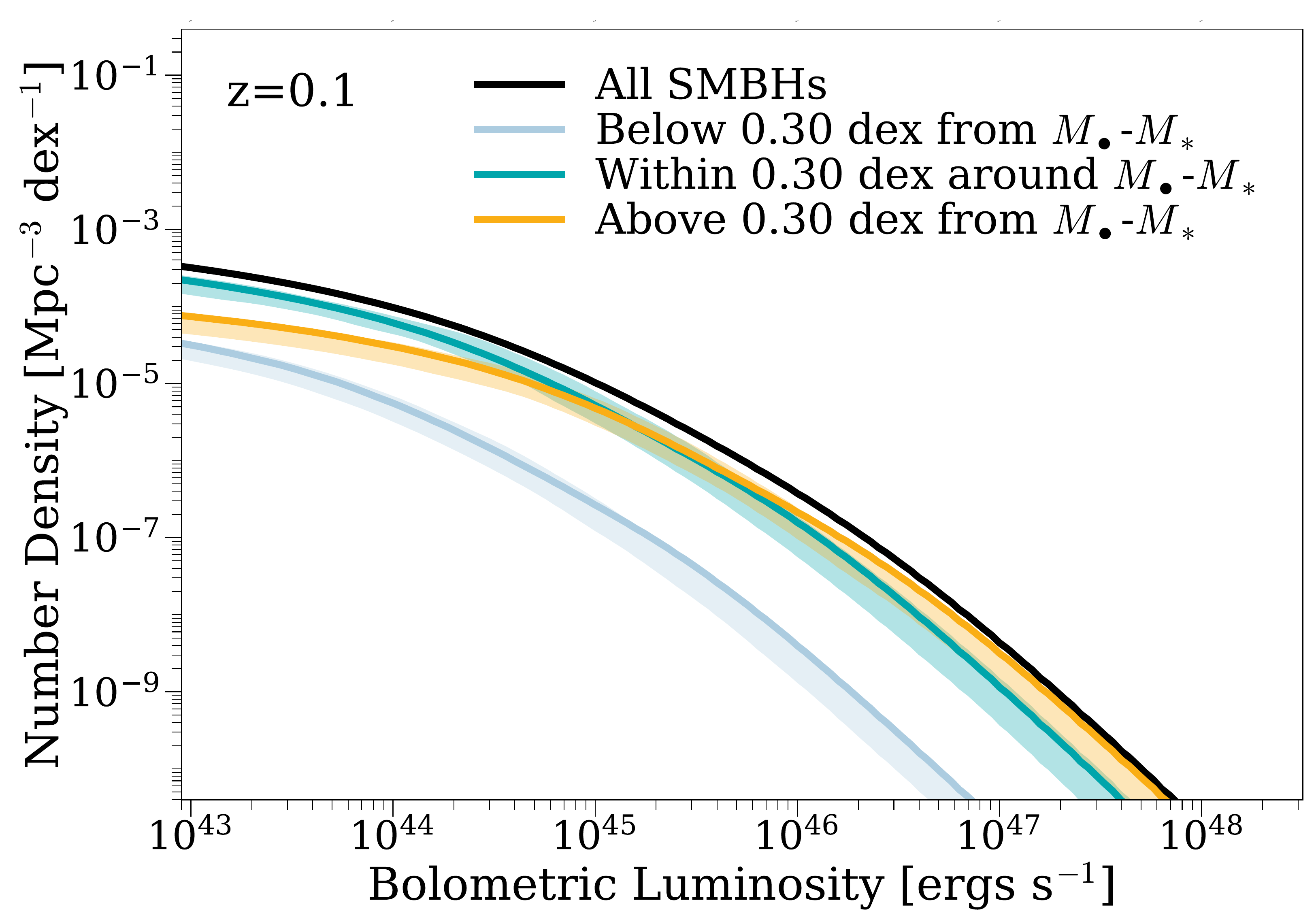}
}
\subfigure{
\includegraphics[width=0.48\textwidth]{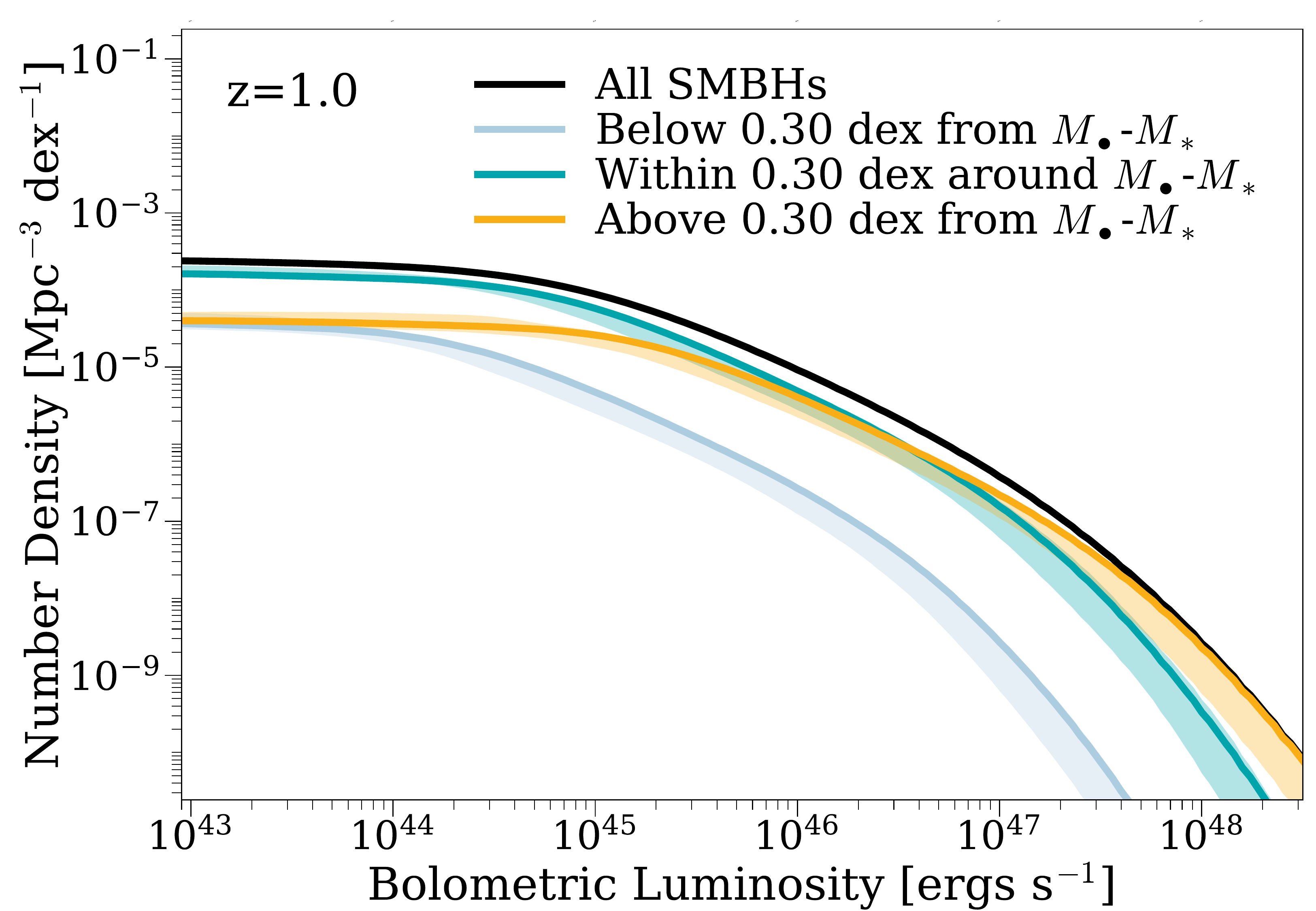}
}
\subfigure{
\includegraphics[width=0.48\textwidth]{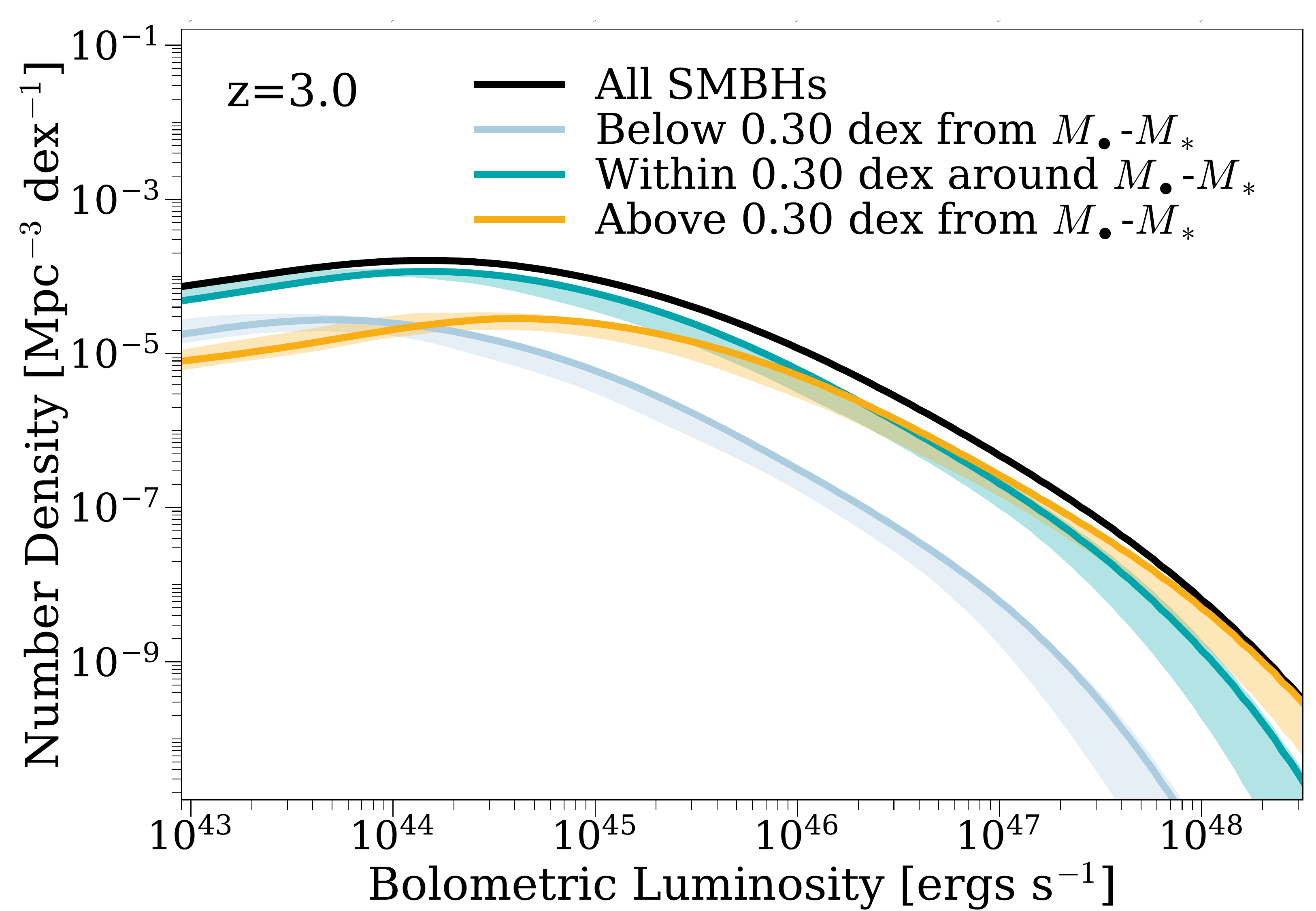}
}
\subfigure{
\includegraphics[width=0.48\textwidth]{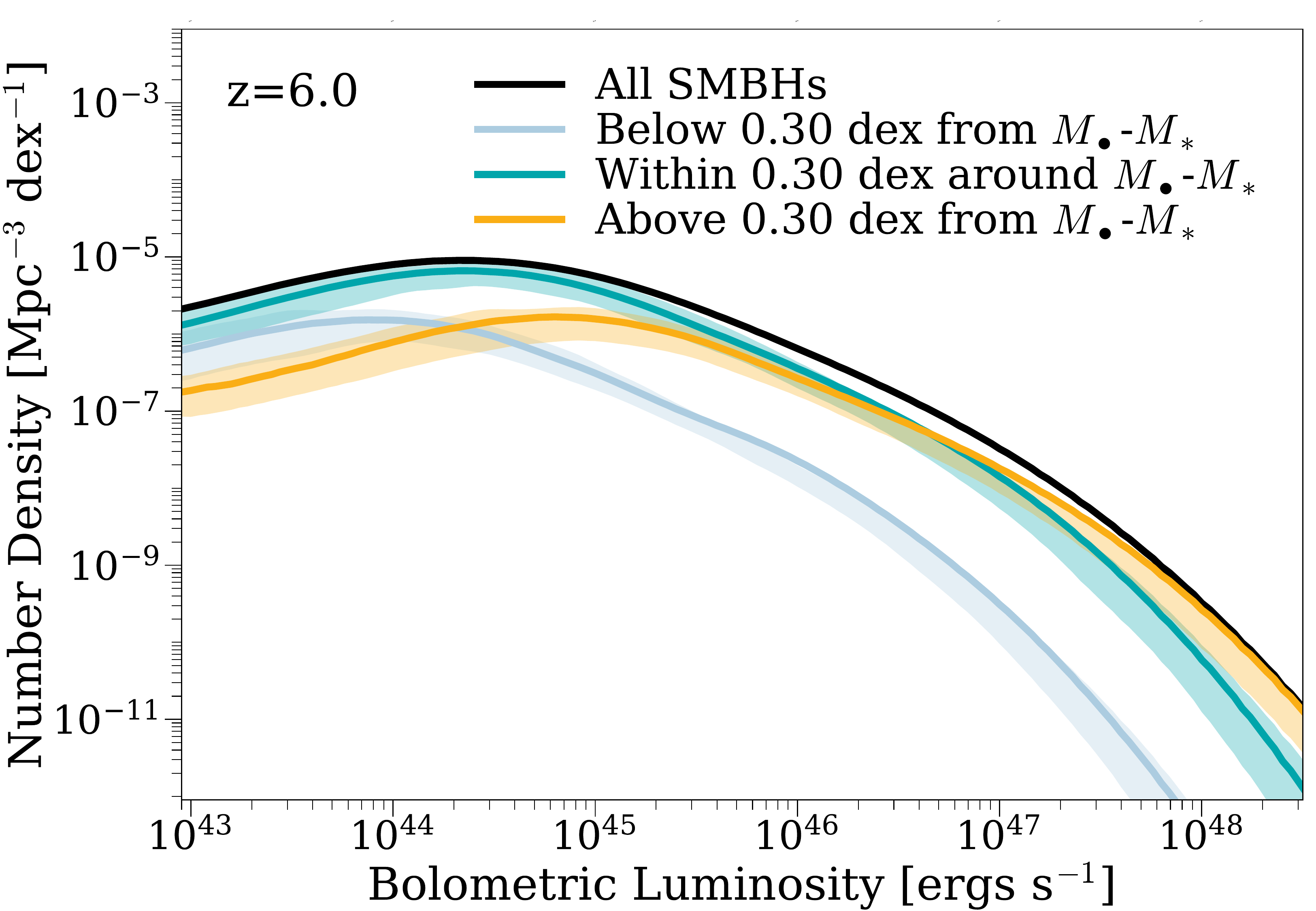}
}
\subfigure{
\includegraphics[width=0.48\textwidth]{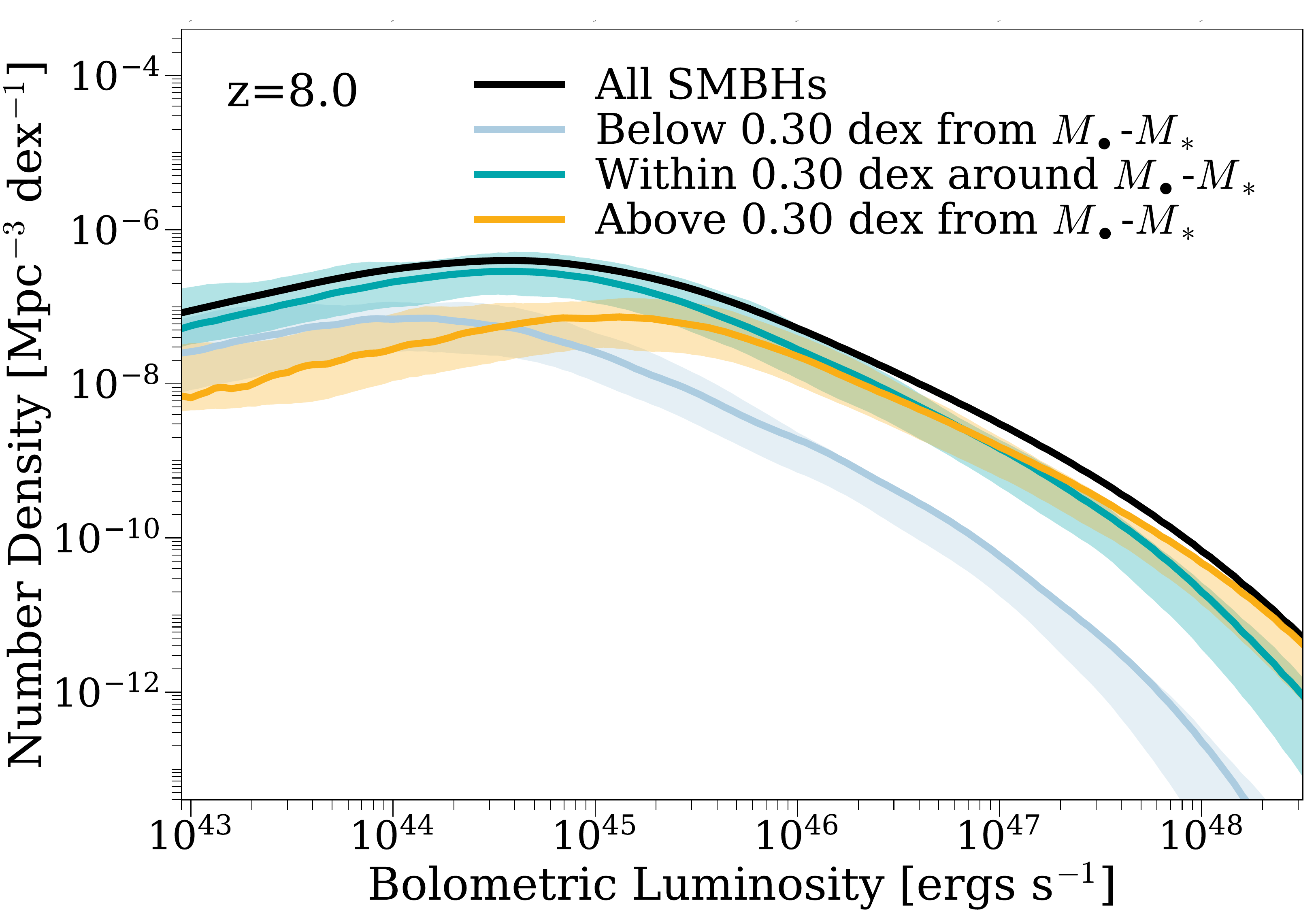}
}
\subfigure{
\includegraphics[width=0.48\textwidth]{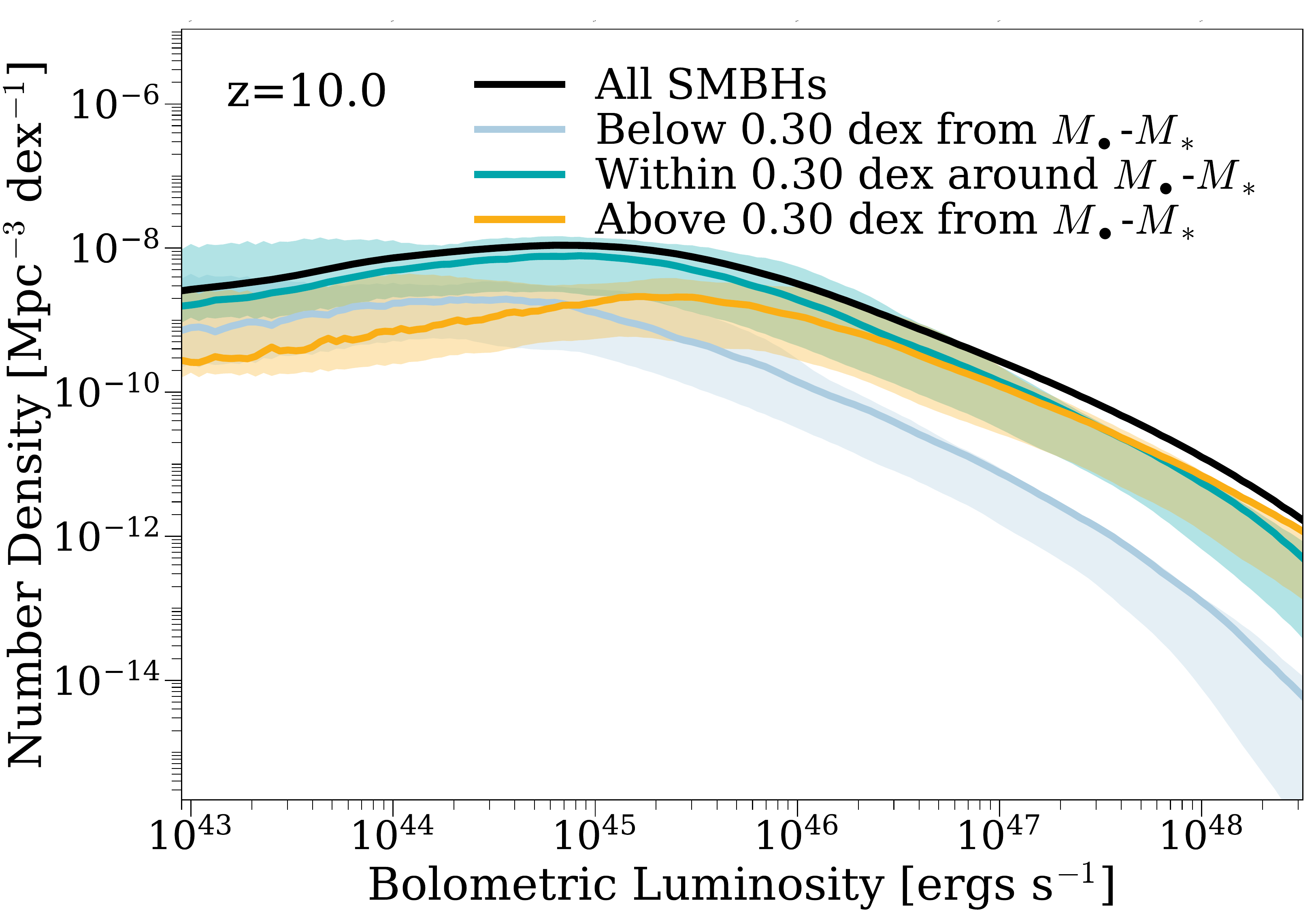}
}
\caption{Quasar luminosity functions in bins of the deviation in SMBH mass from the median \bhsm{} relation at $z=0.1$ (upper left), $z=1.0$ (upper right), $z=3.0$ (middle left), $z=6.0$ (middle right), $z=8.0$ (lower left), and $z=10.0$ (lower right). \shadedregions{} See \S\ref{ss:qlf_mstar_bias}.}
\label{f:qlf_mstar_bias}
\end{figure*}

Fig.\ \ref{f:qlf_mstar_bias} shows QLFs in bins of offset in SMBH mass compared to the median \bhsm{} relation at different redshifts. This figure quantifies the Lauer bias, i.e., the amount by which brighter quasars are more likely to be driven by over-massive SMBHs (compared to their host galaxy mass) than typical SMBHs \citep[][]{Lauer2007}. Observations constrain this effect mainly because extremely massive host galaxies are very rare by themselves, and cannot account for the observed number density of bright quasars (e.g., \citealt{Yang2021}, also see \citealt{Zhang2023b}). According to \textsc{Trinity}, over-massive SMBHs ($>0.3$ dex above the median \bhbm{} relation) contribute significantly to QLFs down to $\log_{10}L_\mathrm{bol} \sim 46-47$ beyond $z\sim 1$. At $z\sim 0$, this limit is lower, $\log_{10}L_\mathrm{bol} \sim 45.3$. This results from the overall decline in SMBH activity level. Typical SMBHs around the intrinsic \bhsm{} relation dominate QLFs at $\log_{10}L_\mathrm{bol} \sim 45$. Such a strong Lauer bias demonstrates the necessity of including faint AGN (i.e., $\log_{10}L_\mathrm{bol} \sim 45$) when studying the redshift evolution of the galaxy--SMBH mass connection, instead of comparing the local \bhsm{} relation with quasar mass estimates.  Under-massive SMBHs are always subdominant contributors to QLFs throughout cosmic time. Given that \citet{Zhang2023b} and \citet{Zhang2023c} have already shown the \bhsm{} relations for luminosity-selected AGNs at $z\geq 6$, we refer readers to these two papers for the plots of such scaling relations at $z\geq 6$, and \href{https://github.com/HaowenZhang/TRINITY/tree/main/plots/Paper3/Additional_BHSM_lbol}{https://github.com/HaowenZhang/TRINITY/tree/main/plots/Paper3/\\Additional\_BHSM\_lbol} for the plots at $z < 6$.

\subsubsection{Quasars binned by Eddington ratio}
\label{ss:qlf_eta}

\begin{figure*}
\subfigure{
\includegraphics[width=0.48\textwidth]{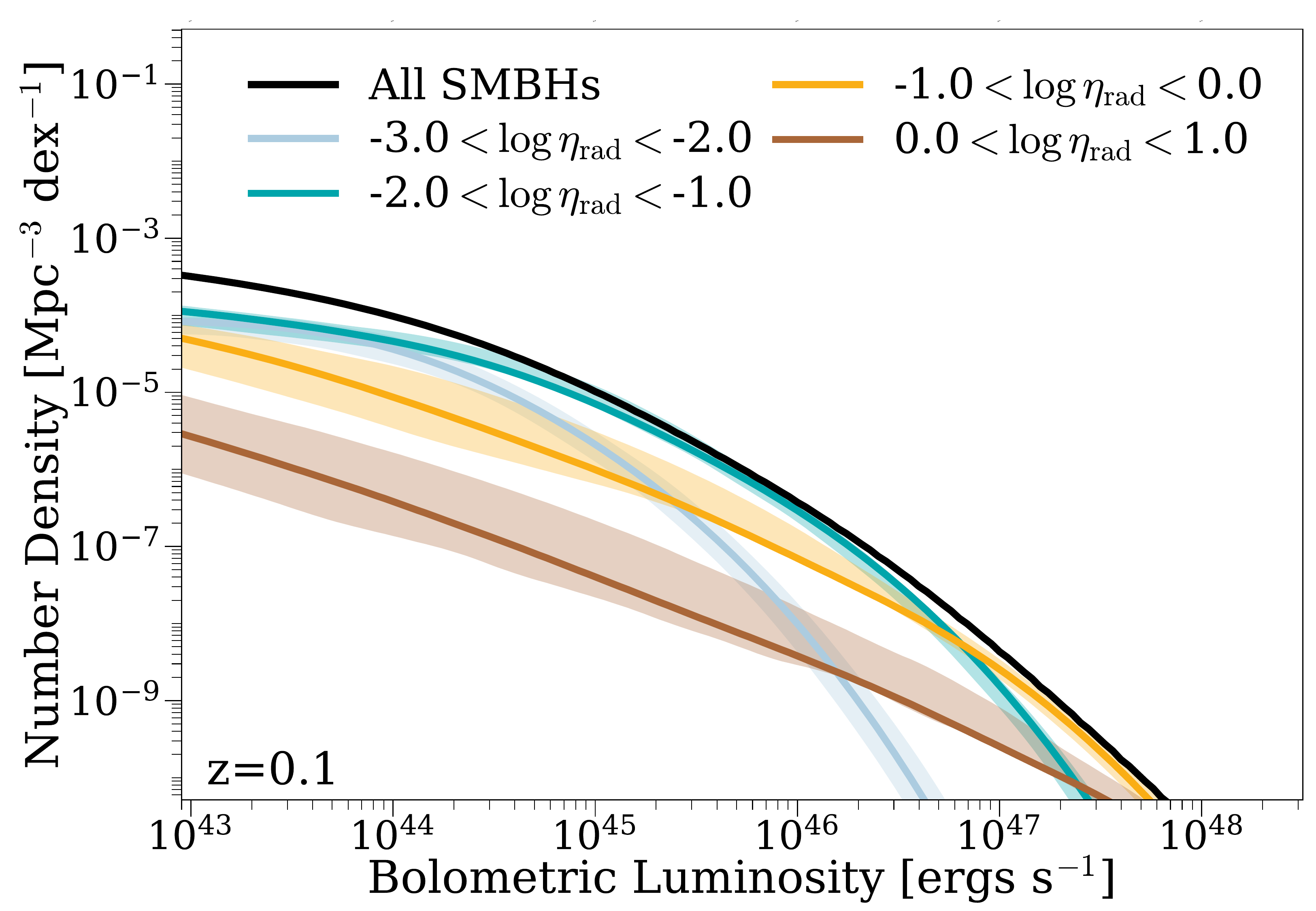}
}
\subfigure{
\includegraphics[width=0.48\textwidth]{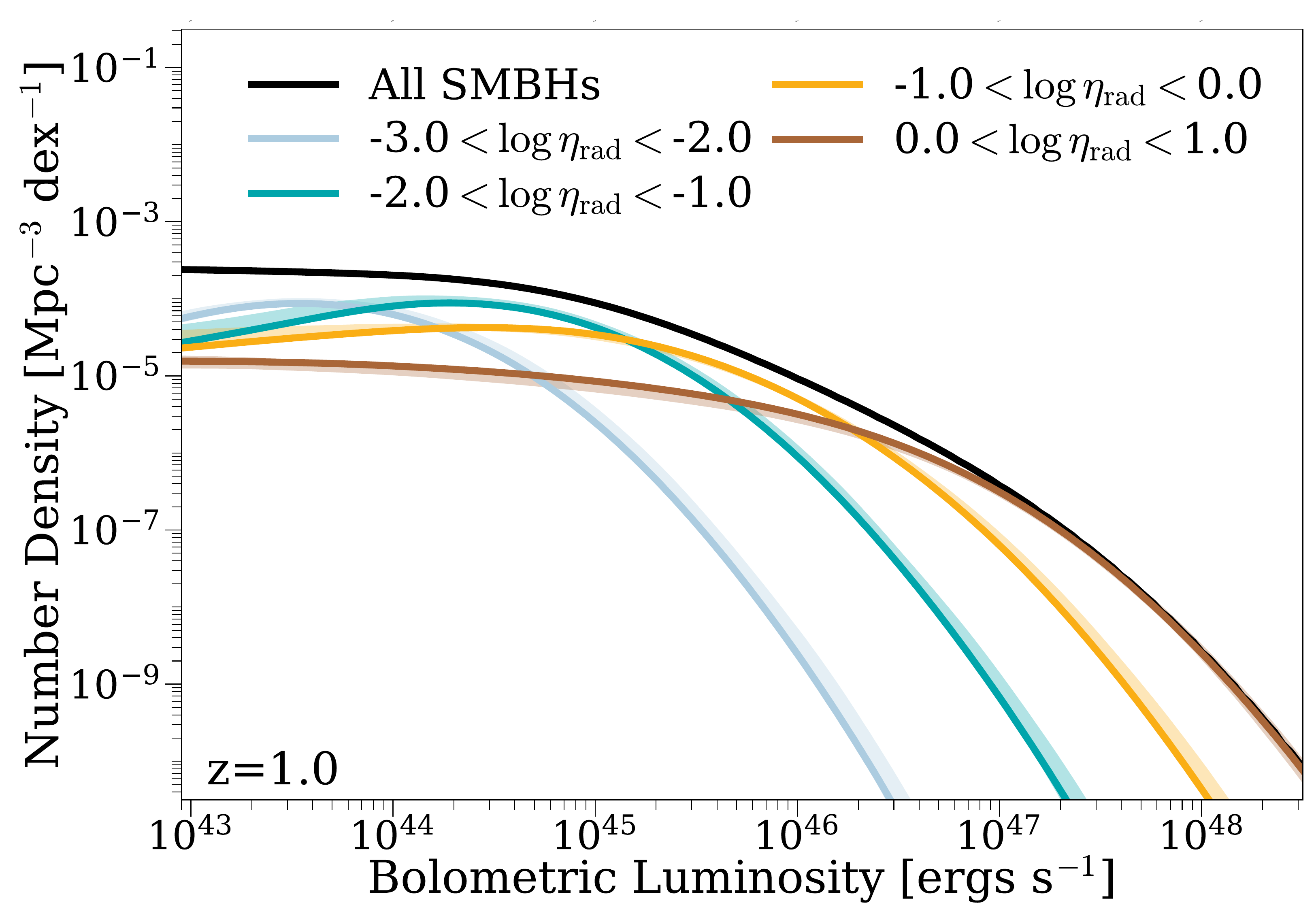}
}
\subfigure{
\includegraphics[width=0.48\textwidth]{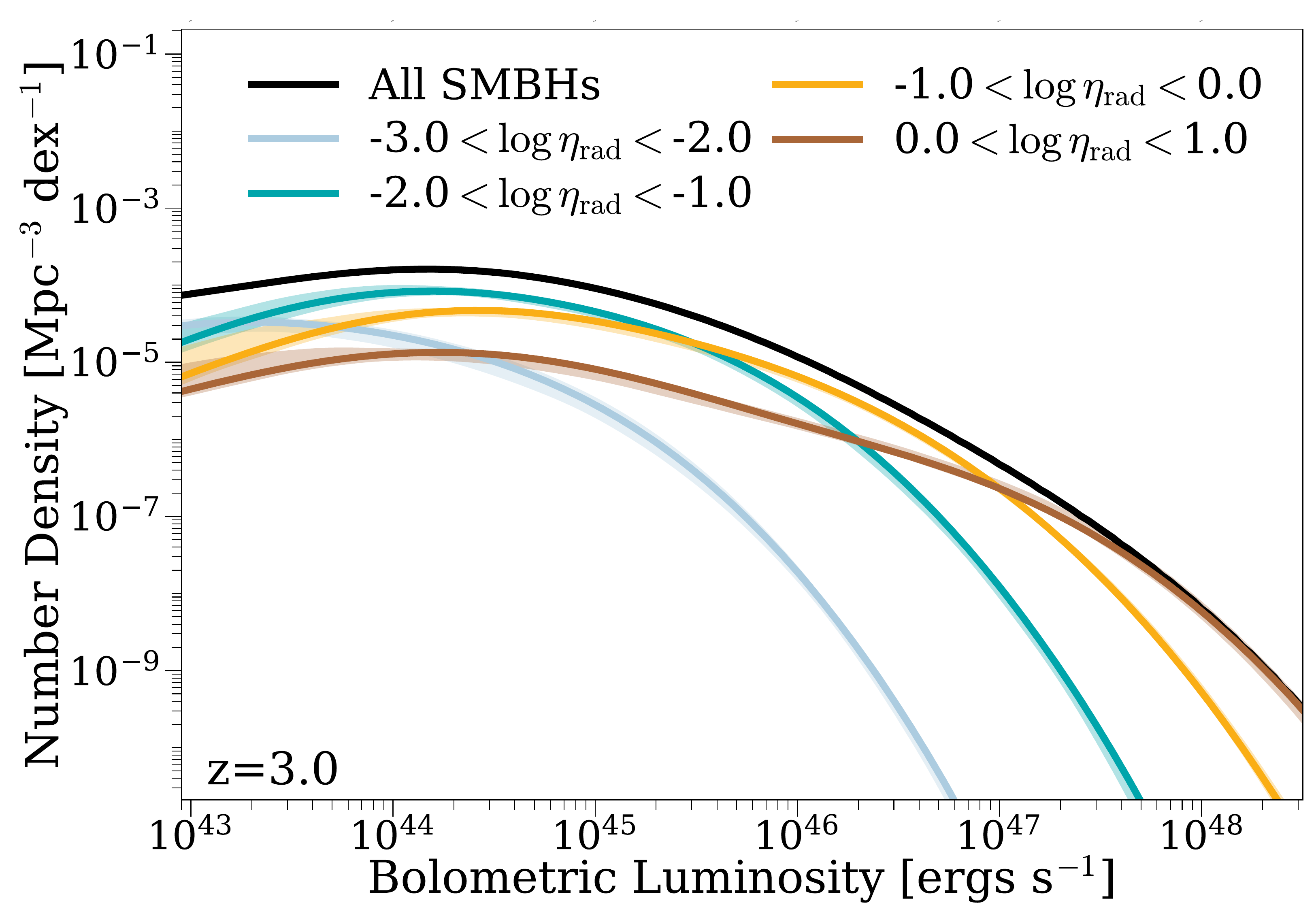}
}
\subfigure{
\includegraphics[width=0.48\textwidth]{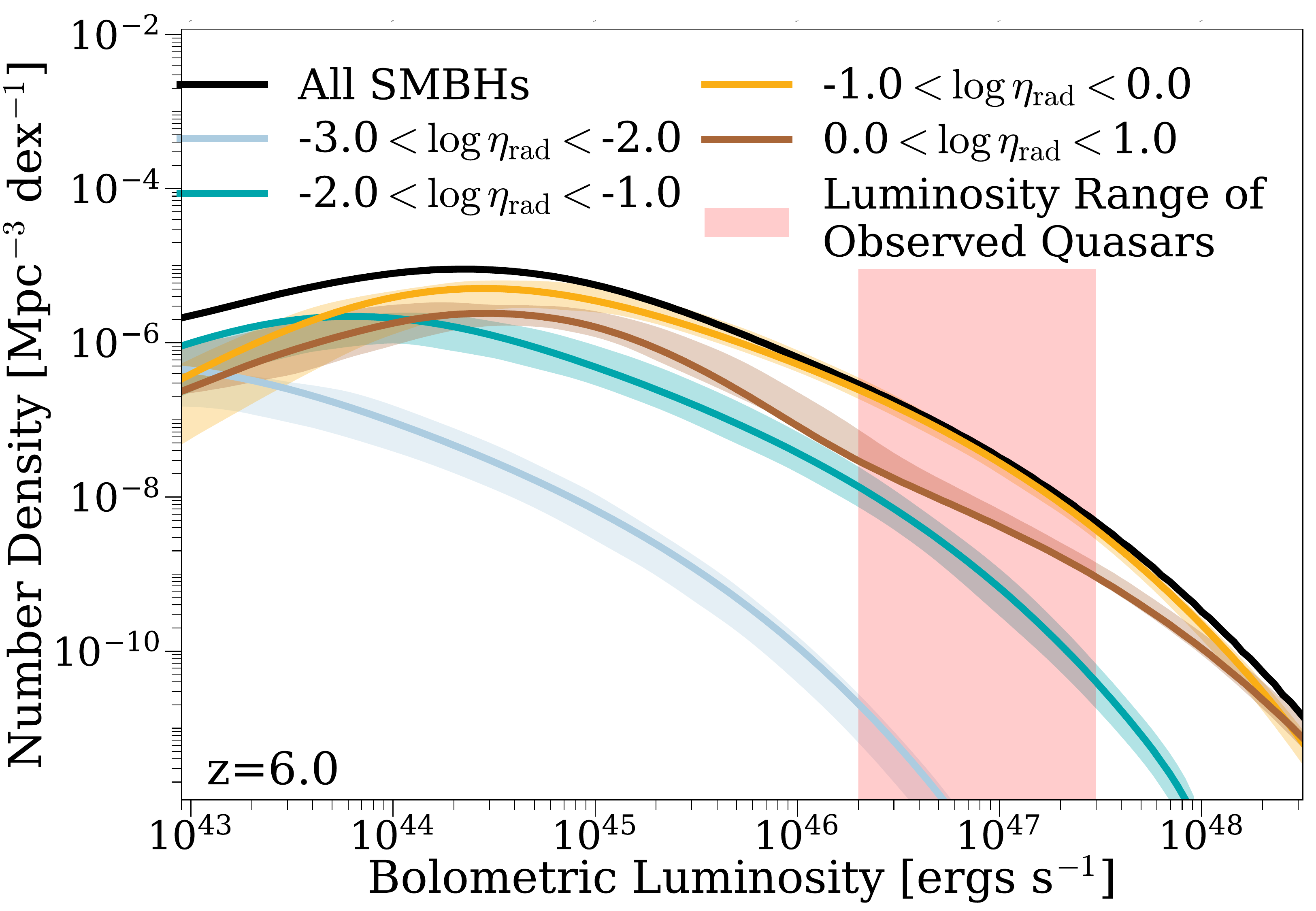}
}
\subfigure{
\includegraphics[width=0.48\textwidth]{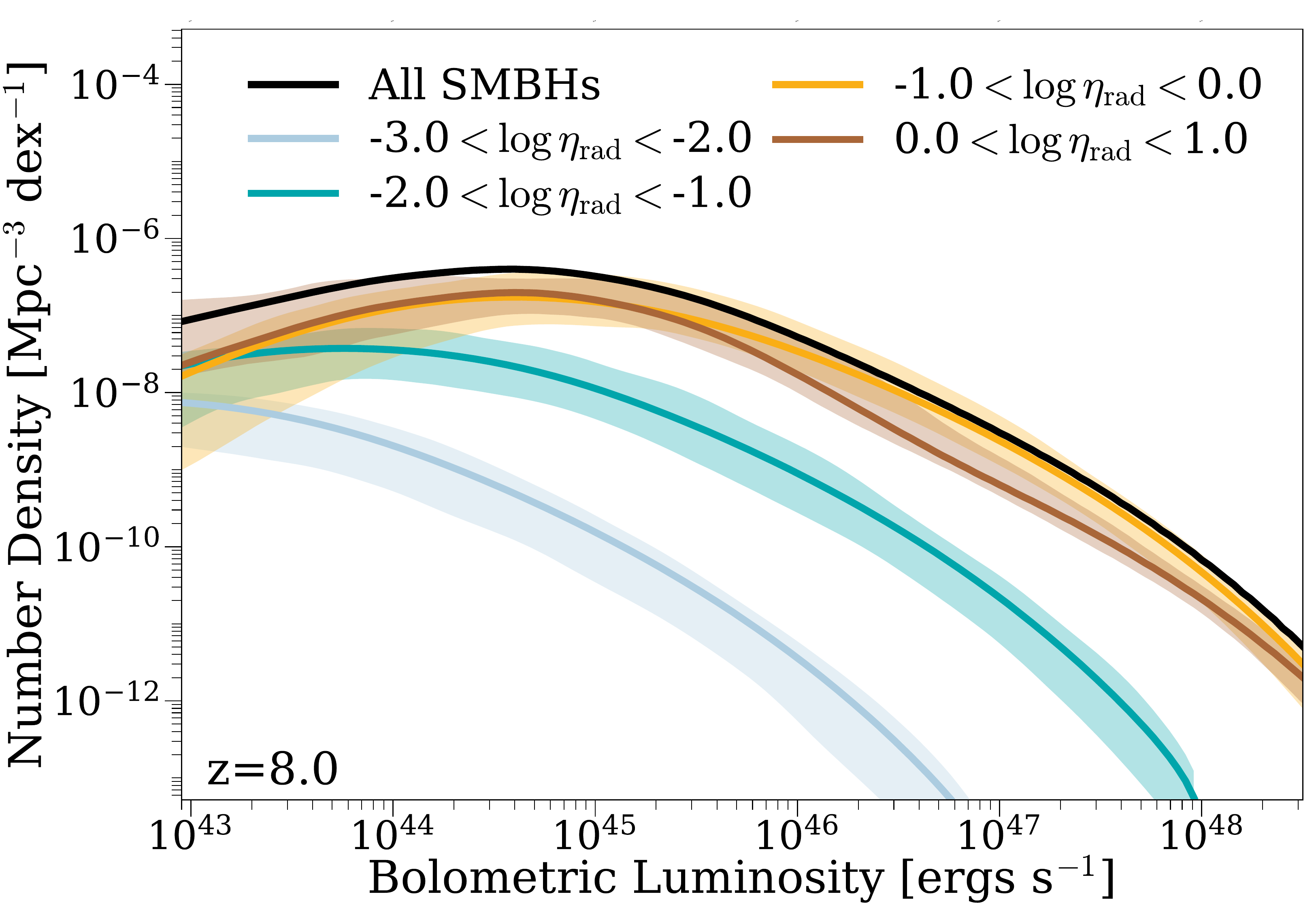}
}
\subfigure{
\includegraphics[width=0.48\textwidth]{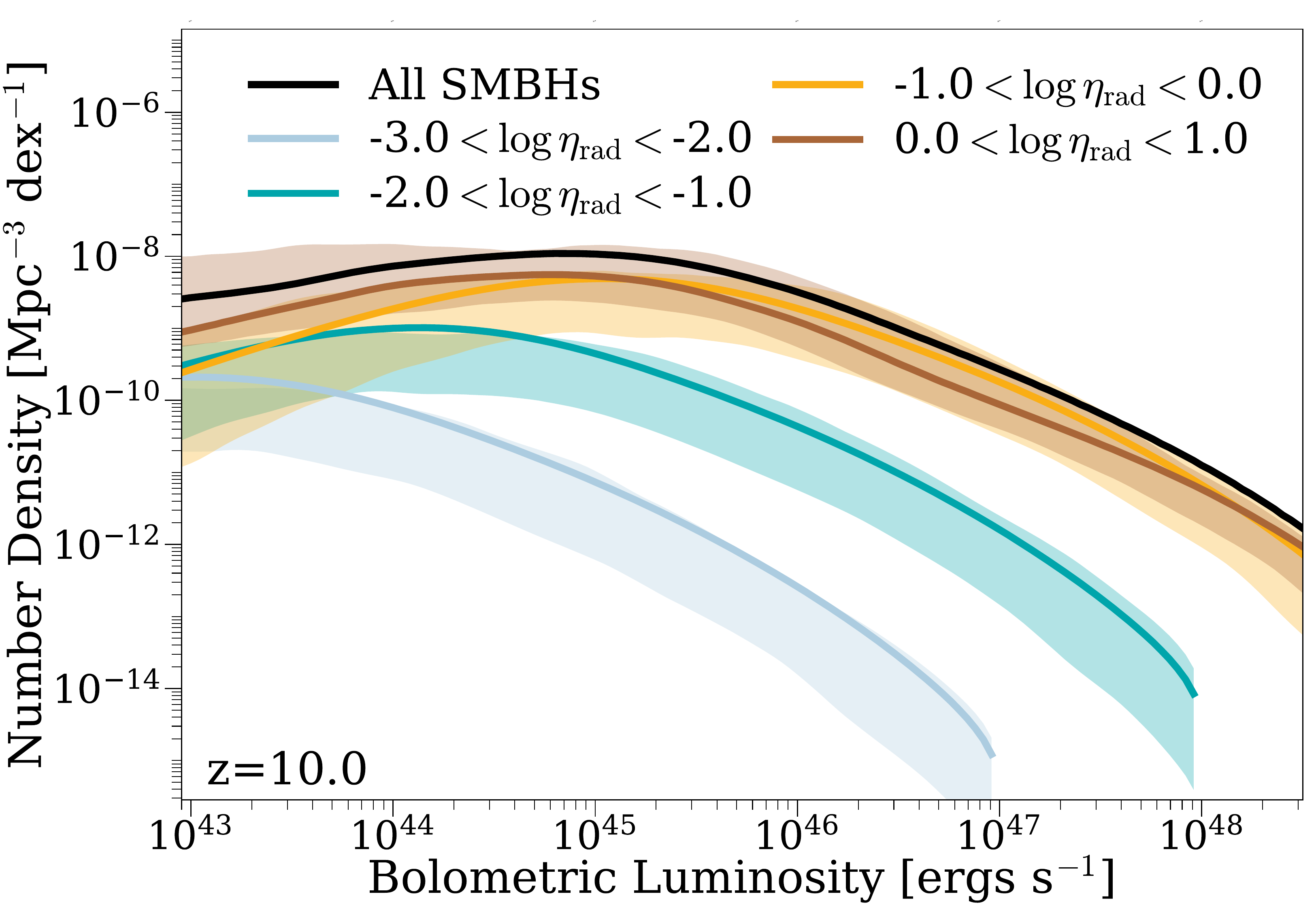}
}
\caption{Quasar luminosity functions split in different Eddington ratio slices at $z=0.1$ (upper left), $z=1.0$ (upper right), $z=3.0$ (middle left), $z=6.0$ (middle right), $z=8.0$ (lower left) and $z=10.0$ (lower right).\shadedregions{} The red shaded regions denotes the typical luminosity range for observed $z\sim 6$ quasars. See \S\ref{ss:qlf_eta}.}
\label{f:qlf_eta}
\end{figure*}

Fig.\ \ref{f:qlf_eta} shows QLFs in Eddington ratio bins at different redshifts. At low redshifts, Eddington ratio distributions are broader. Consequently, quasars with different Eddington ratios dominate different parts of QLFs, with higher-Eddington ratio objects being generally brighter. Between $1\lesssim z \lesssim 3$, super-Eddington AGNs dominate the QLFs at $\log L_\mathrm{bol} \gtrsim 47.5$. This comes from the extrapolation of QPDFs from \citet{Aird2018} towards highest specific black hole accretion rates (sBHARs). In \textsc{Trinity}, we chose not to include QPDFs from \citet{Aird2018} at the high-sBHAR end. This is because they adopted a smoothness prior on modeled QPDFs, which may lead to systematic issues when the sample size is small. A larger sample of massive galaxies with highly active SMBHs is needed for more robust determination of QPDFs in this regime, which will serve as a strong test of this prediction by \textsc{Trinity}. We also note that the MCMC uncertainties for QLFs in different Eddington ratio bins are small between $1\lesssim z \lesssim 3$. This is because the parameterization of the \textsc{Trinity} model is not flexible enough to cover the \emph{full uncertainty ranges} as determined by \citet{Aird2018}. Consequently, our estimates of QLF uncertainties in different Eddington ratio bins are likely lower limits. Towards higher redshifts, SMBHs of all masses become more active. In other words, Eddington ratio distributions become narrower and shift towards higher values. At $6\lesssim z\lesssim 8$, essentially all bright SMBHs are accreting at close to the Eddington rate. As a result, high-redshift QLFs are dominated by quasars with $0.1 < \eta_\mathrm{rad} < 1.0$. At $9\lesssim z\lesssim 10$, the contribution from super-Eddington quasars becomes more significant. The recent detection and measurement of the SMBH in GN-z11 is consistent with this picture \citep{Maiolino2023}, although a larger sample from future observations by, e.g., \textit{JWST}, is needed for a statistical comparison.

\subsection{AGN radiative power densities}
\label{ss:agn_rho_rad}

\begin{figure}
\subfigure{
\includegraphics[width=0.48\textwidth]{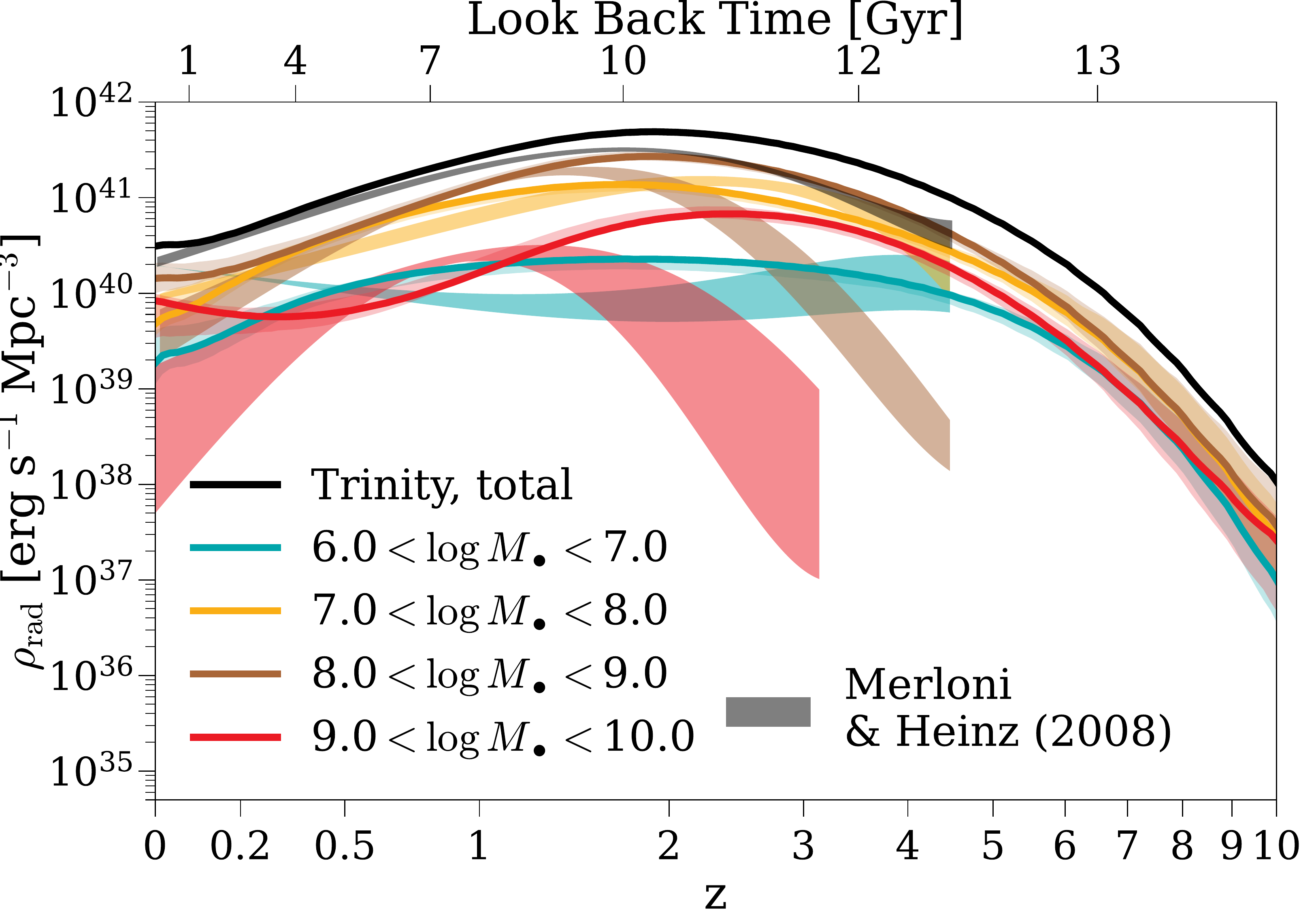}
}
\subfigure{
\includegraphics[width=0.48\textwidth]{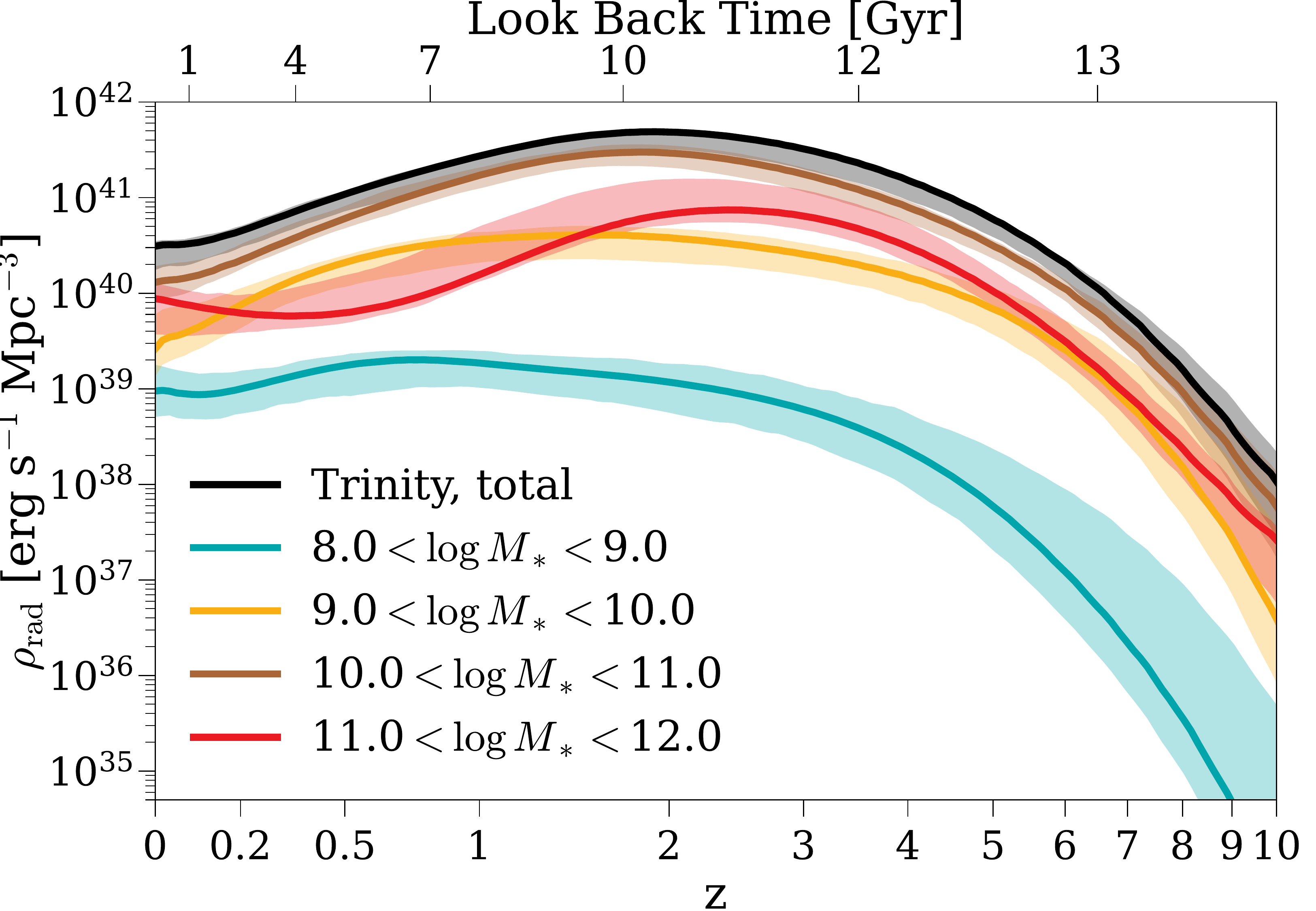}
}
\caption{\textbf{Top Panel:} Cosmic AGN radiative power densities from \textsc{Trinity} and \citet{Merloni2008}. Different colors denote the contribution from black holes of various masses. The shaded regions around the solid curves are the 68\% confidence intervals from \textsc{Trinity}, while other shaded regions denote the predictions from \citet{Merloni2008}. \textbf{Bottom Panel:} The same as the top panel, but different colors denote the contributions from different \emph{stellar mass} bins. See \S\ref{ss:agn_rho_rad}.}
\label{f:rho_rad}
\end{figure}

The top panel of Fig.\ \ref{f:rho_rad} shows the cosmic AGN radiative power densities, $\rho_\mathrm{rad}$, from \textsc{Trinity} and \citet{Merloni2008}. Qualitatively, $\rho_\mathrm{rad}$ (including the total $\rho_\mathrm{rad}$ from different SMBH populations) from \textsc{Trinity} is broadly consistent with that from \citet{Merloni2008} below $z\sim 2$. But black holes in \textsc{Trinity} have much higher Eddington ratios at $z\gtrsim 4$. As a result, we also predict higher radiative power densities for every mass bin compared to \citet{Merloni2008} in the early Universe. Another significant difference is in the low-redshift behavior of black holes with $10^6 M_\odot < M_\bullet < 10^7 M_\odot$: \citet{Merloni2008} predicts a power density that increases with time below $z\sim0.5$, while we find the opposite. One potential explanation is that in addition to luminosity functions, we included the quasar probability distribution functions from \citet{Aird2018}. These data put upper limits on the energy output from low mass black holes in \textsc{Trinity}, which were absent in \citet{Merloni2008}. The bottom panel of Fig.\ \ref{f:rho_rad} shows the same $\rho_\mathrm{rad}$, as well as the contributions from different \emph{galaxy} populations. In terms of redshift evolution, $\rho_\mathrm{rad}$ from different galaxy populations are qualitatively similar to those of different SMBH populations, including the ``AGN downsizing'' effect. There are offsets between curves of the same color from these two panels, especially for lower mass galaxies/SMBHs at higher redshifts. This is mostly driven by the stronger redshift evolution in the \bhsm{} relation at the less-massive end.

\subsection{Black hole duty cycles}
\label{ss:results_duty_cycle_eff}

\begin{figure}
\subfigure{
\includegraphics[width=0.48\textwidth]{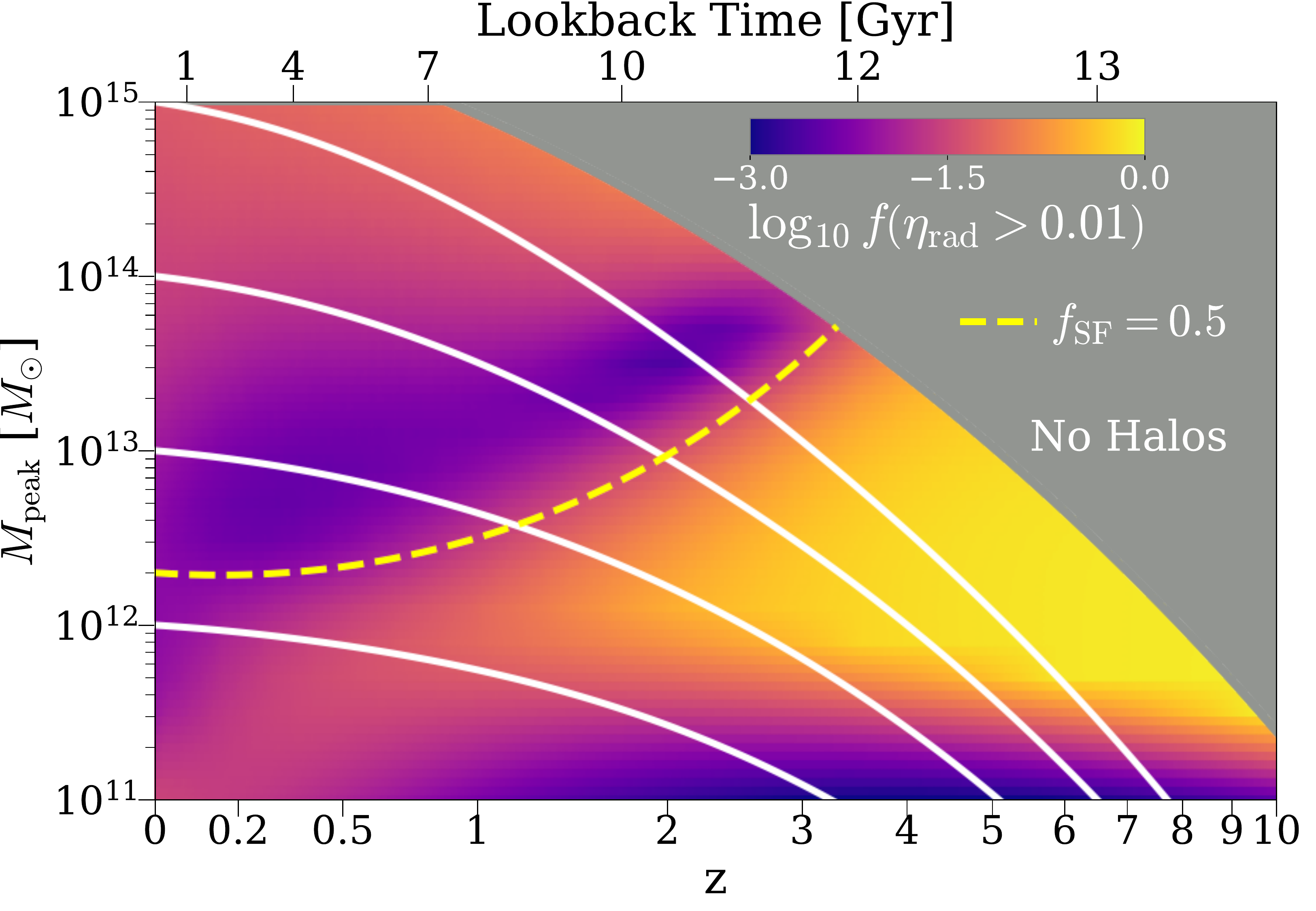}
}
\subfigure{
\includegraphics[width=0.48\textwidth]{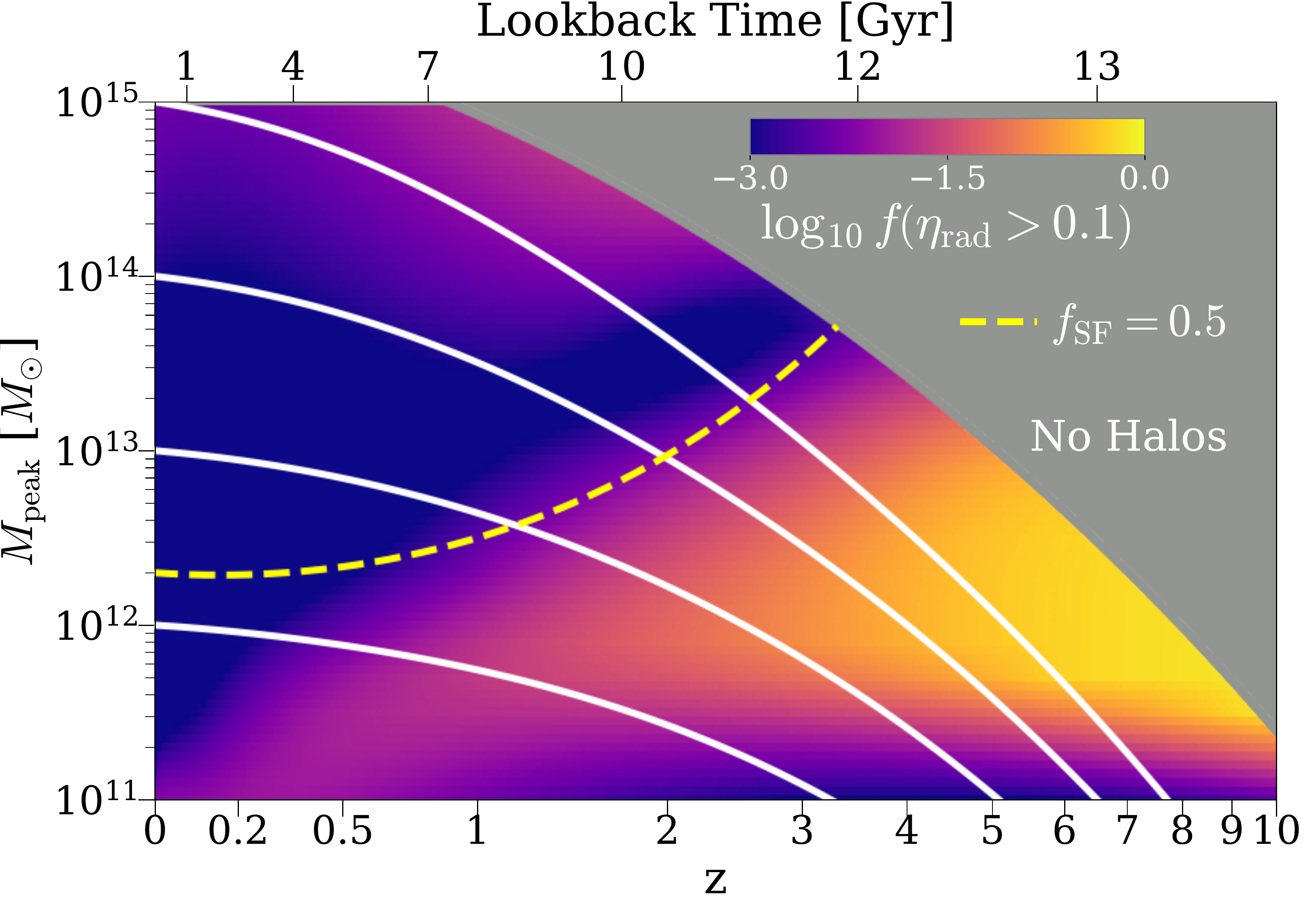}
}
\caption{\textbf{Top Panel:} the fractions of black holes accreting at $\eta_{\rm rad} > 0.01$ as a function of \mpeak{} and $z$. \textbf{Bottom Panel:} the fractions of black holes accreting at $\eta_{\rm rad} > 0.1$ as a function of \mpeak{} and $z$. \halocurves{} \sfcurve{} See \S \ref{ss:results_duty_cycle_eff}.}
\label{f:f_ledd}
\end{figure}

\begin{figure}
\subfigure{
\includegraphics[width=0.48\textwidth]{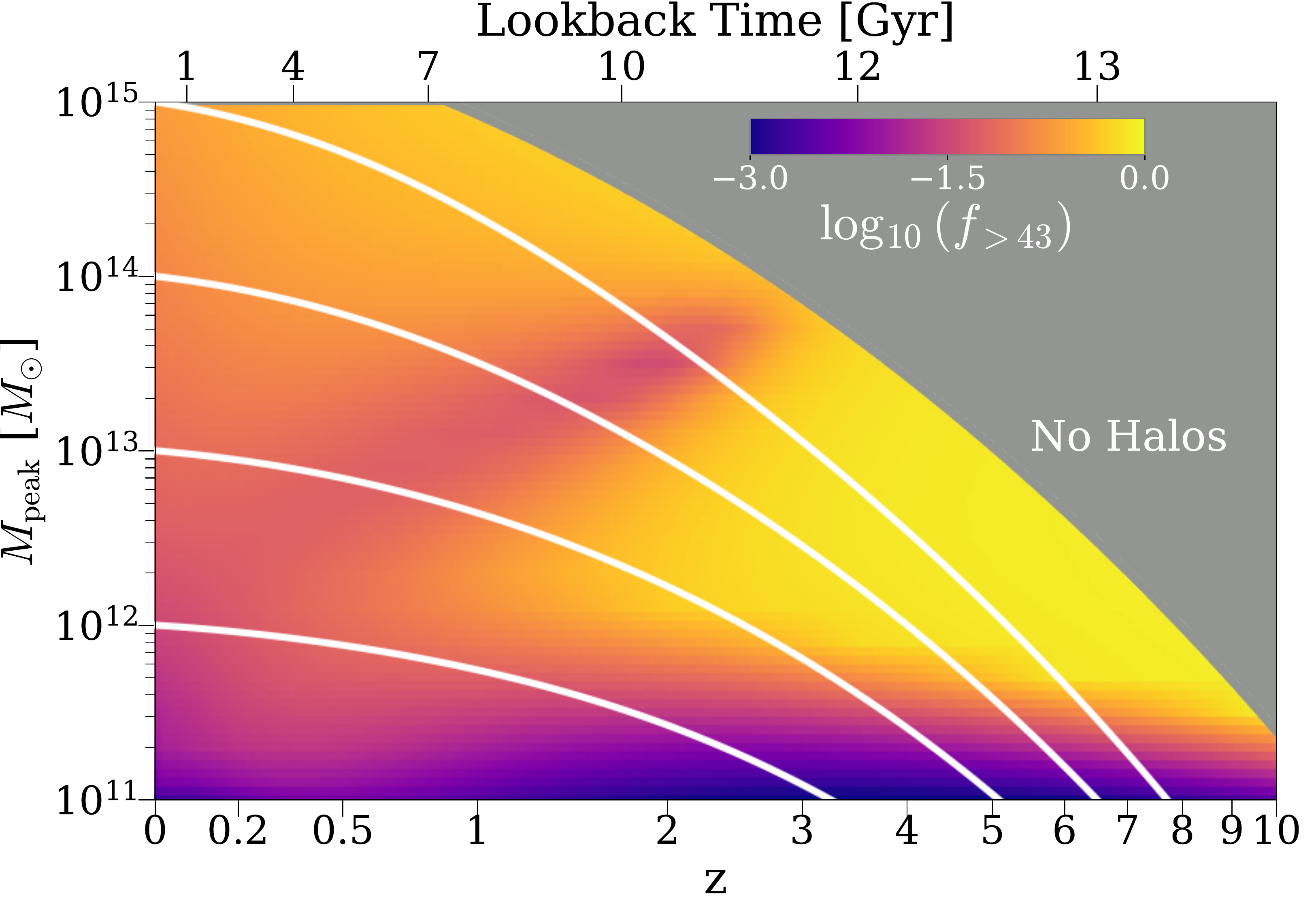}
}
\subfigure{
\includegraphics[width=0.48\textwidth]{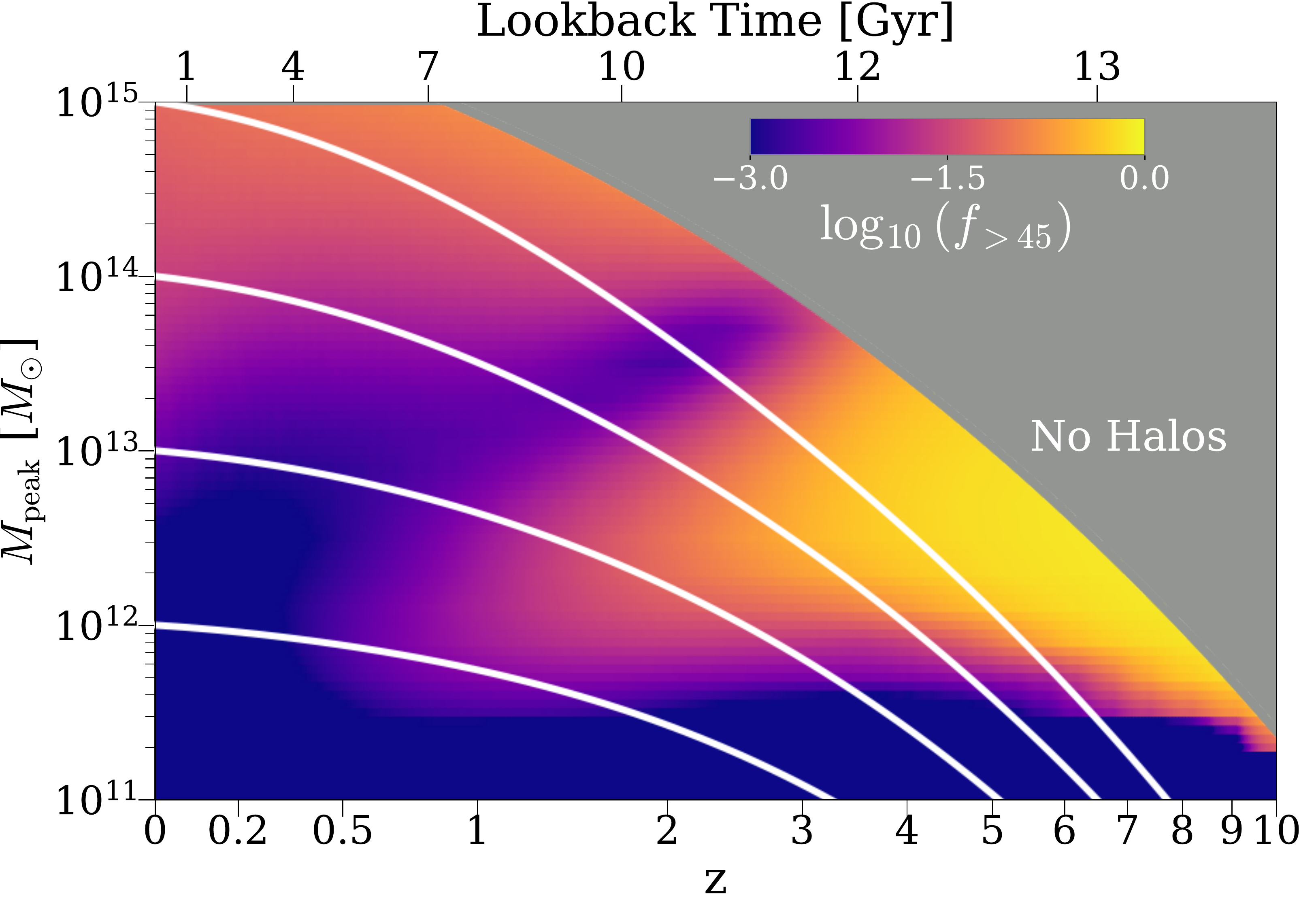}
}
\caption{\textbf{Top Panel:} the fractions of black holes with $L_{\rm bol} > 10^{43}$ erg/s as a function of \mpeak{} and $z$. \textbf{Bottom Panel:} the fractions of black holes with $L_{\rm bol} > 10^{45}$ erg/s as a function of \mpeak{} and $z$. \halocurves{} See \S\ref{ss:results_duty_cycle_eff}. }
\label{f:f_lbol}
\end{figure}

\textsc{Trinity} predicts SMBH Eddington ratio distributions as functions of halo/galaxy/SMBH mass and redshift. This enables us to calculate AGN duty cycles based on observables such as Eddington ratios or luminosities. Fig.\ \ref{f:f_ledd} shows AGN duty cycles when defined as the fractions of black holes accreting at $\eta_{\rm rad}>0.01$ (\fcent, top panel) and $\eta_{\rm rad}>0.1$ (\fdime, bottom panel), as a function of \mpeak{} and $z$. In general, both fractions increase with redshift. In general, \fcent{} and \fdime{} are significantly lower in haloes with $\log M_\mathrm{peak} \lesssim 11.5$. This is constrained by the QPDFs from \citet{Aird2018}, which indicates lower duty cycles for low-mass galaxies. Note that these QPDFs only cover a redshift range of $z \leq 2.5$. Thus, the prediction for low-mass haloes beyond $z\sim 2.5$ are effectively extrapolations by \textsc{Trinity}. Overall, such low AGN duty cycles in low-mass haloes mean that SMBHs in these low-mass haloes/galaxies likely grow in intermittent accretion events. The majority of their growth histories are instead dominated by a dormant/low-activity phase. During this dormant/low-activity phase, any AGN feedback for these dwarf galaxies is likely dominated by kinetic energy, as suggested in Appendix \ref{a:rad_kin_edd}. These low duty cycles in low-mass haloes is also seen in many recent theoretical studies, e.g., \citealt{Dubois2015,Bower2017,Habouzit2017,AnglesAlcazar2017} (also see the summary by \citealt{Tillman2022}). In these models, the gas fuel for SMBH growth are likely removed by supernova feedback in low-mass galaxies. While their host galaxies keep growing, these SMBHs cannot grow substantially via extended and intensive accretion events. SMBHs growth becomes efficient after their host galaxies are too massive for supernova feedback to be efficient, which allows gas reservoir to build up in galaxy centers. The low AGN duty cycles can also solve a common issue in many theoretical models of galaxy formation: many of them are not calibrated against quasar luminosity functions (QLFs). So they tend to overproduce the faint ends of QLFs (e.g., \citealt{Sijacki2015,Volonteri2016,Habouzit2017,Habouzit2022}). By matching both the faint-end QLFs from \citet{Ueda2014} and the QPDFs for low-mass galaxies from \citet{Aird2018}, \textsc{Trinity} shows that it is possible to reproduce the faint-end QLFs by keeping the majority of low-mass SMBHs dormant across cosmic time, and only allow intermittent AGN activity.

Above $z\sim 5$, both \fcent{} and \fdime{} are essentially mass-independent, above $\log M_\mathrm{peak} \sim 11.5$. This is because the Eddington ratio distributions of \emph{all} SMBHs become narrower and shifted to higher values at higher redshifts. Below $z\sim 3$, both fractions decrease towards lower redshifts. More massive black holes experience this decline earlier, known as the ``AGN downsizing'' phenomenon. Note that there is residual AGN activity among massive haloes with $M_\mathrm{peak} \gtrsim 10^{13.5} M_\odot$ between $0.5 \lesssim z \lesssim 2$. This is extrapolated from the QPDFs from \citet{Aird2018}, which implies higher average SMBH accretion rate among more massive galaxies. Overall, the decrease in Eddington ratio also coincides with the the decline of the star-forming fraction among galaxies, as shown by the $f_\mathrm{SF}=0.5$ curve (yellow dashed line). Specifically, the halo mass at which the fractions of star-forming and quiescent galaxies become equal (i.e., the $f_\mathrm{SF}=0.5$ curve) coincides with an average Eddington ratio of $\eta_\mathrm{rad}\sim 0.01$.

We also show AGN duty cycles when defined as the fraction of black holes with bolometric luminosities $L_{\rm bol}$ above a certain limit in Fig.\ \ref{f:f_lbol}. The top panel shows the fraction of black holes with $L_{\rm bol} > 10^{43}$ erg$\cdot$s$^{-1}$ ($f_{>43}$), and the bottom panel shows the fraction of black holes with $L_{\rm bol} > 10^{45}$ erg$\cdot$s$^{-1}$ ($f_{>45}$), as functions of \mpeak{} and $z$. The mass and redshift dependencies of both fractions are similar to the ones in Fig.\ \ref{f:f_ledd}. But, compared to low mass black holes, there are more massive black holes with $L_{\rm bol} > 10^{43}$ erg$\cdot$s$^{-1}$ than $\eta_{\rm rad} > 0.01$ below $z\sim 3$. The underlying reason is simple: luminosities depend on the product of black hole mass and Eddington ratio. Thus, despite lower Eddington ratios, massive black holes can still produce similar luminosities as smaller objects. In general, AGN duty cycles are higher at high redshifts. One implication of this is that the active black hole mass functions (ABHMFs) become closer to total BHMFs (i.e., including both active and dormant SMBHs) at high redshifts.

\begin{figure}
\subfigure{
\includegraphics[width=0.48\textwidth]{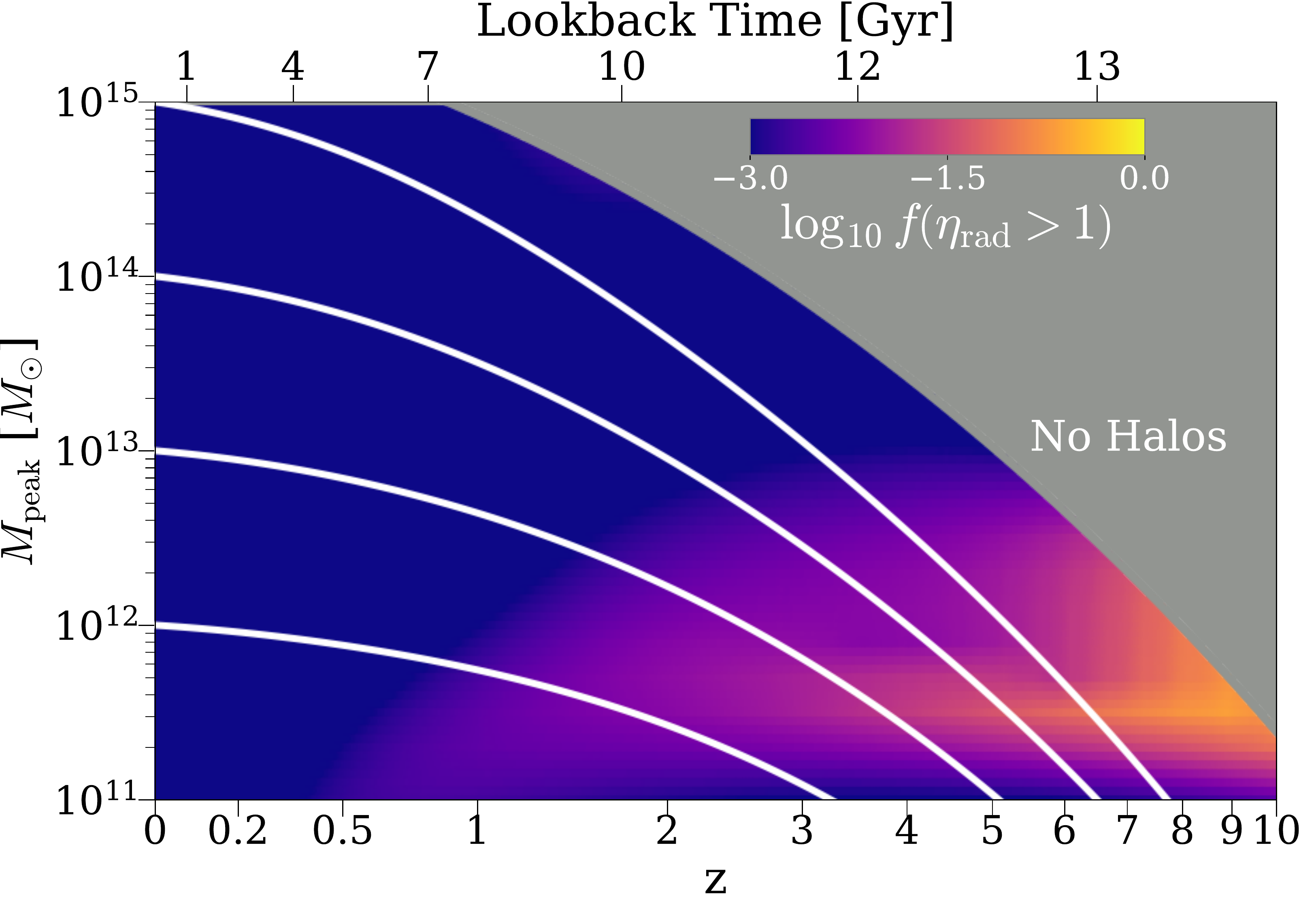}
}

\caption{The fractions of black holes accreting above the Eddington rate as a function of \mpeak{} and $z$. \halocurves{} See \S\ref{ss:results_super_eddington}. }
\label{f:f_super_edd}
\end{figure}

\subsection{Super-Eddington accretion}
\label{ss:results_super_eddington}

The derivation of the Eddington limit means that it represents at most a time-averaged limit on SMBH accretion for spherically-symmetric systems. Indeed, one of the findings of \cite{Zhang2023a} was that existing SMBH observations require super-Eddington accretion to occur some fraction of the time. Specifically, Eddington-limited accretion yields too slow SMBH growth to reproduce the local \bhbm{} relation (see Appendix E1 of \citealt{Zhang2023a}). On the other hand, super-Eddington accretion is also necessary to reproduce sufficient numbers of bright quasars and SMBH mass functions simultaneously.

Fig.\ \ref{f:f_super_edd} shows the fractions of black holes accreting at super-Eddington rates as a function of \mpeak{} and $z$. The mass and redshift dependence of the super-Eddington fraction is qualitatively similar to that shown in Fig.\ \ref{f:f_ledd}. This demonstrates that the decline in black hole activity below $z\sim 6$ happens among black holes at all different activity levels.

\section{Comparison with previous studies and discussion}
\label{s:discussion}

\S\ref{ss:discussions_eff} compares the mass-to-energy conversion efficiency of AGNs in \textsc{Trinity} and other studies, and \S\ref{ss:feedback_mode_edd_ratio} discusses the potential correlation between the decline in AGN Eddington ratio and the transition between AGN feedback modes.

\subsection{AGN energy efficiency}
\label{ss:discussions_eff}

\begin{figure}
\includegraphics[width=0.48\textwidth]{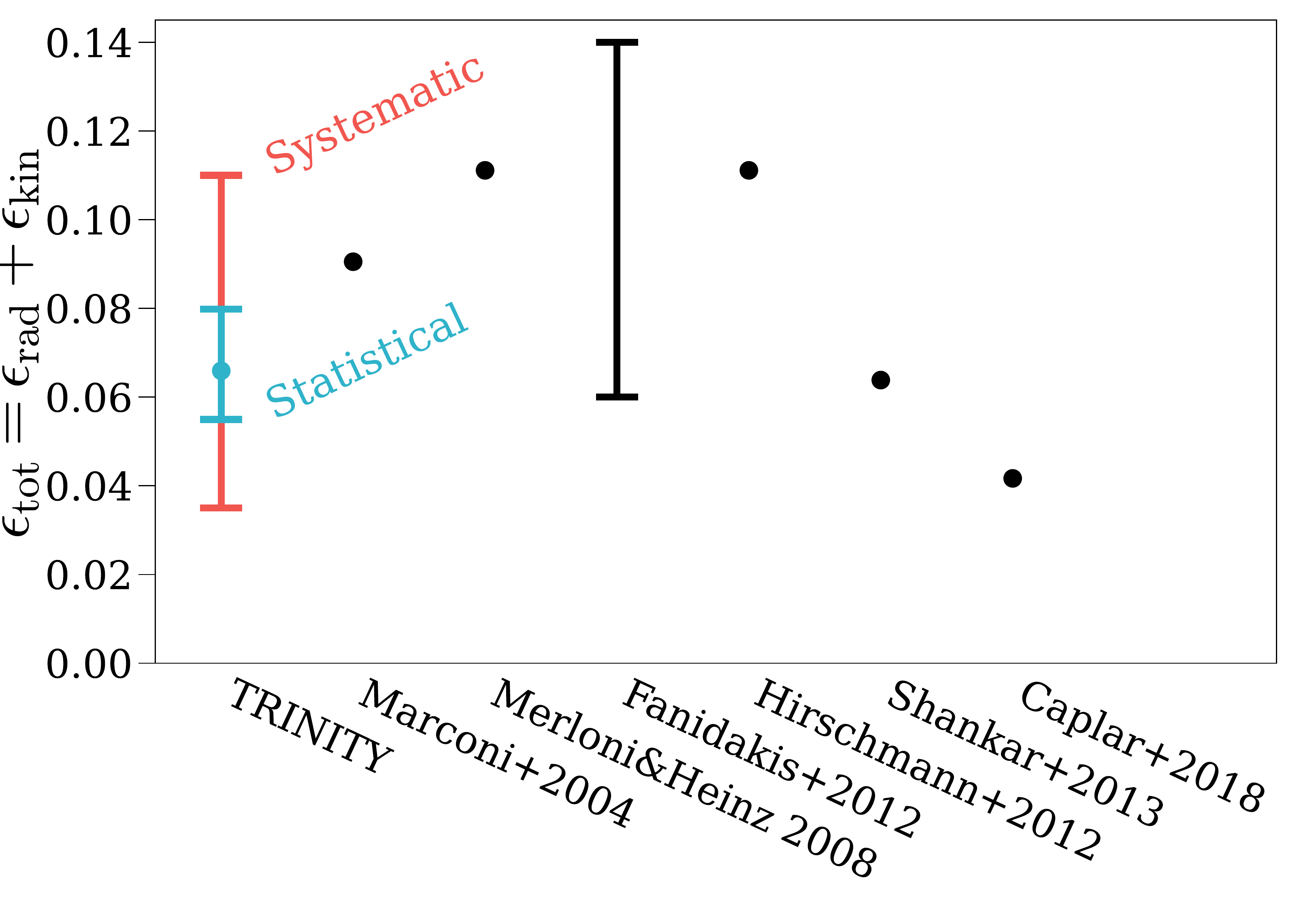}
\caption{Comparison between AGN efficiencies (\eff{}) from different studies. The black error bar is the 10--90$^\mathrm{th}$ percentile range from \citet{Fanidakis2012}. The blue dot and error bar are the best-fitting value and the 68\% confidence intervals for \textsc{Trinity}. The red error bar shows the range of systematic change in \eff{} when relevant input assumptions are changed (for details about this systematic change, see Appendix E2 of \citealt{Zhang2023a}). See \S\ref{ss:discussions_eff}.}
\label{f:eff_tot}
\end{figure}

Fig.\ \ref{f:eff_tot} shows the inferred or assumed AGN \emph{total} radiative$+$kinetic efficiencies, \eff{}$=\epsilon_\mathrm{rad}+\epsilon_\mathrm{kin}$, in the following studies: \textsc{Trinity}, the empirical models by \citet{Marconi2004}, \citet{Merloni2008}, \citet{Shankar2013}, and \citet{Caplar2018}, and the semi-analytical models by \citet{Fanidakis2012} and \citet{Hirschmann2012}. \citet{Fanidakis2012} calculated energy efficiencies of individual black holes from their spins, so here we show the 10--90$^\mathrm{th}$ percentile range that covers their black holes. For \textsc{Trinity}, the blue error bar denotes the 16--84$^\mathrm{th}$ percentile confidence interval from the posterior distribution of the model parameters, and the red error bar shows the range of systematic change in \eff{} when relevant input assumptions are changed, as discussed in Appendix E2 of \citet[][]{Zhang2023a}. The \eff{} from \textsc{Trinity} lies within the range spanned by different studies, but it is also important to keep in mind that the uncertainties in input assumptions to \textsc{Trinity} leads to additional systematic uncertainty of $\sim0.25$ dex in \eff{}. Such uncertainties include the choice of local \bhbm{} relation, the choice of bolometric corrections, the Compton-thick fraction corrections, and the way that the galaxy--SMBH mass connection is parametrized. For further details, we refer readers to Appendix E2 of \citet[][]{Zhang2023a}.

\subsection{AGN feedback mode vs. Eddington ratio}
\label{ss:feedback_mode_edd_ratio}
In the top panel of Fig.\ \ref{f:f_ledd}, we showed that the following regions are coincident with each other: 1) the region where the fraction of star-forming(quenched) galaxies is exactly 50\% (SF/Q equality region hereafter); 2) the region where the fraction of SMBHs with Eddington ratio $\eta_\mathrm{rad} > 0.01$ decreases dramatically (AGN boundary hereafter). Observations require the coincidence of these regions because the stellar mass at which galaxies transition between star-forming and quiescent coincides with strong decreases in specific quasar luminosity distributions from \cite{Aird2018}. Interestingly, $\eta_\mathrm{rad}\sim 0.01$ is also consistent with the boundary between ``radiative mode'' and the ``kinetic mode'' feedback---i.e., the boundary between primarily radiative and primarily kinetic energy output from AGNs. This two-mode feedback scenario is widely adopted in semi-analytical models and hydrodynamical simulations (e.g., \citealt{Bower2006,Croton2006,Sijacki2007,Somerville2008,Dubois2014b,Vogelsberger2014,Sijacki2015,Weinberger2017,Dave2019}). However, it is also worth noting that the fraction of active SMBHs defined with respect to other critical Eddington ratio values (e.g., $\eta_\mathrm{rad}=0.1$, the bottom panel of Fig.\ \ref{f:f_ledd}) also drops sharply close to the SF/Q equality region, so it remains unclear if $\eta_\mathrm{rad}\sim 0.01$ is where the actual transition between radiative mode and kinetic mode feedback occurs. In addition, this overlap of boundaries only indicates \emph{correlation}, rather than \emph{causation}, between SMBH quiescence and galaxy quenching.

\section{Conclusions}
\label{s:conclusions}

In this work, we present the build-up of SMBH populations from \textsc{Trinity}, which is an empirical model that parametrizes the statistical halo--galaxy--SMBH connection (\S\ref{s:method}). Compared to previous studies that are typically focused on one or two kinds of observables, \textsc{Trinity} features the ability to self-consistently match a comprehensive set of observational data for galaxies and SMBHs from $z=0-10$. As a result, \textsc{Trinity} is able to extract information that is only accessible with joint constraints from multiple datasets. In this paper, we investigated quasar demographics; key results are as follows:
\begin{itemize}

     \item From $z\sim 10$ to $z\sim 4$, the normalization of quasar bolometric luminosity functions (QLFs) increases by $\sim 3-4$ dex  (\S\ref{ss:qlf_evol}, Fig.\ \ref{f:high_z_qlf}). This is due to the mass buildup of different SMBH populations in \textsc{Trinity}, which is informed by observed QLFs and quasar probability distribution functions (QPDFs) at $z\leq 5$ and extrapolations at $z\ge 6$.
    
    \item Overall, less massive galaxies and SMBHs become more and more dominant in QLFs towards lower redshifts. This is due to the AGN downsizing effect, where more massive SMBHs become less active at earlier cosmic times. An exception is that at $z\sim 0$, massive haloes/galaxies/SMBHs dominate the bright end of QLFs again. This is largely due to the low Eddington ratios among all SMBHs, which makes massive SMBHs the only population able to support bright AGNs (\S\ref{ss:qlf_mbh}, \ref{ss:qlf_mstar_mh}, Figs.\ \ref{f:qlf_mbh}, \ref{f:qlf_mstar}). These results are informed by observed QPDFs and QLFs.

    \item At $z\gtrsim 1$, most observed quasars live in dark matter haloes with $M_\mathrm{peak}\sim 10^{12} M_\odot$ (\S\ref{ss:qlf_mstar_mh}, Fig.\ \ref{f:qlf_mh}). This is because $M_\mathrm{peak}\sim 10^{12} M_\odot$ haloes are where the bulk of the star formation happens in the Universe, as required by galaxy data constraints, and because SMBH growth closely tracks galaxy growth in these galaxies.
    
    \item Many observed quasars have SMBH masses $\gtrsim 0.3$ dex higher than predicted by the median \bhsm{} relation for all SMBHs, leading to significant Lauer bias in inferred \bhsm{} relations. At brighter luminosities, QLFs are more dominated by over-massive SMBHs compared to their host galaxies. Due to the overall decline in the AGN activity level with time, these over-massive SMBHs dominate QLFs down to lower luminosities at lower redshifts(\S\ref{ss:qlf_mstar_bias}, Fig.\ \ref{f:qlf_mstar_bias}). This magnitude of Lauer bias is determined by the Eddington ratio distribution and the random scatter around the \bhsm{} relation, which are constrained by the shapes of QPDFs, QLFs, and active black hole mass functions (ABHMFs).

    \item At $z\sim 6-8$,  quasar luminosity functions are dominated by AGNs with radiative Eddington ratios in the range $0.1 - 1$, i.e., mildly sub-Eddington to Eddington rates. At $z\sim 9-10$, we predict that super-Eddington AGNs contribute significantly to total QLFs (\S\ref{ss:qlf_eta}, Fig.\ \ref{f:qlf_eta}). This result is driven by extrapolating SMBH growth histories towards $z=10$ and the  observational constraint that most bright quasars at $z\sim 6$ are sub-Eddington.

    \item Below $z\sim 6$, the decline of AGN activity happens at all levels, i.e., the fraction of SMBHs above all Eddington ratio thresholds decrease towards lower redshifts (\S\ref{ss:results_duty_cycle_eff}, \ref{ss:results_super_eddington}, Figs.\ \ref{f:f_ledd}, \ref{f:f_super_edd}). This result is driven by the redshift evolution of QLFs, QPDFs, and ABHMFs.

\end{itemize}

\section*{Data availability}
\label{s:data_availability}

The parallel implementation of \textsc{Trinity}, the compiled datasets (\S\ref{ss:obs_data}), the posterior distribution of model parameters, and the codes to compute the predictions for all the plots in this work are available at \href{https://github.com/HaowenZhang/TRINITY}{https://github.com/HaowenZhang/TRINITY}.

\section*{Acknowledgements}
\label{s:acknowledgements}

We thank the anonymous referee for their valuable and constructive comments, as well as Gurtina Besla, Haley Bowden, Arjun Dey, Sandy Faber, Dan Foreman-Mackey, Richard Green, Jenny Greene, Melanie Habouzit, Andrew Hearin, Raphael Hviding, Takuma Izumi, David Koo, Pavel Kroupa, Tod Lauer, Junyao Li, Jianwei Lyu, Alessandro Peca, Joel Primack, Yujing Qin, George Rieke, Marcia Rieke, Jan-Torge Schindler, John Silverman, Xuejian Shen, Yue Shen, Rachel Somerville, Fengwu Sun, Wei-Leong Tee, Yoshihiro Ueda, Feige Wang, and Minghao Yue for very valuable discussions.

Support for this research came partially via program number HST-AR-15631.001-A, provided through a grant from the Space Telescope Science Institute under NASA contract NAS5-26555. PB was partially funded by a Packard Fellowship, Grant \#2019-69646. PB was also partially supported by a Giacconi Fellowship from the Space Telescope Science Institute.  Finally, PB was also partially supported through program number HST-HF2-51353.001-A, provided by NASA through a Hubble Fellowship grant from the Space Telescope Science Institute, under NASA contract NAS5-26555.

Data compilations from many studies used in this paper were made much more accurate and efficient by the online \textsc{WebPlotDigitizer} code.\footnote{\url{https://apps.automeris.io/wpd/}} This research has made extensive use of the arXiv and NASA's Astrophysics Data System.

This research used the Ocelote supercomputer of the University of Arizona. The allocation of computer time from the UA Research
Computing High Performance Computing (HPC) at the University
of Arizona is gratefully acknowledged. The Bolshoi-Planck simulation was performed by Anatoly Klypin within the Bolshoi project of the University of California High-Performance AstroComputing Center (UC-HiPACC; PI Joel Primack).


\bsp	
\label{lastpage}


\bibliography{trinity}

\begin{thebibliography}{}
\makeatletter
\relax
\def\mn@urlcharsother{\let\do\@makeother \do\$\do\&\do\#\do\^\do\_\do\%\do\~}
\def\mn@doi{\begingroup\mn@urlcharsother \@ifnextchar [ {\mn@doi@}
  {\mn@doi@[]}}
\def\mn@doi@[#1]#2{\def\@tempa{#1}\ifx\@tempa\@empty \href
  {http://dx.doi.org/#2} {doi:#2}\else \href {http://dx.doi.org/#2} {#1}\fi
  \endgroup}
\def\mn@eprint#1#2{\mn@eprint@#1:#2::\@nil}
\def\mn@eprint@arXiv#1{\href {http://arxiv.org/abs/#1} {{\tt arXiv:#1}}}
\def\mn@eprint@dblp#1{\href {http://dblp.uni-trier.de/rec/bibtex/#1.xml}
  {dblp:#1}}
\def\mn@eprint@#1:#2:#3:#4\@nil{\def\@tempa {#1}\def\@tempb {#2}\def\@tempc
  {#3}\ifx \@tempc \@empty \let \@tempc \@tempb \let \@tempb \@tempa \fi \ifx
  \@tempb \@empty \def\@tempb {arXiv}\fi \@ifundefined
  {mn@eprint@\@tempb}{\@tempb:\@tempc}{\expandafter \expandafter \csname
  mn@eprint@\@tempb\endcsname \expandafter{\@tempc}}}

\bibitem[\protect\citeauthoryear{{Aird} et~al.,}{{Aird}
  et~al.}{2010}]{Aird2010}
{Aird} J.,  et~al., 2010, \mn@doi [\mnras] {10.1111/j.1365-2966.2009.15829.x},
  \href {https://ui.adsabs.harvard.edu/abs/2010MNRAS.401.2531A} {401, 2531}

\bibitem[\protect\citeauthoryear{{Aird}, {Coil}  \& {Georgakakis}}{{Aird}
  et~al.}{2018}]{Aird2018}
{Aird} J.,  {Coil} A.~L.,   {Georgakakis} A.,  2018, \mn@doi [\mnras]
  {10.1093/mnras/stx2700}, \href
  {https://ui.adsabs.harvard.edu/abs/2018MNRAS.474.1225A} {474, 1225}

\bibitem[\protect\citeauthoryear{{Alexander} \& {Hickox}}{{Alexander} \&
  {Hickox}}{2012}]{Alexander2012}
{Alexander} D.~M.,  {Hickox} R.~C.,  2012, \mn@doi [\nar]
  {10.1016/j.newar.2011.11.003}, \href
  {https://ui.adsabs.harvard.edu/abs/2012NewAR..56...93A} {56, 93}

\bibitem[\protect\citeauthoryear{{Angl{\'e}s-Alc{\'a}zar},
  {Faucher-Gigu{\`e}re}, {Quataert}, {Hopkins}, {Feldmann}, {Torrey}, {Wetzel}
  \& {Kere{\v{s}}}}{{Angl{\'e}s-Alc{\'a}zar} et~al.}{2017}]{AnglesAlcazar2017}
{Angl{\'e}s-Alc{\'a}zar} D.,  {Faucher-Gigu{\`e}re} C.-A.,  {Quataert} E.,
  {Hopkins} P.~F.,  {Feldmann} R.,  {Torrey} P.,  {Wetzel} A.,   {Kere{\v{s}}}
  D.,  2017, \mn@doi [\mnras] {10.1093/mnrasl/slx161}, \href
  {https://ui.adsabs.harvard.edu/abs/2017MNRAS.472L.109A} {472, L109}

\bibitem[\protect\citeauthoryear{{Behroozi}, {Wechsler}  \&
  {Conroy}}{{Behroozi} et~al.}{2013}]{Behroozi2013}
{Behroozi} P.~S.,  {Wechsler} R.~H.,   {Conroy} C.,  2013, \mn@doi [\apj]
  {10.1088/0004-637X/770/1/57}, \href
  {https://ui.adsabs.harvard.edu/abs/2013ApJ...770...57B} {770, 57}

\bibitem[\protect\citeauthoryear{{Behroozi}, {Wechsler}, {Hearin}  \&
  {Conroy}}{{Behroozi} et~al.}{2019}]{Behroozi2019}
{Behroozi} P.,  {Wechsler} R.~H.,  {Hearin} A.~P.,   {Conroy} C.,  2019,
  \mn@doi [\mnras] {10.1093/mnras/stz1182}, \href
  {https://ui.adsabs.harvard.edu/abs/2019MNRAS.488.3143B} {488, 3143}

\bibitem[\protect\citeauthoryear{{Bouwens}, {Stefanon}, {Oesch}, {Illingworth},
  {Nanayakkara}, {Roberts-Borsani}, {Labb{\'e}}  \& {Smit}}{{Bouwens}
  et~al.}{2019}]{Bouwens2019}
{Bouwens} R.~J.,  {Stefanon} M.,  {Oesch} P.~A.,  {Illingworth} G.~D.,
  {Nanayakkara} T.,  {Roberts-Borsani} G.,  {Labb{\'e}} I.,   {Smit} R.,  2019,
  \mn@doi [\apj] {10.3847/1538-4357/ab24c5}, \href
  {https://ui.adsabs.harvard.edu/abs/2019ApJ...880...25B} {880, 25}

\bibitem[\protect\citeauthoryear{{Bower}, {Benson}, {Malbon}, {Helly}, {Frenk},
  {Baugh}, {Cole}  \& {Lacey}}{{Bower} et~al.}{2006}]{Bower2006}
{Bower} R.~G.,  {Benson} A.~J.,  {Malbon} R.,  {Helly} J.~C.,  {Frenk} C.~S.,
  {Baugh} C.~M.,  {Cole} S.,   {Lacey} C.~G.,  2006, \mn@doi [\mnras]
  {10.1111/j.1365-2966.2006.10519.x}, \href
  {https://ui.adsabs.harvard.edu/abs/2006MNRAS.370..645B} {370, 645}

\bibitem[\protect\citeauthoryear{{Bower}, {Schaye}, {Frenk}, {Theuns},
  {Schaller}, {Crain}  \& {McAlpine}}{{Bower} et~al.}{2017}]{Bower2017}
{Bower} R.~G.,  {Schaye} J.,  {Frenk} C.~S.,  {Theuns} T.,  {Schaller} M.,
  {Crain} R.~A.,   {McAlpine} S.,  2017, \mn@doi [\mnras]
  {10.1093/mnras/stw2735}, \href
  {https://ui.adsabs.harvard.edu/abs/2017MNRAS.465...32B} {465, 32}

\bibitem[\protect\citeauthoryear{{Brandt} \& {Alexander}}{{Brandt} \&
  {Alexander}}{2015}]{Brandt2015}
{Brandt} W.~N.,  {Alexander} D.~M.,  2015, \mn@doi [\aapr]
  {10.1007/s00159-014-0081-z}, \href
  {https://ui.adsabs.harvard.edu/abs/2015A&ARv..23....1B} {23, 1}

\bibitem[\protect\citeauthoryear{{Bruzual} \& {Charlot}}{{Bruzual} \&
  {Charlot}}{2003}]{Bruzual2003}
{Bruzual} G.,  {Charlot} S.,  2003, \mn@doi [\mnras]
  {10.1046/j.1365-8711.2003.06897.x}, \href
  {https://ui.adsabs.harvard.edu/abs/2003MNRAS.344.1000B} {344, 1000}

\bibitem[\protect\citeauthoryear{{Bryan} \& {Norman}}{{Bryan} \&
  {Norman}}{1998}]{Bryan1998}
{Bryan} G.~L.,  {Norman} M.~L.,  1998, \mn@doi [\apj] {10.1086/305262}, \href
  {https://ui.adsabs.harvard.edu/abs/1998ApJ...495...80B} {495, 80}

\bibitem[\protect\citeauthoryear{{Calzetti}, {Armus}, {Bohlin}, {Kinney},
  {Koornneef}  \& {Storchi-Bergmann}}{{Calzetti} et~al.}{2000}]{Calzetti2000}
{Calzetti} D.,  {Armus} L.,  {Bohlin} R.~C.,  {Kinney} A.~L.,  {Koornneef} J.,
   {Storchi-Bergmann} T.,  2000, \mn@doi [\apj] {10.1086/308692}, \href
  {https://ui.adsabs.harvard.edu/abs/2000ApJ...533..682C} {533, 682}

\bibitem[\protect\citeauthoryear{{Caplar}, {Lilly}  \& {Trakhtenbrot}}{{Caplar}
  et~al.}{2015}]{Caplar2015}
{Caplar} N.,  {Lilly} S.~J.,   {Trakhtenbrot} B.,  2015, \mn@doi [\apj]
  {10.1088/0004-637X/811/2/148}, \href
  {https://ui.adsabs.harvard.edu/abs/2015ApJ...811..148C} {811, 148}

\bibitem[\protect\citeauthoryear{{Caplar}, {Lilly}  \& {Trakhtenbrot}}{{Caplar}
  et~al.}{2018}]{Caplar2018}
{Caplar} N.,  {Lilly} S.~J.,   {Trakhtenbrot} B.,  2018, \mn@doi [\apj]
  {10.3847/1538-4357/aae691}, \href
  {https://ui.adsabs.harvard.edu/abs/2018ApJ...867..148C} {867, 148}

\bibitem[\protect\citeauthoryear{{Cavagnolo}, {McNamara}, {Nulsen}, {Carilli},
  {Jones}  \& {B{\^\i}rzan}}{{Cavagnolo} et~al.}{2010}]{Cavagnolo2010}
{Cavagnolo} K.~W.,  {McNamara} B.~R.,  {Nulsen} P.~E.~J.,  {Carilli} C.~L.,
  {Jones} C.,   {B{\^\i}rzan} L.,  2010, \mn@doi [\apj]
  {10.1088/0004-637X/720/2/1066}, \href
  {https://ui.adsabs.harvard.edu/abs/2010ApJ...720.1066C} {720, 1066}

\bibitem[\protect\citeauthoryear{{Chabrier}}{{Chabrier}}{2003}]{Chabrier2003}
{Chabrier} G.,  2003, \mn@doi [\pasp] {10.1086/376392}, \href
  {https://ui.adsabs.harvard.edu/abs/2003PASP..115..763C} {115, 763}

\bibitem[\protect\citeauthoryear{{Choi}, {Ostriker}, {Naab}, {Somerville},
  {Hirschmann}, {N{\'u}{\~n}ez}, {Hu}  \& {Oser}}{{Choi}
  et~al.}{2017}]{Choi2017}
{Choi} E.,  {Ostriker} J.~P.,  {Naab} T.,  {Somerville} R.~S.,  {Hirschmann}
  M.,  {N{\'u}{\~n}ez} A.,  {Hu} C.-Y.,   {Oser} L.,  2017, \mn@doi [\apj]
  {10.3847/1538-4357/aa7849}, \href
  {https://ui.adsabs.harvard.edu/abs/2017ApJ...844...31C} {844, 31}

\bibitem[\protect\citeauthoryear{{Cirasuolo}, {Magliocchetti}  \&
  {Celotti}}{{Cirasuolo} et~al.}{2005}]{CMC05}
{Cirasuolo} M.,  {Magliocchetti} M.,   {Celotti} A.,  2005, \mn@doi [\mnras]
  {10.1111/j.1365-2966.2005.08729.x}, \href
  {https://ui.adsabs.harvard.edu/abs/2005MNRAS.357.1267C} {357, 1267}

\bibitem[\protect\citeauthoryear{{Conroy} \& {White}}{{Conroy} \&
  {White}}{2013}]{Conroy2013}
{Conroy} C.,  {White} M.,  2013, \mn@doi [\apj] {10.1088/0004-637X/762/2/70},
  \href {https://ui.adsabs.harvard.edu/abs/2013ApJ...762...70C} {762, 70}

\bibitem[\protect\citeauthoryear{{Costa}, {Sijacki}  \& {Haehnelt}}{{Costa}
  et~al.}{2014}]{Costa2014}
{Costa} T.,  {Sijacki} D.,   {Haehnelt} M.~G.,  2014, \mn@doi [\mnras]
  {10.1093/mnras/stu1632}, \href
  {https://ui.adsabs.harvard.edu/abs/2014MNRAS.444.2355C} {444, 2355}

\bibitem[\protect\citeauthoryear{{Croton} et~al.,}{{Croton}
  et~al.}{2006}]{Croton2006}
{Croton} D.~J.,  et~al., 2006, \mn@doi [\mnras]
  {10.1111/j.1365-2966.2005.09675.x}, \href
  {https://ui.adsabs.harvard.edu/abs/2006MNRAS.365...11C} {365, 11}

\bibitem[\protect\citeauthoryear{{Dav{\'e}}, {Angl{\'e}s-Alc{\'a}zar},
  {Narayanan}, {Li}, {Rafieferantsoa}  \& {Appleby}}{{Dav{\'e}}
  et~al.}{2019}]{Dave2019}
{Dav{\'e}} R.,  {Angl{\'e}s-Alc{\'a}zar} D.,  {Narayanan} D.,  {Li} Q.,
  {Rafieferantsoa} M.~H.,   {Appleby} S.,  2019, \mn@doi [\mnras]
  {10.1093/mnras/stz937}, \href
  {https://ui.adsabs.harvard.edu/abs/2019MNRAS.486.2827D} {486, 2827}

\bibitem[\protect\citeauthoryear{{Delvecchio} et~al.,}{{Delvecchio}
  et~al.}{2014}]{Delvecchio2014}
{Delvecchio} I.,  et~al., 2014, \mn@doi [\mnras] {10.1093/mnras/stu130}, \href
  {https://ui.adsabs.harvard.edu/abs/2014MNRAS.439.2736D} {439, 2736}

\bibitem[\protect\citeauthoryear{{Dubois}, {Volonteri}  \& {Silk}}{{Dubois}
  et~al.}{2014}]{Dubois2014b}
{Dubois} Y.,  {Volonteri} M.,   {Silk} J.,  2014, \mn@doi [\mnras]
  {10.1093/mnras/stu373}, \href
  {https://ui.adsabs.harvard.edu/abs/2014MNRAS.440.1590D} {440, 1590}

\bibitem[\protect\citeauthoryear{{Dubois}, {Volonteri}, {Silk}, {Devriendt},
  {Slyz}  \& {Teyssier}}{{Dubois} et~al.}{2015}]{Dubois2015}
{Dubois} Y.,  {Volonteri} M.,  {Silk} J.,  {Devriendt} J.,  {Slyz} A.,
  {Teyssier} R.,  2015, \mn@doi [\mnras] {10.1093/mnras/stv1416}, \href
  {https://ui.adsabs.harvard.edu/abs/2015MNRAS.452.1502D} {452, 1502}

\bibitem[\protect\citeauthoryear{{Fanidakis} et~al.,}{{Fanidakis}
  et~al.}{2012}]{Fanidakis2012}
{Fanidakis} N.,  et~al., 2012, \mn@doi [\mnras]
  {10.1111/j.1365-2966.2011.19931.x}, \href
  {https://ui.adsabs.harvard.edu/abs/2012MNRAS.419.2797F} {419, 2797}

\bibitem[\protect\citeauthoryear{{Ferrarese} \& {Merritt}}{{Ferrarese} \&
  {Merritt}}{2000}]{Ferrarese2000}
{Ferrarese} L.,  {Merritt} D.,  2000, \mn@doi [\apjl] {10.1086/312838}, \href
  {https://ui.adsabs.harvard.edu/abs/2000ApJ...539L...9F} {539, L9}

\bibitem[\protect\citeauthoryear{{Gebhardt} et~al.,}{{Gebhardt}
  et~al.}{2000}]{Gebhardt2000}
{Gebhardt} K.,  et~al., 2000, \mn@doi [\apjl] {10.1086/312840}, \href
  {https://ui.adsabs.harvard.edu/abs/2000ApJ...539L..13G} {539, L13}

\bibitem[\protect\citeauthoryear{{G{\"u}ltekin} et~al.,}{{G{\"u}ltekin}
  et~al.}{2009}]{Gultekin2009}
{G{\"u}ltekin} K.,  et~al., 2009, \mn@doi [\apj] {10.1088/0004-637X/698/1/198},
  \href {https://ui.adsabs.harvard.edu/abs/2009ApJ...698..198G} {698, 198}

\bibitem[\protect\citeauthoryear{{Haario}, {Saksman}  \& {Tamminen}}{{Haario}
  et~al.}{2001}]{Haario2001}
{Haario} H.,  {Saksman} E.,   {Tamminen} J.,  2001, Bernoulli, \href
  {https://projecteuclid.org/euclid.bj/1080222083} {7, 223}

\bibitem[\protect\citeauthoryear{{Habouzit}, {Volonteri}  \&
  {Dubois}}{{Habouzit} et~al.}{2017}]{Habouzit2017}
{Habouzit} M.,  {Volonteri} M.,   {Dubois} Y.,  2017, \mn@doi [\mnras]
  {10.1093/mnras/stx666}, \href
  {https://ui.adsabs.harvard.edu/abs/2017MNRAS.468.3935H} {468, 3935}

\bibitem[\protect\citeauthoryear{{Habouzit} et~al.,}{{Habouzit}
  et~al.}{2022}]{Habouzit2022}
{Habouzit} M.,  et~al., 2022, \mn@doi [\mnras] {10.1093/mnras/stab3147}, \href
  {https://ui.adsabs.harvard.edu/abs/2022MNRAS.509.3015H} {509, 3015}

\bibitem[\protect\citeauthoryear{{Harikane}, {Nakajima}, {Ouchi}, {Umeda},
  {Isobe}, {Ono}, {Xu}  \& {Zhang}}{{Harikane} et~al.}{2023}]{Harikane2023b}
{Harikane} Y.,  {Nakajima} K.,  {Ouchi} M.,  {Umeda} H.,  {Isobe} Y.,  {Ono}
  Y.,  {Xu} Y.,   {Zhang} Y.,  2023, \mn@doi [arXiv e-prints]
  {10.48550/arXiv.2304.06658}, \href
  {https://ui.adsabs.harvard.edu/abs/2023arXiv230406658H} {p. arXiv:2304.06658}

\bibitem[\protect\citeauthoryear{{H{\"a}ring} \& {Rix}}{{H{\"a}ring} \&
  {Rix}}{2004}]{Haring2004}
{H{\"a}ring} N.,  {Rix} H.-W.,  2004, \mn@doi [\apjl] {10.1086/383567}, \href
  {https://ui.adsabs.harvard.edu/abs/2004ApJ...604L..89H} {604, L89}

\bibitem[\protect\citeauthoryear{{Heckman} \& {Best}}{{Heckman} \&
  {Best}}{2014}]{Heckman2014}
{Heckman} T.~M.,  {Best} P.~N.,  2014, \mn@doi [\araa]
  {10.1146/annurev-astro-081913-035722}, \href
  {https://ui.adsabs.harvard.edu/abs/2014ARA&A..52..589H} {52, 589}

\bibitem[\protect\citeauthoryear{{Hirschmann}, {Somerville}, {Naab}  \&
  {Burkert}}{{Hirschmann} et~al.}{2012}]{Hirschmann2012}
{Hirschmann} M.,  {Somerville} R.~S.,  {Naab} T.,   {Burkert} A.,  2012,
  \mn@doi [\mnras] {10.1111/j.1365-2966.2012.21626.x}, \href
  {https://ui.adsabs.harvard.edu/abs/2012MNRAS.426..237H} {426, 237}

\bibitem[\protect\citeauthoryear{{Hlavacek-Larrondo}
  et~al.,}{{Hlavacek-Larrondo} et~al.}{2015}]{Hlavacek-Larrondo2015}
{Hlavacek-Larrondo} J.,  et~al., 2015, \mn@doi [\apj]
  {10.1088/0004-637X/805/1/35}, \href
  {https://ui.adsabs.harvard.edu/abs/2015ApJ...805...35H} {805, 35}

\bibitem[\protect\citeauthoryear{{Ho}}{{Ho}}{2002}]{Ho2002}
{Ho} L.~C.,  2002, \mn@doi [\apj] {10.1086/324399}, \href
  {https://ui.adsabs.harvard.edu/abs/2002ApJ...564..120H} {564, 120}

\bibitem[\protect\citeauthoryear{{Ho}}{{Ho}}{2008}]{Ho2008}
{Ho} L.~C.,  2008, \mn@doi [\araa] {10.1146/annurev.astro.45.051806.110546},
  \href {https://ui.adsabs.harvard.edu/abs/2008ARA&A..46..475H} {46, 475}

\bibitem[\protect\citeauthoryear{{Hopkins}, {Bundy}, {Hernquist}  \&
  {Ellis}}{{Hopkins} et~al.}{2007}]{Hopkins2006}
{Hopkins} P.~F.,  {Bundy} K.,  {Hernquist} L.,   {Ellis} R.~S.,  2007, \mn@doi
  [\apj] {10.1086/512091}, \href
  {https://ui.adsabs.harvard.edu/abs/2007ApJ...659..976H} {659, 976}

\bibitem[\protect\citeauthoryear{{Ishigaki}, {Kawamata}, {Ouchi}, {Oguri},
  {Shimasaku}  \& {Ono}}{{Ishigaki} et~al.}{2018}]{Ishigaki2018}
{Ishigaki} M.,  {Kawamata} R.,  {Ouchi} M.,  {Oguri} M.,  {Shimasaku} K.,
  {Ono} Y.,  2018, \mn@doi [\apj] {10.3847/1538-4357/aaa544}, \href
  {https://ui.adsabs.harvard.edu/abs/2018ApJ...854...73I} {854, 73}

\bibitem[\protect\citeauthoryear{{Izumi} et~al.,}{{Izumi}
  et~al.}{2019}]{Izumi2019}
{Izumi} T.,  et~al., 2019, \mn@doi [\pasj] {10.1093/pasj/psz096}, \href
  {https://ui.adsabs.harvard.edu/abs/2019PASJ...71..111I} {71, 111}

\bibitem[\protect\citeauthoryear{{Jarosik} et~al.,}{{Jarosik}
  et~al.}{2011}]{Jarosik2011}
{Jarosik} N.,  et~al., 2011, \mn@doi [\apjs] {10.1088/0067-0049/192/2/14},
  \href {https://ui.adsabs.harvard.edu/abs/2011ApJS..192...14J} {192, 14}

\bibitem[\protect\citeauthoryear{{Kelly} \& {Shen}}{{Kelly} \&
  {Shen}}{2013}]{Kelly2013}
{Kelly} B.~C.,  {Shen} Y.,  2013, \mn@doi [\apj] {10.1088/0004-637X/764/1/45},
  \href {https://ui.adsabs.harvard.edu/abs/2013ApJ...764...45K} {764, 45}

\bibitem[\protect\citeauthoryear{{Klypin}, {Trujillo-Gomez}  \&
  {Primack}}{{Klypin} et~al.}{2011}]{Klypin2011}
{Klypin} A.~A.,  {Trujillo-Gomez} S.,   {Primack} J.,  2011, \mn@doi [\apj]
  {10.1088/0004-637X/740/2/102}, \href
  {https://ui.adsabs.harvard.edu/abs/2011ApJ...740..102K} {740, 102}

\bibitem[\protect\citeauthoryear{{Kormendy} \& {Ho}}{{Kormendy} \&
  {Ho}}{2013}]{Kormendy2013}
{Kormendy} J.,  {Ho} L.~C.,  2013, \mn@doi [\araa]
  {10.1146/annurev-astro-082708-101811}, \href
  {https://ui.adsabs.harvard.edu/abs/2013ARA&A..51..511K} {51, 511}

\bibitem[\protect\citeauthoryear{{Kormendy} \& {Richstone}}{{Kormendy} \&
  {Richstone}}{1995}]{Kormendy1995}
{Kormendy} J.,  {Richstone} D.,  1995, \mn@doi [\araa]
  {10.1146/annurev.aa.33.090195.003053}, \href
  {https://ui.adsabs.harvard.edu/abs/1995ARA&A..33..581K} {33, 581}

\bibitem[\protect\citeauthoryear{{La Franca}, {Melini}  \& {Fiore}}{{La Franca}
  et~al.}{2010}]{LaFranca2010}
{La Franca} F.,  {Melini} G.,   {Fiore} F.,  2010, \mn@doi [\apj]
  {10.1088/0004-637X/718/1/368}, \href
  {https://ui.adsabs.harvard.edu/abs/2010ApJ...718..368L} {718, 368}

\bibitem[\protect\citeauthoryear{{Lang} et~al.,}{{Lang}
  et~al.}{2014}]{Lang2014}
{Lang} P.,  et~al., 2014, \mn@doi [\apj] {10.1088/0004-637X/788/1/11}, \href
  {https://ui.adsabs.harvard.edu/abs/2014ApJ...788...11L} {788, 11}

\bibitem[\protect\citeauthoryear{{Lauer}, {Tremaine}, {Richstone}  \&
  {Faber}}{{Lauer} et~al.}{2007}]{Lauer2007}
{Lauer} T.~R.,  {Tremaine} S.,  {Richstone} D.,   {Faber} S.~M.,  2007, \mn@doi
  [\apj] {10.1086/522083}, \href
  {https://ui.adsabs.harvard.edu/abs/2007ApJ...670..249L} {670, 249}

\bibitem[\protect\citeauthoryear{{Magorrian} et~al.,}{{Magorrian}
  et~al.}{1998}]{Magorrian1998}
{Magorrian} J.,  et~al., 1998, \mn@doi [\aj] {10.1086/300353}, \href
  {https://ui.adsabs.harvard.edu/abs/1998AJ....115.2285M} {115, 2285}

\bibitem[\protect\citeauthoryear{{Maiolino} et~al.,}{{Maiolino}
  et~al.}{2023}]{Maiolino2023}
{Maiolino} R.,  et~al., 2023, \mn@doi [arXiv e-prints]
  {10.48550/arXiv.2305.12492}, \href
  {https://ui.adsabs.harvard.edu/abs/2023arXiv230512492M} {p. arXiv:2305.12492}

\bibitem[\protect\citeauthoryear{{Marconi}, {Risaliti}, {Gilli}, {Hunt},
  {Maiolino}  \& {Salvati}}{{Marconi} et~al.}{2004}]{Marconi2004}
{Marconi} A.,  {Risaliti} G.,  {Gilli} R.,  {Hunt} L.~K.,  {Maiolino} R.,
  {Salvati} M.,  2004, \mn@doi [\mnras] {10.1111/j.1365-2966.2004.07765.x},
  \href {https://ui.adsabs.harvard.edu/abs/2004MNRAS.351..169M} {351, 169}

\bibitem[\protect\citeauthoryear{{McConnell} \& {Ma}}{{McConnell} \&
  {Ma}}{2013}]{McConnell2013}
{McConnell} N.~J.,  {Ma} C.-P.,  2013, \mn@doi [\apj]
  {10.1088/0004-637X/764/2/184}, \href
  {https://ui.adsabs.harvard.edu/abs/2013ApJ...764..184M} {764, 184}

\bibitem[\protect\citeauthoryear{{McDonald}, {McNamara}, {Calzadilla}, {Chen},
  {Gaspari}, {Hickox}, {Kara}  \& {Korchagin}}{{McDonald}
  et~al.}{2021}]{McDonald2021}
{McDonald} M.,  {McNamara} B.~R.,  {Calzadilla} M.~S.,  {Chen} C.-T.,
  {Gaspari} M.,  {Hickox} R.~C.,  {Kara} E.,   {Korchagin} I.,  2021, \mn@doi
  [\apj] {10.3847/1538-4357/abd47f}, \href
  {https://ui.adsabs.harvard.edu/abs/2021ApJ...908...85M} {908, 85}

\bibitem[\protect\citeauthoryear{{Mendel}, {Simard}, {Palmer}, {Ellison}  \&
  {Patton}}{{Mendel} et~al.}{2014}]{Mendel2014}
{Mendel} J.~T.,  {Simard} L.,  {Palmer} M.,  {Ellison} S.~L.,   {Patton} D.~R.,
   2014, \mn@doi [\apjs] {10.1088/0067-0049/210/1/3}, \href
  {https://ui.adsabs.harvard.edu/abs/2014ApJS..210....3M} {210, 3}

\bibitem[\protect\citeauthoryear{{Merloni} \& {Heinz}}{{Merloni} \&
  {Heinz}}{2007}]{Merloni2007}
{Merloni} A.,  {Heinz} S.,  2007, \mn@doi [\mnras]
  {10.1111/j.1365-2966.2007.12253.x}, \href
  {https://ui.adsabs.harvard.edu/abs/2007MNRAS.381..589M} {381, 589}

\bibitem[\protect\citeauthoryear{{Merloni} \& {Heinz}}{{Merloni} \&
  {Heinz}}{2008}]{Merloni2008}
{Merloni} A.,  {Heinz} S.,  2008, \mn@doi [\mnras]
  {10.1111/j.1365-2966.2008.13472.x}, \href
  {https://ui.adsabs.harvard.edu/abs/2008MNRAS.388.1011M} {388, 1011}

\bibitem[\protect\citeauthoryear{{Merloni}, {Rudnick}  \& {Di
  Matteo}}{{Merloni} et~al.}{2004}]{Merloni2004}
{Merloni} A.,  {Rudnick} G.,   {Di Matteo} T.,  2004, \mn@doi [\mnras]
  {10.1111/j.1365-2966.2004.08382.x}, \href
  {https://ui.adsabs.harvard.edu/abs/2004MNRAS.354L..37M} {354, L37}

\bibitem[\protect\citeauthoryear{{Mineshige}, {Kawaguchi}, {Takeuchi}  \&
  {Hayashida}}{{Mineshige} et~al.}{2000}]{Mineshige2000}
{Mineshige} S.,  {Kawaguchi} T.,  {Takeuchi} M.,   {Hayashida} K.,  2000,
  \mn@doi [\pasj] {10.1093/pasj/52.3.499}, \href
  {https://ui.adsabs.harvard.edu/abs/2000PASJ...52..499M} {52, 499}

\bibitem[\protect\citeauthoryear{{Moster}, {Naab}  \& {White}}{{Moster}
  et~al.}{2013}]{Moster2013}
{Moster} B.~P.,  {Naab} T.,   {White} S. D.~M.,  2013, \mn@doi [\mnras]
  {10.1093/mnras/sts261}, \href
  {https://ui.adsabs.harvard.edu/abs/2013MNRAS.428.3121M} {428, 3121}

\bibitem[\protect\citeauthoryear{{Moster}, {Naab}  \& {White}}{{Moster}
  et~al.}{2018}]{Moster2018}
{Moster} B.~P.,  {Naab} T.,   {White} S. D.~M.,  2018, \mn@doi [\mnras]
  {10.1093/mnras/sty655}, \href
  {https://ui.adsabs.harvard.edu/abs/2018MNRAS.477.1822M} {477, 1822}

\bibitem[\protect\citeauthoryear{{Nagar}, {Falcke}  \& {Wilson}}{{Nagar}
  et~al.}{2005}]{Nagar2005}
{Nagar} N.~M.,  {Falcke} H.,   {Wilson} A.~S.,  2005, \mn@doi [\aap]
  {10.1051/0004-6361:20042277}, \href
  {https://ui.adsabs.harvard.edu/abs/2005A&A...435..521N} {435, 521}

\bibitem[\protect\citeauthoryear{{Narayan} \& {Yi}}{{Narayan} \&
  {Yi}}{1994}]{Narayan1994}
{Narayan} R.,  {Yi} I.,  1994, \mn@doi [\apjl] {10.1086/187381}, \href
  {https://ui.adsabs.harvard.edu/abs/1994ApJ...428L..13N} {428, L13}

\bibitem[\protect\citeauthoryear{{Oesch}, {Bouwens}, {Illingworth}, {Labb{\'e}}
   \& {Stefanon}}{{Oesch} et~al.}{2018}]{Oesch2018}
{Oesch} P.~A.,  {Bouwens} R.~J.,  {Illingworth} G.~D.,  {Labb{\'e}} I.,
  {Stefanon} M.,  2018, \mn@doi [\apj] {10.3847/1538-4357/aab03f}, \href
  {https://ui.adsabs.harvard.edu/abs/2018ApJ...855..105O} {855, 105}

\bibitem[\protect\citeauthoryear{{Peca} et~al.,}{{Peca}
  et~al.}{2023}]{Peca2023}
{Peca} A.,  et~al., 2023, \mn@doi [\apj] {10.3847/1538-4357/acac28}, \href
  {https://ui.adsabs.harvard.edu/abs/2023ApJ...943..162P} {943, 162}

\bibitem[\protect\citeauthoryear{{Planck Collaboration} et~al.,}{{Planck
  Collaboration} et~al.}{2016}]{Planck2016}
{Planck Collaboration} et~al., 2016, \mn@doi [\aap]
  {10.1051/0004-6361/201525830}, \href
  {https://ui.adsabs.harvard.edu/abs/2016A&A...594A..13P} {594, A13}

\bibitem[\protect\citeauthoryear{{Reines} \& {Volonteri}}{{Reines} \&
  {Volonteri}}{2015}]{Reines2015}
{Reines} A.~E.,  {Volonteri} M.,  2015, \mn@doi [\apj]
  {10.1088/0004-637X/813/2/82}, \href
  {https://ui.adsabs.harvard.edu/abs/2015ApJ...813...82R} {813, 82}

\bibitem[\protect\citeauthoryear{{Rodr{\'\i}guez-Puebla}, {Behroozi},
  {Primack}, {Klypin}, {Lee}  \& {Hellinger}}{{Rodr{\'\i}guez-Puebla}
  et~al.}{2016}]{RodriguezPuebla2016}
{Rodr{\'\i}guez-Puebla} A.,  {Behroozi} P.,  {Primack} J.,  {Klypin} A.,  {Lee}
  C.,   {Hellinger} D.,  2016, \mn@doi [\mnras] {10.1093/mnras/stw1705}, \href
  {https://ui.adsabs.harvard.edu/abs/2016MNRAS.462..893R} {462, 893}

\bibitem[\protect\citeauthoryear{{Savorgnan}, {Graham}, {Marconi}  \&
  {Sani}}{{Savorgnan} et~al.}{2016}]{Savorgnan2016}
{Savorgnan} G. A.~D.,  {Graham} A.~W.,  {Marconi} A.~r.,   {Sani} E.,  2016,
  \mn@doi [\apj] {10.3847/0004-637X/817/1/21}, \href
  {https://ui.adsabs.harvard.edu/abs/2016ApJ...817...21S} {817, 21}

\bibitem[\protect\citeauthoryear{{Schulze} \& {Wisotzki}}{{Schulze} \&
  {Wisotzki}}{2010}]{Schulze2010}
{Schulze} A.,  {Wisotzki} L.,  2010, \mn@doi [\aap]
  {10.1051/0004-6361/201014193}, \href
  {https://ui.adsabs.harvard.edu/abs/2010A&A...516A..87S} {516, A87}

\bibitem[\protect\citeauthoryear{{Schulze} et~al.,}{{Schulze}
  et~al.}{2015}]{Schulze2015}
{Schulze} A.,  et~al., 2015, \mn@doi [\mnras] {10.1093/mnras/stu2549}, \href
  {https://ui.adsabs.harvard.edu/abs/2015MNRAS.447.2085S} {447, 2085}

\bibitem[\protect\citeauthoryear{{Shakura} \& {Sunyaev}}{{Shakura} \&
  {Sunyaev}}{1973}]{Shakura1973}
{Shakura} N.~I.,  {Sunyaev} R.~A.,  1973, \aap, \href
  {https://ui.adsabs.harvard.edu/abs/1973A&A....24..337S} {24, 337}

\bibitem[\protect\citeauthoryear{{Shankar}, {Cavaliere}, {Cirasuolo}  \&
  {Maraschi}}{{Shankar} et~al.}{2008}]{Shankar2008}
{Shankar} F.,  {Cavaliere} A.,  {Cirasuolo} M.,   {Maraschi} L.,  2008, \mn@doi
  [\apj] {10.1086/528836}, \href
  {https://ui.adsabs.harvard.edu/abs/2008ApJ...676..131S} {676, 131}

\bibitem[\protect\citeauthoryear{{Shankar}, {Weinberg}  \&
  {Miralda-Escud{\'e}}}{{Shankar} et~al.}{2009}]{Shankar2009}
{Shankar} F.,  {Weinberg} D.~H.,   {Miralda-Escud{\'e}} J.,  2009, \mn@doi
  [\apj] {10.1088/0004-637X/690/1/20}, \href
  {https://ui.adsabs.harvard.edu/abs/2009ApJ...690...20S} {690, 20}

\bibitem[\protect\citeauthoryear{{Shankar}, {Weinberg}  \&
  {Miralda-Escud{\'e}}}{{Shankar} et~al.}{2013}]{Shankar2013}
{Shankar} F.,  {Weinberg} D.~H.,   {Miralda-Escud{\'e}} J.,  2013, \mn@doi
  [\mnras] {10.1093/mnras/sts026}, \href
  {https://ui.adsabs.harvard.edu/abs/2013MNRAS.428..421S} {428, 421}

\bibitem[\protect\citeauthoryear{{Shankar} et~al.,}{{Shankar}
  et~al.}{2020}]{Shankar2020}
{Shankar} F.,  et~al., 2020, \mn@doi [\mnras] {10.1093/mnras/stz3522}, \href
  {https://ui.adsabs.harvard.edu/abs/2020MNRAS.493.1500S} {493, 1500}

\bibitem[\protect\citeauthoryear{{Shen}}{{Shen}}{2013}]{Shen2013}
{Shen} Y.,  2013, Bulletin of the Astronomical Society of India, \href
  {https://ui.adsabs.harvard.edu/abs/2013BASI...41...61S} {41, 61}

\bibitem[\protect\citeauthoryear{{Shen}, {Hopkins}, {Faucher-Gigu{\`e}re},
  {Alexander}, {Richards}, {Ross}  \& {Hickox}}{{Shen} et~al.}{2020}]{Shen2020}
{Shen} X.,  {Hopkins} P.~F.,  {Faucher-Gigu{\`e}re} C.-A.,  {Alexander} D.~M.,
  {Richards} G.~T.,  {Ross} N.~P.,   {Hickox} R.~C.,  2020, \mn@doi [\mnras]
  {10.1093/mnras/staa1381}, \href
  {https://ui.adsabs.harvard.edu/abs/2020MNRAS.495.3252S} {495, 3252}

\bibitem[\protect\citeauthoryear{{Sijacki}, {Springel}, {Di Matteo}  \&
  {Hernquist}}{{Sijacki} et~al.}{2007}]{Sijacki2007}
{Sijacki} D.,  {Springel} V.,  {Di Matteo} T.,   {Hernquist} L.,  2007, \mn@doi
  [\mnras] {10.1111/j.1365-2966.2007.12153.x}, \href
  {https://ui.adsabs.harvard.edu/abs/2007MNRAS.380..877S} {380, 877}

\bibitem[\protect\citeauthoryear{{Sijacki}, {Vogelsberger}, {Genel},
  {Springel}, {Torrey}, {Snyder}, {Nelson}  \& {Hernquist}}{{Sijacki}
  et~al.}{2015}]{Sijacki2015}
{Sijacki} D.,  {Vogelsberger} M.,  {Genel} S.,  {Springel} V.,  {Torrey} P.,
  {Snyder} G.~F.,  {Nelson} D.,   {Hernquist} L.,  2015, \mn@doi [\mnras]
  {10.1093/mnras/stv1340}, \href
  {https://ui.adsabs.harvard.edu/abs/2015MNRAS.452..575S} {452, 575}

\bibitem[\protect\citeauthoryear{{Silk} \& {Rees}}{{Silk} \&
  {Rees}}{1998}]{Silk1998}
{Silk} J.,  {Rees} M.~J.,  1998, \aap, \href
  {https://ui.adsabs.harvard.edu/abs/1998A&A...331L...1S} {331, L1}

\bibitem[\protect\citeauthoryear{{Silverman} et~al.,}{{Silverman}
  et~al.}{2008}]{Silverman2008}
{Silverman} J.~D.,  et~al., 2008, \mn@doi [\apj] {10.1086/529572}, \href
  {https://ui.adsabs.harvard.edu/abs/2008ApJ...679..118S} {679, 118}

\bibitem[\protect\citeauthoryear{{Smol{\v{c}}i{\'c}}
  et~al.,}{{Smol{\v{c}}i{\'c}} et~al.}{2017}]{Smolcic2017}
{Smol{\v{c}}i{\'c}} V.,  et~al., 2017, \mn@doi [\aap]
  {10.1051/0004-6361/201730685}, \href
  {https://ui.adsabs.harvard.edu/abs/2017A&A...602A...6S} {602, A6}

\bibitem[\protect\citeauthoryear{{So\l{}tan}}{{So\l{}tan}}{1982}]{Soltan1982}
{So\l{}tan} A.,  1982, \mn@doi [\mnras] {10.1093/mnras/200.1.115}, \href
  {https://ui.adsabs.harvard.edu/abs/1982MNRAS.200..115S} {200, 115}

\bibitem[\protect\citeauthoryear{{Somerville}, {Hopkins}, {Cox}, {Robertson}
  \& {Hernquist}}{{Somerville} et~al.}{2008}]{Somerville2008}
{Somerville} R.~S.,  {Hopkins} P.~F.,  {Cox} T.~J.,  {Robertson} B.~E.,
  {Hernquist} L.,  2008, \mn@doi [\mnras] {10.1111/j.1365-2966.2008.13805.x},
  \href {https://ui.adsabs.harvard.edu/abs/2008MNRAS.391..481S} {391, 481}

\bibitem[\protect\citeauthoryear{{Tillman}, {Wellons}, {Faucher-Gigu{\`e}re},
  {Kelley}  \& {Angl{\'e}s-Alc{\'a}zar}}{{Tillman} et~al.}{2022}]{Tillman2022}
{Tillman} M.~T.,  {Wellons} S.,  {Faucher-Gigu{\`e}re} C.-A.,  {Kelley} L.~Z.,
   {Angl{\'e}s-Alc{\'a}zar} D.,  2022, \mn@doi [\mnras]
  {10.1093/mnras/stac398}, \href
  {https://ui.adsabs.harvard.edu/abs/2022MNRAS.511.5756T} {511, 5756}

\bibitem[\protect\citeauthoryear{{Tinker}, {Kravtsov}, {Klypin}, {Abazajian},
  {Warren}, {Yepes}, {Gottl{\"o}ber}  \& {Holz}}{{Tinker}
  et~al.}{2008}]{Tinker2008}
{Tinker} J.,  {Kravtsov} A.~V.,  {Klypin} A.,  {Abazajian} K.,  {Warren} M.,
  {Yepes} G.,  {Gottl{\"o}ber} S.,   {Holz} D.~E.,  2008, \mn@doi [\apj]
  {10.1086/591439}, \href
  {https://ui.adsabs.harvard.edu/abs/2008ApJ...688..709T} {688, 709}

\bibitem[\protect\citeauthoryear{{Tremaine} et~al.,}{{Tremaine}
  et~al.}{2002}]{Tremaine2002}
{Tremaine} S.,  et~al., 2002, \mn@doi [\apj] {10.1086/341002}, \href
  {https://ui.adsabs.harvard.edu/abs/2002ApJ...574..740T} {574, 740}

\bibitem[\protect\citeauthoryear{{Tremmel}}{{Tremmel}}{2017}]{Tremmel2017PhDT}
{Tremmel} M.,  2017, PhD thesis, University of Washington, Seattle

\bibitem[\protect\citeauthoryear{{Ueda}, {Akiyama}, {Hasinger}, {Miyaji}  \&
  {Watson}}{{Ueda} et~al.}{2014}]{Ueda2014}
{Ueda} Y.,  {Akiyama} M.,  {Hasinger} G.,  {Miyaji} T.,   {Watson} M.~G.,
  2014, \mn@doi [\apj] {10.1088/0004-637X/786/2/104}, \href
  {https://ui.adsabs.harvard.edu/abs/2014ApJ...786..104U} {786, 104}

\bibitem[\protect\citeauthoryear{{Vogelsberger} et~al.,}{{Vogelsberger}
  et~al.}{2014}]{Vogelsberger2014}
{Vogelsberger} M.,  et~al., 2014, \mn@doi [\mnras] {10.1093/mnras/stu1536},
  \href {https://ui.adsabs.harvard.edu/abs/2014MNRAS.444.1518V} {444, 1518}

\bibitem[\protect\citeauthoryear{{Volonteri}, {Dubois}, {Pichon}  \&
  {Devriendt}}{{Volonteri} et~al.}{2016}]{Volonteri2016}
{Volonteri} M.,  {Dubois} Y.,  {Pichon} C.,   {Devriendt} J.,  2016, \mn@doi
  [\mnras] {10.1093/mnras/stw1123}, \href
  {https://ui.adsabs.harvard.edu/abs/2016MNRAS.460.2979V} {460, 2979}

\bibitem[\protect\citeauthoryear{{Weinberger} et~al.,}{{Weinberger}
  et~al.}{2017}]{Weinberger2017}
{Weinberger} R.,  et~al., 2017, \mn@doi [\mnras] {10.1093/mnras/stw2944}, \href
  {https://ui.adsabs.harvard.edu/abs/2017MNRAS.465.3291W} {465, 3291}

\bibitem[\protect\citeauthoryear{{Yang} et~al.,}{{Yang}
  et~al.}{2018}]{Yang2018}
{Yang} G.,  et~al., 2018, \mn@doi [\mnras] {10.1093/mnras/stx2805}, \href
  {https://ui.adsabs.harvard.edu/abs/2018MNRAS.475.1887Y} {475, 1887}

\bibitem[\protect\citeauthoryear{{Yang} et~al.,}{{Yang}
  et~al.}{2021}]{Yang2021}
{Yang} J.,  et~al., 2021, \mn@doi [\apj] {10.3847/1538-4357/ac2b32}, \href
  {https://ui.adsabs.harvard.edu/abs/2021ApJ...923..262Y} {923, 262}

\bibitem[\protect\citeauthoryear{{Yu} \& {Tremaine}}{{Yu} \&
  {Tremaine}}{2002}]{Yu2002}
{Yu} Q.,  {Tremaine} S.,  2002, \mn@doi [\mnras]
  {10.1046/j.1365-8711.2002.05532.x}, \href
  {https://ui.adsabs.harvard.edu/abs/2002MNRAS.335..965Y} {335, 965}

\bibitem[\protect\citeauthoryear{{Zhang}, {Behroozi}, {Volonteri}, {Silk},
  {Fan}, {Hopkins}, {Yang}  \& {Aird}}{{Zhang} et~al.}{2023a}]{Zhang2023b}
{Zhang} H.,  {Behroozi} P.,  {Volonteri} M.,  {Silk} J.,  {Fan} X.,  {Hopkins}
  P.~F.,  {Yang} J.,   {Aird} J.,  2023a, \mn@doi [arXiv e-prints]
  {10.48550/arXiv.2303.08150}, \href
  {https://ui.adsabs.harvard.edu/abs/2023arXiv230308150Z} {p. arXiv:2303.08150}

\bibitem[\protect\citeauthoryear{{Zhang} et~al.,}{{Zhang}
  et~al.}{2023b}]{Zhang2023c}
{Zhang} H.,  et~al., 2023b, \mn@doi [arXiv e-prints]
  {10.48550/arXiv.2309.07210}, \href
  {https://ui.adsabs.harvard.edu/abs/2023arXiv230907210Z} {p. arXiv:2309.07210}

\bibitem[\protect\citeauthoryear{{Zhang}, {Behroozi}, {Volonteri}, {Silk},
  {Fan}, {Hopkins}, {Yang}  \& {Aird}}{{Zhang} et~al.}{2023c}]{Zhang2023a}
{Zhang} H.,  {Behroozi} P.,  {Volonteri} M.,  {Silk} J.,  {Fan} X.,  {Hopkins}
  P.~F.,  {Yang} J.,   {Aird} J.,  2023c, \mn@doi [\mnras]
  {10.1093/mnras/stac2633}, \href
  {https://ui.adsabs.harvard.edu/abs/2023MNRAS.518.2123Z} {518, 2123}

\makeatother
\end{thebibliography}

\appendix

\section{Radiative and total efficiencies}
\label{a:rad_kin_edd}
\subsection{The fiducial scaling relation between $\epsilon_\mathrm{rad}$ and $\epsilon_\mathrm{tot}$}
\label{aa:fiducial}
\begin{figure}
\includegraphics[width=0.48\textwidth]{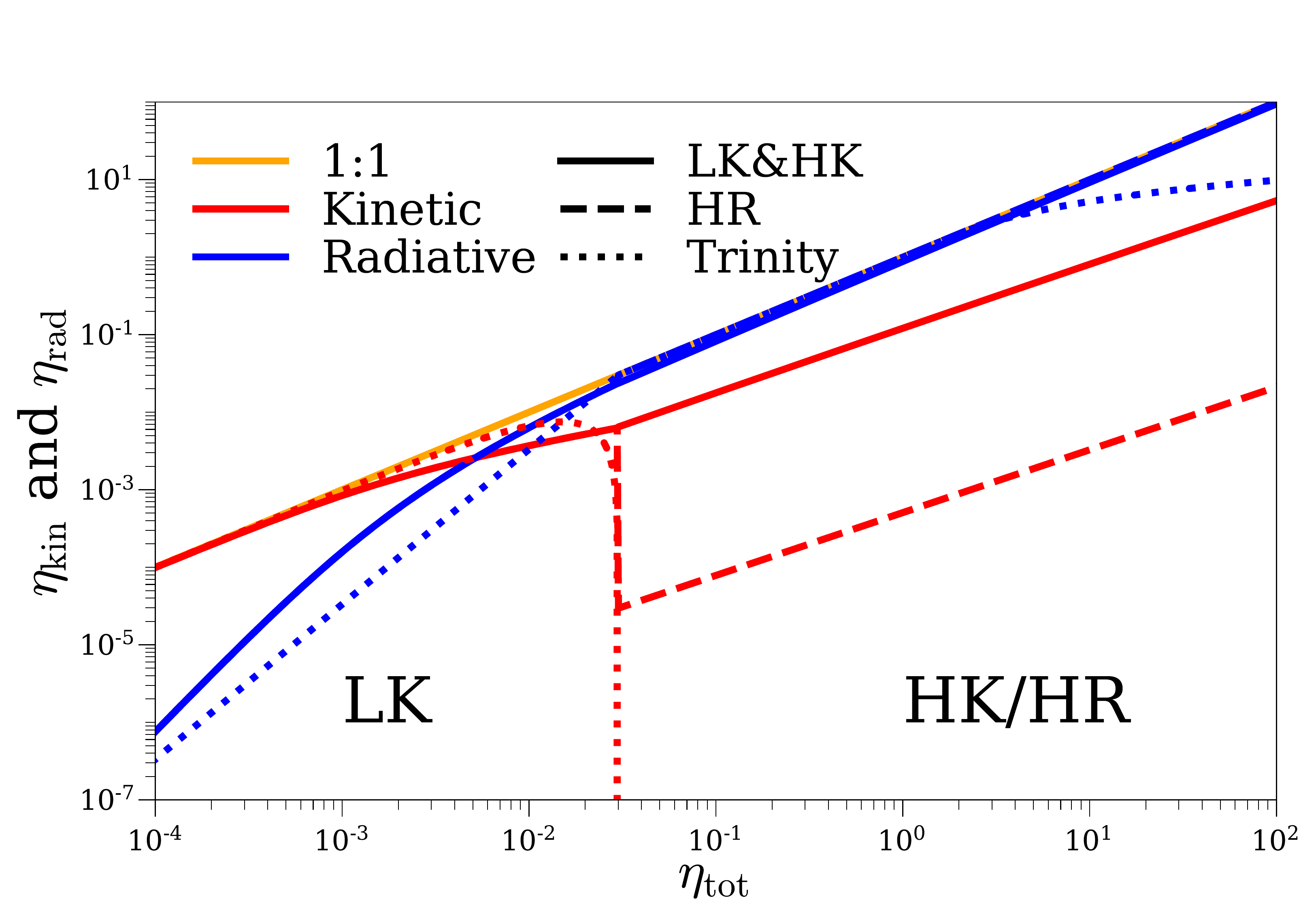}
\caption{The comparison between the $\eta_\mathrm{rad}$-$\eta_\mathrm{kin}$-$\eta_\mathrm{tot}$ scaling relations used in \textsc{Trinity} and \citet[][]{Merloni2008}. The orange solid line is the 1:1 equality line. Three AGN accretion modes are defined by \citet{Merloni2008}: the ``low kinetic'' (LK, solid lines) mode with $\eta_\mathrm{tot} \leq 0.03$, the ``high kinetic'' (HK, solid lines) mode, and the ``high radiative'' (HR, dashed lines) mode. An AGN with $\eta_\mathrm{tot} > 0.03$ is in either the HK or HR mode. $\eta_\mathrm{rad}$ and $\eta_\mathrm{kin}$ are shown in blue and red curves, respectively. The scaling relations used in \textsc{Trinity} are shown in dotted lines.}
\label{f:compare_mer}
\end{figure}

In \textsc{Trinity}, we adopt a non-linear scaling relation between radiative and kinetic components of AGN energy efficiency, $\epsilon_\mathrm{rad}$ and $\epsilon_\mathrm{kin}$, and the total energy efficiency, $\epsilon_\mathrm{tot}$. This is motivated by the fact that active SMBHs convert gravitational energy not only into radiation, but also into kinetic energy in the form of jets and/or outflows. The scaling relation is as follows:

\begin{equation}
\label{e:eff_tot_rad}
    \epsilon_{\rm rad} = \epsilon_\mathrm{tot}\times
    \begin{cases}
      \eta/\eta_\mathrm{crit}, & \eta \leq \eta_\mathrm{crit}\\
      1, & \eta_\mathrm{crit} < \eta \leq 2\\
      2/\eta \left[1 + \ln(\eta/2)\right], & \eta > 2\\
      
    \end{cases}\ ,
\end{equation}
and:
\begin{equation}
\label{e:eff_kin_rad}
    \epsilon_{\rm kin} = 
    \begin{cases}
        \epsilon_\mathrm{tot}(1 - \eta/0.03), & \eta \leq \eta_\mathrm{crit}\\
        0, & \eta > \eta_\mathrm{crit}\\
    \end{cases}\ ,
\end{equation}
where $\eta_\mathrm{crit}=0.03$. This is also equivalent to defining radiative and kinetic components of Eddington ratio, $\eta_\mathrm{rad}$ and $\eta_\mathrm{kin}$, so that $\eta = \eta_{\rm rad} + \eta_{\rm kin}$:

\begin{equation}
\label{e:eta_tot_rad}
    \eta_{\rm rad} = 
    \begin{cases}
      \eta^2/\eta_\mathrm{crit}, & \eta \leq \eta_\mathrm{crit}\\
      \eta, & \eta_\mathrm{crit} < \eta \leq 2\\
      2\left[1 + \ln(\eta/2)\right], & \eta > 2\\
      
    \end{cases}\ ,
\end{equation}
and
\begin{equation}
\label{e:erdf_kin_rad}
    \eta_{\rm kin} = 
    \begin{cases}
                \eta - \eta^2 / \eta_\mathrm{crit}\ ,\ &\eta \leq \eta_\mathrm{crit}\\ 
                0, &\eta > \eta_\mathrm{crit}\\
    \end{cases}\ .
\end{equation}
As is shown in Fig.\ \ref{f:compare_mer}, the scaling between $\eta_\mathrm{rad}$ and $\eta$ is similar to the one used by \citet{Merloni2008} at $\eta \leq 2$, which was based on the observed scaling found by \citet{Merloni2007}. But \citet{Merloni2008} adopted multiple scaling relations between $\eta$, $\eta_\mathrm{rad}$, and $\eta_\mathrm{kin}$, each describing a different AGN accretion and feedback mode. At $\eta > \eta_\mathrm{crit}$ the ``high kinetic'' (HK, solid lines) mode features similar kinetic energy and radiation outputs, and radiation dominates over kinetic energy in the ``high radiative'' mode (HR, dashed lines). \citet[][]{Merloni2008} further assumed that the ratio of HK and HR quasars is 1:9. The $\eta_\mathrm{rad}$-$\eta$ relations were derived with observed scaling relations between AGN radio and X-ray luminosities, and SMBH masses, which came with substantial scatter. In light of this, we choose to stick to the simpler scaling in Eq.\ \ref{e:eta_tot_rad}. For $\eta \geq 2$, we adopt a logarithmic function to account for the part of the outgoing radiation is trapped by the thickened accretion disk \citep{Mineshige2000}. At a given $\eta<\eta_\mathrm{crit}$ (``low kinetic'' mode, or LK in \citealt{Merloni2008}), Eq.\ \ref{e:erdf_kin_rad} produces more kinetic energy than \citet{Merloni2008}. We also ignore the kinetic energy output from active SMBHs with $\eta > \eta_\mathrm{crit}$, due to the lack of observational constraints. 

\subsection{Variants of the $\epsilon_\mathrm{rad}$--$\epsilon_\mathrm{tot}$ relation}
\label{aa:variants}
To account for the systematic uncertainties induced by our assumed $\epsilon_\mathrm{rad}$--$\epsilon_\mathrm{tot}$ relation, we also reran \textsc{Trinity} with two different variants of the $\epsilon_\mathrm{rad}$--$\epsilon_\mathrm{tot}$ relation: 1) the one where $\epsilon_\mathrm{rad}\equiv \epsilon_\mathrm{tot}$, i.e., no kinetic energy is produced from SMBH accretion (``no kinetic'' hereafter), and 2) the one where $\eta_\mathrm{crit}=0.1$, i.e., more kinetic energy is generated in SMBH accretion (``more kinetic'' hereafter). These two variants represent the two extremes of kinetic energy output in SMBH accretion. All the qualitative conclusions remain unchanged with both variant models. To save space, we uploaded Figs.\ \ref{f:high_z_qlf}-\ref{f:f_super_edd} from these two variant models \href{https://github.com/HaowenZhang/TRINITY/tree/main/plots/Paper3/Model\_Variants}{https://github.com/HaowenZhang/TRINITY/tree/main/plots/Paper3/Model\\\_Variants.}

\end{document}